\documentclass[twocolumn]{aastex61}
\usepackage{comment}

\usepackage{float}

\newcommand{\unit}[1]{\ensuremath{\, \mathrm{#1}}}
\newcommand\iab{i_{\mathrm{AB}}}

\received{December 21, 2018}
\revised{January 30, 2019}
\accepted{February 2, 2019}
\submitjournal{Astronomical Journal}

\shorttitle{The PAU Camera}
\shortauthors{The PAUCam Collaboration}

\begin{document}

\title{The Physics of the Accelerating Universe Camera}

\correspondingauthor{Crist\'obal Padilla}
\email{Cristobal.Padilla@ifae.es}

\author{Crist\'obal Padilla}
\affiliation{Institut de F\'\i sica d'Altes Energies (IFAE), The Barcelona Institute of Science and Technology, 08193 Bellaterra (Barcelona), Spain}

\author{Francisco J. Castander}
\affiliation{Institute of Space Sciences (ICE, CSIC), Campus UAB, Carrer de Can Magrans, s/n, 08193 Bellaterra (Barcelona), Spain}
\affiliation{Institut d'Estudis Espacials de Catalunya (IEEC), 08034 Barcelona, Spain}

\author{Alex Alarc\'on}
\affiliation{Institute of Space Sciences (ICE, CSIC), Campus UAB, Carrer de Can Magrans, s/n, 08193 Bellaterra (Barcelona), Spain}
\affiliation{Institut d'Estudis Espacials de Catalunya (IEEC), 08034 Barcelona, Spain}

\author{Jelena Aleksic}
\affiliation{Institut de F\'\i sica d'Altes Energies (IFAE), The Barcelona Institute of Science and Technology, 08193 Bellaterra (Barcelona), Spain}

\author{Otger Ballester}
\affiliation{Institut de F\'\i sica d'Altes Energies (IFAE), The Barcelona Institute of Science and Technology, 08193 Bellaterra (Barcelona), Spain}

\author{Laura Cabayol}
\affiliation{Institut de F\'\i sica d'Altes Energies (IFAE), The Barcelona Institute of Science and Technology, 08193 Bellaterra (Barcelona), Spain}

\author{Laia Cardiel-Sas}
\affiliation{Institut de F\'\i sica d'Altes Energies (IFAE), The Barcelona Institute of Science and Technology, 08193 Bellaterra (Barcelona), Spain}

\author{Jorge Carretero}
\altaffiliation{Also at Port d'Informaci\'o Cient\'\i fica (PIC), Campus UAB, C. Albareda s/n, 08193 Bellaterra (Barcelona), Spain}
\affiliation{Institut de F\'\i sica d'Altes Energies (IFAE), The Barcelona Institute of Science and Technology, 08193 Bellaterra (Barcelona), Spain}

\author{Ricard Casas}
\affiliation{Institute of Space Sciences (ICE, CSIC), Campus UAB, Carrer de Can Magrans, s/n, 08193 Bellaterra (Barcelona), Spain}
\affiliation{Institut d'Estudis Espacials de Catalunya (IEEC), 08034 Barcelona, Spain}

\author{Javier Castilla}
\affiliation{Centro de Investigaciones Energ\'eticas, Medioambientales y Tecnol\'ogicas (CIEMAT), Avenida Complutense 40, 28040 Madrid, Spain}

\author{Martin Crocce}
\affiliation{Institute of Space Sciences (ICE, CSIC), Campus UAB, Carrer de Can Magrans, s/n, 08193 Bellaterra (Barcelona), Spain}
\affiliation{Institut d'Estudis Espacials de Catalunya (IEEC), 08034 Barcelona, Spain}

\author{Manuel Delfino}
\altaffiliation{Also at Port d'Informaci\'o Cient\'\i fica (PIC), Campus UAB, C. Albareda s/n, 08193 Bellaterra (Barcelona), Spain}
\affiliation{Institut de F\'\i sica d'Altes Energies (IFAE), The Barcelona Institute of Science and Technology, 08193 Bellaterra (Barcelona), Spain}

\author{Carlos D\'iaz}
\affiliation{Centro de Investigaciones Energ\'eticas, Medioambientales y Tecnol\'ogicas (CIEMAT), Avenida Complutense 40, 28040 Madrid, Spain}

\author{Martin Eriksen}
\altaffiliation{Also at Port d'Informaci\'o Cient\'\i fica (PIC), Campus UAB, C. Albareda s/n, 08193 Bellaterra (Barcelona), Spain}
\affiliation{Institut de F\'\i sica d'Altes Energies (IFAE), The Barcelona Institute of Science and Technology, 08193 Bellaterra (Barcelona), Spain}

\author{Enrique Fern\'andez}
\affiliation{Institut de F\'\i sica d'Altes Energies (IFAE), The Barcelona Institute of Science and Technology, 08193 Bellaterra (Barcelona), Spain}

\author{Pablo Fosalba}
\affiliation{Institute of Space Sciences (ICE, CSIC), Campus UAB, Carrer de Can Magrans, s/n, 08193 Bellaterra (Barcelona), Spain}
\affiliation{Institut d'Estudis Espacials de Catalunya (IEEC), 08034 Barcelona, Spain}

\author{Juan Garc\'ia-Bellido}
\affiliation{Instituto de F\'\i sica Te\'orica (IFT-UAM/CSIC), Universidad Aut\'onoma de Madrid, 28049 Madrid, Spain}

\author{Enrique Gazta\~naga}
\affiliation{Institute of Space Sciences (ICE, CSIC), Campus UAB, Carrer de Can Magrans, s/n, 08193 Bellaterra (Barcelona), Spain}
\affiliation{Institut d'Estudis Espacials de Catalunya (IEEC), 08034 Barcelona, Spain}

\author{Javier Gaweda}
\affiliation{Institut de F\'\i sica d'Altes Energies (IFAE), The Barcelona Institute of Science and Technology, 08193 Bellaterra (Barcelona), Spain}

\author{Ferran Gra\~nena}
\affiliation{Institut de F\'\i sica d'Altes Energies (IFAE), The Barcelona Institute of Science and Technology, 08193 Bellaterra (Barcelona), Spain}

\author{Jos\'e Mar\'\i a \'Illa}
\affiliation{Institut de F\'\i sica d'Altes Energies (IFAE), The Barcelona Institute of Science and Technology, 08193 Bellaterra (Barcelona), Spain}

\author{Jorge Jim\'enez}
\affiliation{Institut de F\'\i sica d'Altes Energies (IFAE), The Barcelona Institute of Science and Technology, 08193 Bellaterra (Barcelona), Spain}

\author{Luis L\'opez}
\altaffiliation{Now at European XFEL GmbH, Holzkoppel, 4, 22869 Schenefeld,
Germany}
\affiliation{Institut de F\'\i sica d'Altes Energies (IFAE), The Barcelona Institute of Science and Technology, 08193 Bellaterra (Barcelona), Spain}

\author{Pol Mart\'\i}
\affiliation{Institut de F\'\i sica d'Altes Energies (IFAE), The Barcelona Institute of Science and Technology, 08193 Bellaterra (Barcelona), Spain}

\author{Ramon Miquel}
\affiliation{Institut de F\'\i sica d'Altes Energies (IFAE), The Barcelona Institute of Science and Technology, 08193 Bellaterra (Barcelona), Spain}
\affiliation{Instituci\'o Catalana de Recerca i Estudis Avan\c cats (ICREA), 08010 Barcelona, Spain}

\author{Christian Neissner}
\altaffiliation{Also at Port d'Informaci\'o Cient\'\i fica (PIC), Campus UAB, C. Albareda s/n, 08193 Bellaterra (Barcelona), Spain}
\affiliation{Institut de F\'\i sica d'Altes Energies (IFAE), The Barcelona Institute of Science and Technology, 08193 Bellaterra (Barcelona), Spain}

\author{Crist\'obal P\'io}
\affiliation{Institut de F\'\i sica d'Altes Energies (IFAE), The Barcelona Institute of Science and Technology, 08193 Bellaterra (Barcelona), Spain}

\author{Eusebio S\'anchez}
\affiliation{Centro de Investigaciones Energ\'eticas, Medioambientales y Tecnol\'ogicas (CIEMAT), Avenida Complutense 40, 28040 Madrid, Spain}

\author{Santiago Serrano}
\affiliation{Institute of Space Sciences (ICE, CSIC), Campus UAB, Carrer de Can Magrans, s/n, 08193 Bellaterra (Barcelona), Spain}
\affiliation{Institut d'Estudis Espacials de Catalunya (IEEC), 08034 Barcelona, Spain}

\author{Ignacio Sevilla-Noarbe}
\affiliation{Centro de Investigaciones Energ\'eticas, Medioambientales y Tecnol\'ogicas (CIEMAT), Avenida Complutense 40, 28040 Madrid, Spain}

\author{Pau Tallada}
\altaffiliation{Also at Port d'Informaci\'o Cient\'\i fica (PIC), Campus UAB, C. Albareda s/n, 08193 Bellaterra (Barcelona), Spain}
\affiliation{Centro de Investigaciones Energ\'eticas, Medioambientales y Tecnol\'ogicas (CIEMAT), Avenida Complutense 40, 28040 Madrid, Spain}

\author{Nadia Tonello}
\altaffiliation{Also at Port d'Informaci\'o Cient\'\i fica (PIC), Campus UAB, C. Albareda s/n, 08193 Bellaterra (Barcelona), Spain}
\affiliation{Institut de F\'\i sica d'Altes Energies (IFAE), The Barcelona Institute of Science and Technology, 08193 Bellaterra (Barcelona), Spain}

\author{Juan de Vicente}
\affiliation{Centro de Investigaciones Energ\'eticas, Medioambientales y Tecnol\'ogicas (CIEMAT), Avenida Complutense 40, 28040 Madrid, Spain}



\begin{abstract}
The PAU (Physics of the Accelerating Universe) Survey 
goal is to obtain photometric redshifts (photo-z) and Spectral Energy Distribution (SED) of astronomical objects with a resolution roughly one order of magnitude better than current broad band photometric surveys. To accomplish this, a new large field of view camera (PAUCam) has been designed, built, commissioned and is now operated at the William Herschel Telescope (WHT). With the current WHT Prime Focus corrector, the camera covers  $\sim$1~degree diameter Field of View (FoV), of which, only the inner $\sim$40~arcmin diameter are  unvignetted. The focal plane consists of a mosaic of 18 2k$\times$4k Hamamatsu fully depleted CCDs, with high quantum efficiency up to $1 \unit{\micron}$ in wavelength. To maximize the detector coverage within the FoV, filters are placed in front of the CCDs inside the camera cryostat (made out of carbon fiber) using a challenging movable tray system. The camera uses a set of 40 narrow band filters ranging from $\sim$4500 to $\sim$8500 Angstroms complemented with six standard broad-band filters, $ugrizY$. The PAU Survey aims to cover roughly 100 square degrees over fields with existing deep photometry and galaxy shapes to obtain accurate photometric redshifts for galaxies down to $\iab\!\sim\! 22.5$, detecting also galaxies down to  $\iab\!\sim\! 24$ with less precision in redshift. With this data set we will be able to measure intrinsic alignments, galaxy clustering and perform galaxy evolution studies in a new range of densities and redshifts. Here, we describe the PAU camera, its first commissioning results and performance. 
\end{abstract}

\keywords{Instrumentation: photometers ---techniques: photometric --- surveys --- dark energy --- large-scale structure of universe}



\section{INTRODUCTION} 
\label{sec:intro}


There has been a tremendous observational effort to carry out large wide field surveys to better understand our universe in the last years. For cosmological studies, one needs to locate the exact position of the galaxy tracers to unveil the large scale structure of the universe. These wide field surveys can be performed imaging the sky or with spectroscopy. Wide field imaging surveys (e.g., 
SDSS\footnote{\label{note1}\url{http://www.sdss.org/}}, DES\footnote{\url{http://www.darkenergysurvey.org/}}, KiDS\footnote{\url{http://kids.strw.leidenuniv.nl/}}) are normally conducted with wide field imaging cameras using a given set of broad band filters. They are very efficient in the sense that they can cover large areas of the sky and sample large numbers of galaxies down to faint magnitudes. However, they only sample the spectral energy distribution (SED) of galaxies very coarsely and therefore cannot infer their redshifts with precision, thus hampering the exact 3D location of the galaxy. Typical broad band filters, reaching a resolution of $R\sim5$, can achieve precisions of the order of a few percent in the galaxy photometric redshift determination. On the other hand, spectroscopic surveys (e.g., SDSS, 2dFGRS\footnote{\url{http://www.2dfgrs.net/}}, GAMA\footnote{\url{http://www.gama-survey.org/}}, VIPERS\footnote{\url{http://vipers.inaf.it/}}, VVDS\footnote{\url{http://cesam.lam.fr/vvds/}}, DEEP2\footnote{\url{http://deep.ucolick.org/}}, zCOSMOS\footnote{\url{http://cesam.lam.fr/zCosmos/}}) sample the galaxies SEDs with higher resolution and can determine their redshifts with high precision. However, given that the light is dispersed, they cannot reach the depths of imaging surveys. Moreover, although the multiplexing capabilities of large spectrographs have increased considerably recently, still the number of objects sampled is limited to those that are targeted. The exception being spectroscopic sliless surveys (e.g., PRIMUS\footnote{\url{https://primus.ucsd.edu/}}) where all objects are dispersed  at the expense of a much larger background that limits the depth that can be reached.

Wide surveys with narrow band filters can bridge the gap between these two observing modes. Narrow band photometry can provide similar number of observed objects as those in broad band photometry, but sampling the SEDs with higher resolution and therefore improving the photo-z precision by an order of magnitude although covering a smaller area for a given observation time. There are many cosmological studies that do not require high redhsift determination precision but need large numbers to reach the statistical significance needed to sample the underlying cosmic structures and their evolution. Examples of scientific problems that can greatly profit from the advantages of narrow band photometry include the study of the intrinsic alignment of galaxies \citep{ref:IA1, ref:IA2} and galaxy clustering at intermediate scales \citep{ref:Cluster1, ref:Cluster2}. Narrow band imaging has been explored previously with medium band observation by  the COMBO-17 \citep{ref:combo-17_1, ref:combo-17_2, ref:combo-17_3} and ALHAMBRA \citep{ref:Alhambra1, ref:Alhambra2} surveys.

To push this technique one step further, we have built a new narrow band photometer: the Physics of the Accelerating Universe (PAU) Camera (PAUCam). PAUcam has been designed to operate at the prime focus of 4.2-meter William Herschel Telescope (WHT) of the Isaac Newton Group (ING) at the Observatory of the Roque de los Muchachos in the Canary Islands. PAUCam is a wide field camera covering the field of view (FoV) delivered by the prime focus corrector of the WHT. It can image in both broad band and narrow band filters. PAUCam was designed to carry out the PAU Survey\footnote{\url{http://www.pausurvey.org/}} \citep{gazta12}, but can also be used to reach other scientific goals. Besides its scientific capabilities, PAUCam is a unique instrument in several technological areas. In particular, it has a filter exchange movable system inside the cryostat vessel, which is made of carbon fiber. To our knowledge, this is the only wide field optical camera made with this material. PAUCam saw first light in June 2015. Since then, it has been periodically taking observations at the WHT. In this paper, we describe the PAUcam instrument, its performance and the PAU Survey operation. 

This paper is outlined as follows. Section~\ref{sec:science} describes the scientific goals of the PAU Survey. Section~\ref{sec:designspecs} details what are the instrument features needed to accomplish its scientific goals. In Section~\ref{sec:Camera},  the elements of the PAU Camera are described as we traverse it following the light path. The next three sections focus on the three software aspects that any instrument needs to function adequately: the Slow Control system (Section~\ref{sec:slowcontrol}), the general control and Data Acquisition (Section~\ref{sec:PAUControlSystem}) and the online analysis and monitoring software (Section~\ref{sec:OnlineAnalysisSw}). Section~\ref{sec:datareduction} describes the Data Reduction software used after images have been taken. Section~\ref{sec:commissioning} describes the commissioning of the camera and its results during the first light night. Section~\ref{sec:operation} details how we implement the survey strategy while operating the camera. The camera performance is summarized in Section~\ref{sec:performance}. Finally, in Section~\ref{sec:summary} we present a summary of the whole paper.

\section{SCIENCE GOALS}
\label{sec:science}

\begin{figure}[tb]
\epsscale{1.13}
\plotone{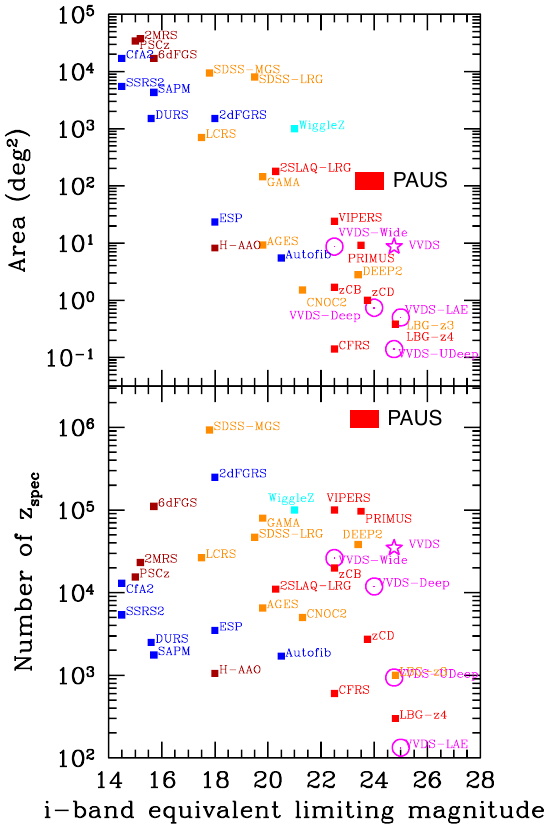}
\caption{ \label{fig:iband} 
Comparison of the covered area (top) and the number of measured spectroscopic redshifts (bottom) versus depth expressed as the equivalent i-band limiting magnitude, between the VVDS (magenta) and different spectroscopic redshift surveys. PAUS will be a few magnitudes deeper (over 10 times denser) than current completed large area flux limited surveys (such as Gama) and 10 times larger than current completed deep flux limited surveys (such as VVDS/VIPERS/Deep2)\citep{ref:Fevre}. }
\end{figure}

The PAU Camera was designed to carry out the PAU Survey (PAUS). The PAU survey wants to take advantage of the capabilities enabled by narrow band imaging in a wide field, reaching sub-percent photometric redshift precision for deep magnitude-limited galaxy samples. 
PAUS aims to observe $\sim 100\unit{deg}^2$ down to $\iab< 22.5$ reaching a volume of $0.3 \unit{(Gpc/h)^3}$ with a few million redshifts. This will allow us to reach sub-percent statistical errors on density measurements out to $10 \unit{Mpc/h}$ on both transverse and radial distances. To expand the science reach of PAUS, whilst making optimal use of existing high-quality data, we plan to survey 3 to 4 wide fields (of several tens of $\unit{deg^2}$ each), in which shape and deep Broad Band (BB) photometric measurements have already been obtained by CFHTLenS \citep{miller13, erben13}, KiDS \citep{dejong17} and/or DES \citep{drlicawagner18, zuntz18}, and in which spectroscopic redshifts for calibration are available from SDSS \citep{york00, abolfathi18}, GAMA \citep{ref:GAMA, baldry18}, VIPERS \citep{ref:VIPERS, scodeggio18}, DEEP2 \citep{ref:DEEP2} and zCOSMOS \citep{lilly09} redshift surveys. As shown in Figure \ref{fig:iband} this will result in a unique survey that will fill a competitive gap compared to current surveys. PAUS will measure 3 million redshifts to  $\iab < 22.5$ (30 times more than currently available). PAUS will provide one of the most detailed studies of intermediate-scale cosmic structure analysis ever undertaken. For most studies the photometric redshift accuracy that we can achieve is sufficient, because it is not necessary to know whether the systems are gravitationally bound, but rather which galaxies are physically close to each other. For instance, the intrinsic alignments galaxies, which arise from local tidal fields, complicate a simple interpretation of the observed alignments of galaxies in cosmic shear surveys such as KiDS \citep{kilbinger13}, DES \citep{troxel18} and Euclid \citep{laureijs11}. Even though PAUS will cover a modest area compared to large wide imaging surveys, PAUS will increase the number density of galaxies with redshifts by nearly two orders of magnitude to tens of thousands of redshifts per square degree and thus enable a first precise measurement of intrinsic alignments at $z \sim 0.75$. 

The PAUS program avoids the target selection issues present in every spectroscopic survey, enabling us to study and calibrate target selection incompleteness of other surveys. Therefore, the PAUS data can be used to improve and model target selection for DESI \citep{desi16} and WEAVE \citep{weave18} and can improve photo-z estimates for KiDS, DES, LSST \citep{lsst08} and Euclid. 

Errors in redshift translate into luminosity errors, given the flux measurements. All other physical quantities are also affected by the error in the measured distances inferred from the measured redshifts. At $z = 0.5$ the typical broadband photo-z error of $\sigma_{\rm 68}/ (1+z)  \simeq 0.05$  translates into $40\%$ error in the luminosity (or 171 Mpc/h in distance) while the PAUS photo-z error $\sigma_{ \rm 68}/ (1+z)  \simeq 0.0035$
corresponds to $2.5\%$ (or $12 \unit{Mpc/h}$ in distance), comparable to other sources of errors (e.g. errors in flux or calibration). Typical spectroscopic errors are three times smaller which are better than needed, given current photometric and calibration errors of few percent. Moreover PAUS has the potential to estimate redshifts in the full sample, without the need of the targeting selection required in spectroscopic surveys. All this makes PAUS unique to study clustering in the weakly non-linear regime and galaxy evolution.

Another advantage of PAUS is the ability to calibrate Spectral Energy Distribution (SED) fluxes, a task that is more complicated (and often not attempted) in spectra. This allows for measurements of total and relative flux in SED features, such as the 4000~\AA\, break and emission lines. Physical modeling and template modeling of SED properties of galaxies and stars as a function of redshift will help improve photo-z codes and will enable environment and morphology studies of galaxy evolution. 

During a good observing night, PAUS is able to provide low-resolution spectra ($\sigma_{\rm 68}(z) \simeq 0.0035(1+z)$) for more than 30,000 galaxies, 5,000 stars, 1,000 quasars, 10 galaxy clusters. With a small number of nights one could measure spectral energy distributions (SED) of moderate resolution for a very large sample of galaxies.

\section{DESIGN SPECIFICATIONS}
\label{sec:designspecs}

\begin{figure}[t]
\epsscale{1.17}
\plotone{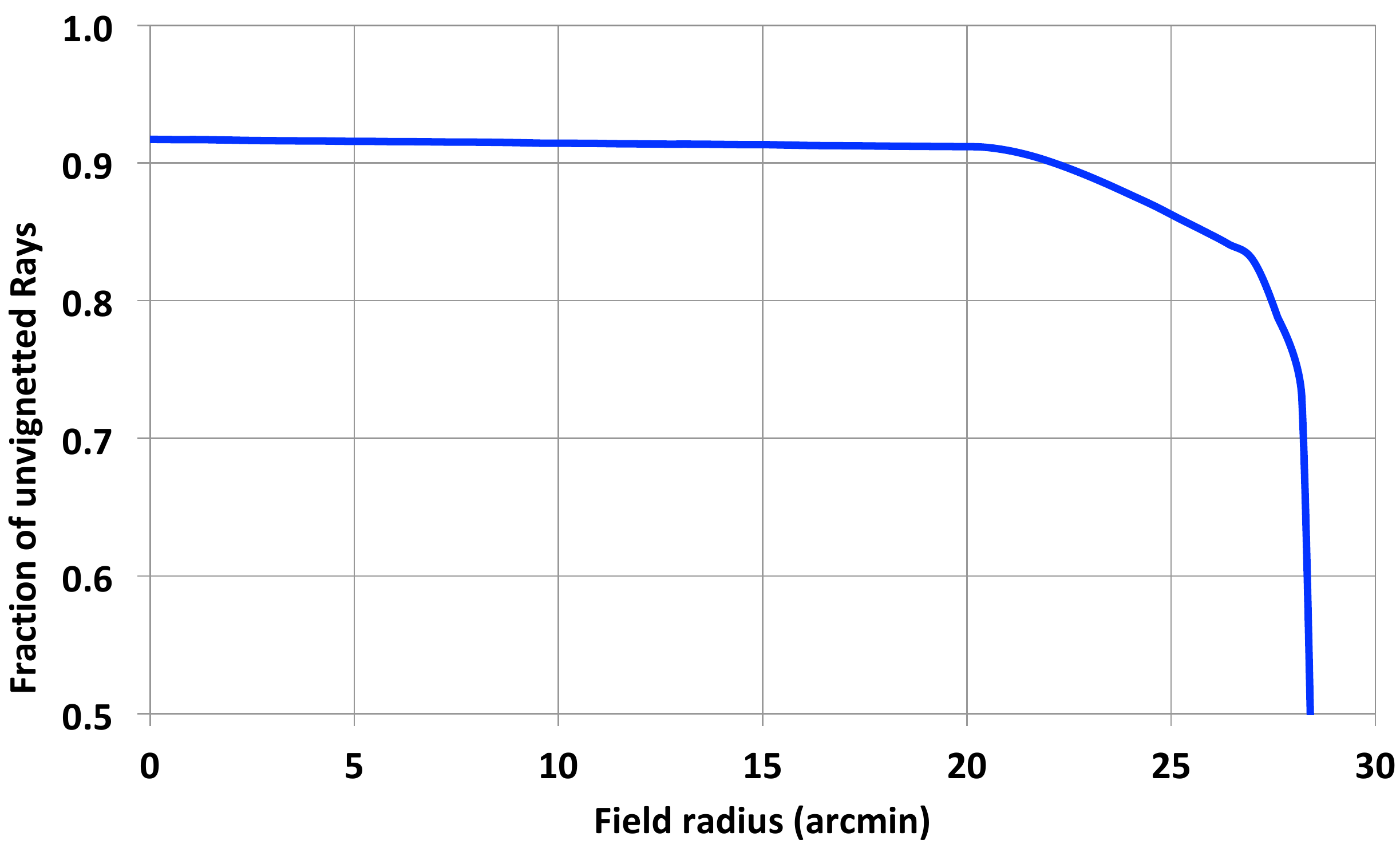}
\caption{ \label{fig:vignetting} Vignetting profile of the WHT prime focus corrector.}
\end{figure}

The PAU Survey, like most cosmological science studies, benefits reaching the largest volume possible. This requirement drove the design of the PAU Camera, given the constraints imposed by the telescope and the expected amount of time available for the survey. To maximize the area coverage, PAUCam covers the whole field of view delivered by the current prime focus corrector, that is, $\sim$1~degree diameter FoV. Unfortunately, the current corrector was not designed to take advantage of the full field of view that the telescope could delivered and offers only a $\sim$40~arcmin diameter unvignetted FoV (see Figure~\ref{fig:vignetting}).  The new WHT prime focus corrector being constructed for the WEAVE instrument will deliver a $\sim$2~degree diameter FoV. PAUCam will increase its surveying capabilities if it could use that new corrector once installed.


The choice of the CCD detectors to populate the focal plane took into consideration maximizing the wavelength coverage and having good quantum efficiency in the red to be able to reach to the highest redshifts possible. We also tried to maximize the size of the detectors to avoid inefficies with gaps when tiling the focal plane with CCDs. After a comprehensive search and given budgetary constraints we settle for the 2k x 4k fully depleted Hamamatsu detectors, with a pixel size of $15\unit{\micron}$. At the pixel scale delivered by the telescope, each CCD covers $\sim 9 \times 18\, arcmin^2$. Given this size, the optimal way to cover the unvignetted central part of the focal plane is with eight CCDs. Eighteen CCDs are needed in total to cover the whole FoV. Having the full FoV covered by our detectors implied that the autoguidibng system of the WHT could not be used, and therefore we decided to add the guiding capability to PAUCam allowing two of the CCDs in the corners to be used for both imaging and guiding. This requirement implied a special design of the electronics (see \ref{sec:readout}).

\begin{figure}[t]
\epsscale{1.15}
\plotone{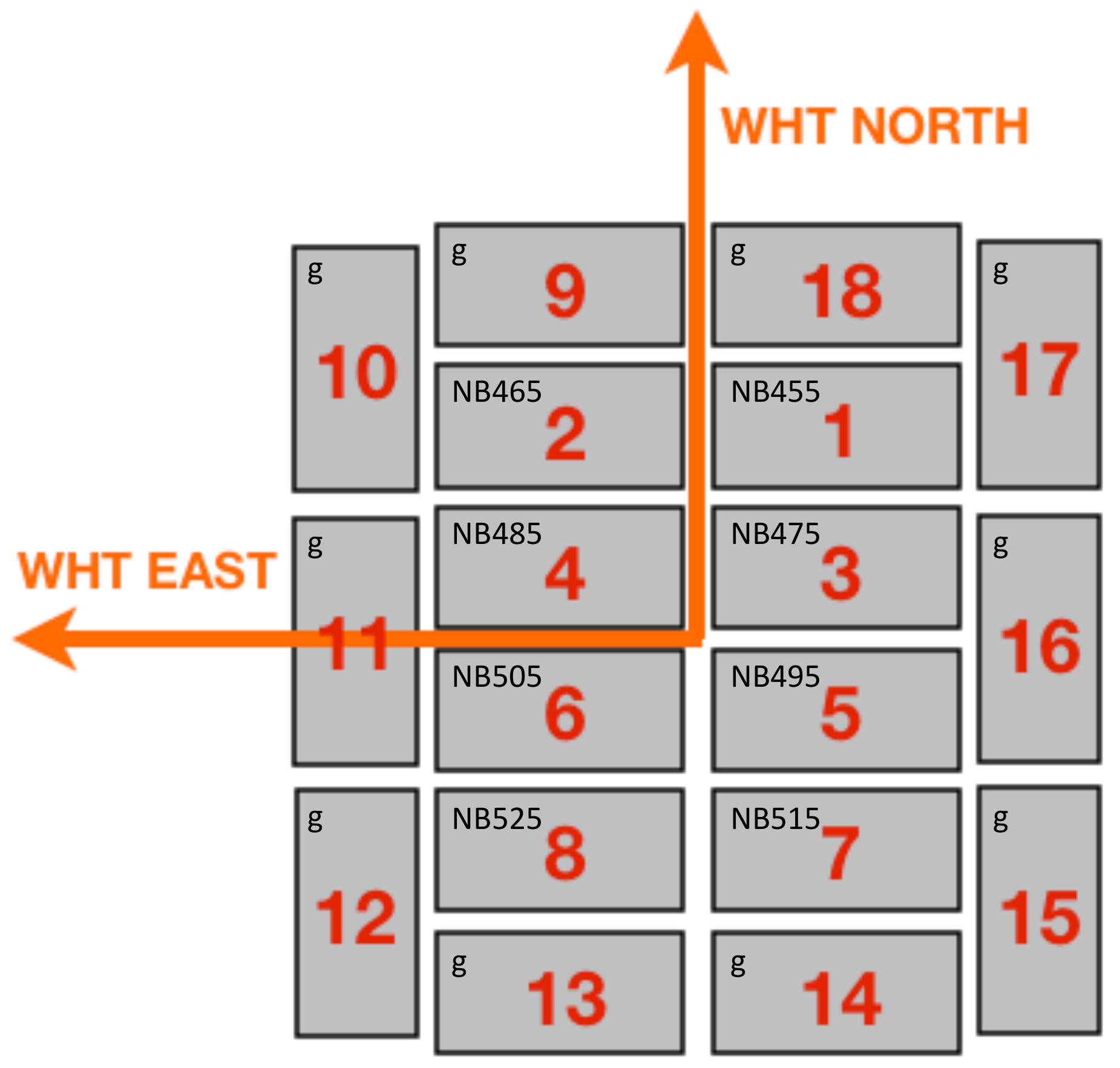}
\caption{ \label{fig:focalplane}  The focal plane of the PAU Camera with its 18 CCDs. Red numbers show the CCD numbering scheme and its position as seen in the sky at the WHT. Black numbers show the filters that are on top of each CCD when one of the filter trays is inserted. The eight central CCDs are covered by a Narrow Band filter while the external CCCs are covered by Broad Band filters.}
\end{figure}

The redshift precision requirement of PAUS is $\sigma(z)/((1+z) \sim 0.0035$ \citep{gazta12, benitez09a}. \cite{marti14} show that this requirement can be achived with a filter set of 40 narrow band filters $\sim 100$ {\AA}-wide, covering the $4500-8500$ {\AA} wavelength range. No vendor could produce this set of filters covering the full FoV. Besides, it would have been impossible to equip the camera with so many large filters and we would have needed to continuously change the filters mounted in the camera. Given these considerations and to optimize costs, we decided to design the system such that every CCD is covered by a single small-size (that is, the size of the CCDs) filter. These filters are mounted in filter trays that can be exchanged and placed in front of the CCDs. With 40 filters to cover the unvignetted focal plane of the eight central CCDs, a total of five filter trays of 8 filters each are needed to observe the survey fields with every filter. To cover the extra 10 CCDs located in the periphery of the focal plane, each tray is filled with 10 additional CCD-sized Broad Band (BB) filters. Figure~\ref{fig:focalplane} shows the CCD numbering scheme of the PAU Camera focal plane and its orientation as seen in the sky at the WHT together with the filters seen by each CCD when one of the trays is inserted. The filters are different for each filter tray. In addition, we decided to also have a standard $ugrizY$ BB filter set (6 large filters, the size of the whole focal plane) that could be used either for the PAUS calibration or by external observers with other scientific goals. Adding up these design goals amounts to a total of at least 11 filter trays. Since the camera has other operational needs, it was decided to design the filter exchange system with a total of 14 tray slots. Further explanation of the filter sets can be found in section \ref{sec:filters}.

Because of the large vignetting in the external part of the focal plane, maximizing the efficiency of the survey required minimizing the dead gap between CCDs in every exposure. To accomplish this goal, we decided to position the filters as close as possible to the focal plane, therefore requiring the filter interchange system to function inside the camera vessel and to have a movable system inside vacuum and cryogenic conditions.

PAUCam is designed as a survey instrument to cover wide areas of the sky. Nevertheless, it can also cover small regions with larger overheads. To observe single objects or a portion of the sky the size of a CCD with all 40 NB filters, an observer would need to point the camera eight times for each of the five trays, requiring 40 exposures. However, for larger fields, like the ones being targeted by the PAU Survey, every image also contains observations of the other seven filters in the same tray and the number of exposures needed to cover a certain large area asymptotically approaches 5, one per tray, as the area increases. When observing a large area, one needs to break each exposure into several dithered pointings in order to cover the CCD gaps. There are several ways of defining a tiling strategy depending on the observer’s preferences.  In the PAU Survey (see Section~\ref{sec:operation}), our strategy is to use three ditherings.

As in any camera aiming at observing distant faint objects, the control of the electronics noise is crucial. This is more relevant for NB observations where the background is lower, especially in the bluer filters, and therefore the contribution of the read-out electronics noise to the overall error budget is larger. PAUCam opted to have also the CCD pre-amplifiers inside the camera cryostat and the rest of the front-end electronics, including all the analog to digital conversion, installed in the camera. Only digital optical fibers transport the data to the computers outside the camera.

PAUCam also had engineering constraints in its design. The WHT prime focus does not allow to install an instrument weighing more than $\sim 270 \unit{Kg}$. This was a stringent limitation with all the services needed in the camera, including the vacuum pumps, the cryogenic system, the services and crates needed for the front end electronics and the motors needed for the filter exchange system. In order to be compliant with this weight limitation, the camera vessel and many parts of the filter exchange system were manufactured with carbon fiber. To our knowledge, this is the only astronomical camera vessel made of carbon fiber. The rest of the services, like all the power supplies, the slow control system and the cryogenic compressors are installed in one of the WHT Nasmyth platforms. Being at the prime focus, the camera could also not vignet the field of view, imposing restictions on its size.

The final constraint came from the fact that PAUCam is not permanently installed at the WHT. The camera is one of many instruments that are available for observations at the telescope. Therefore, the entire system is designed to allow the installation of the camera within hours and to have it operational before the night starts. As we will see, in addition to the mechanical and electronics interfaces, this required a dual cooling system to allow cooling the focal plane to its operational temperature within four hours.

\begin{figure}[t]
\epsscale{1.15}
\plotone{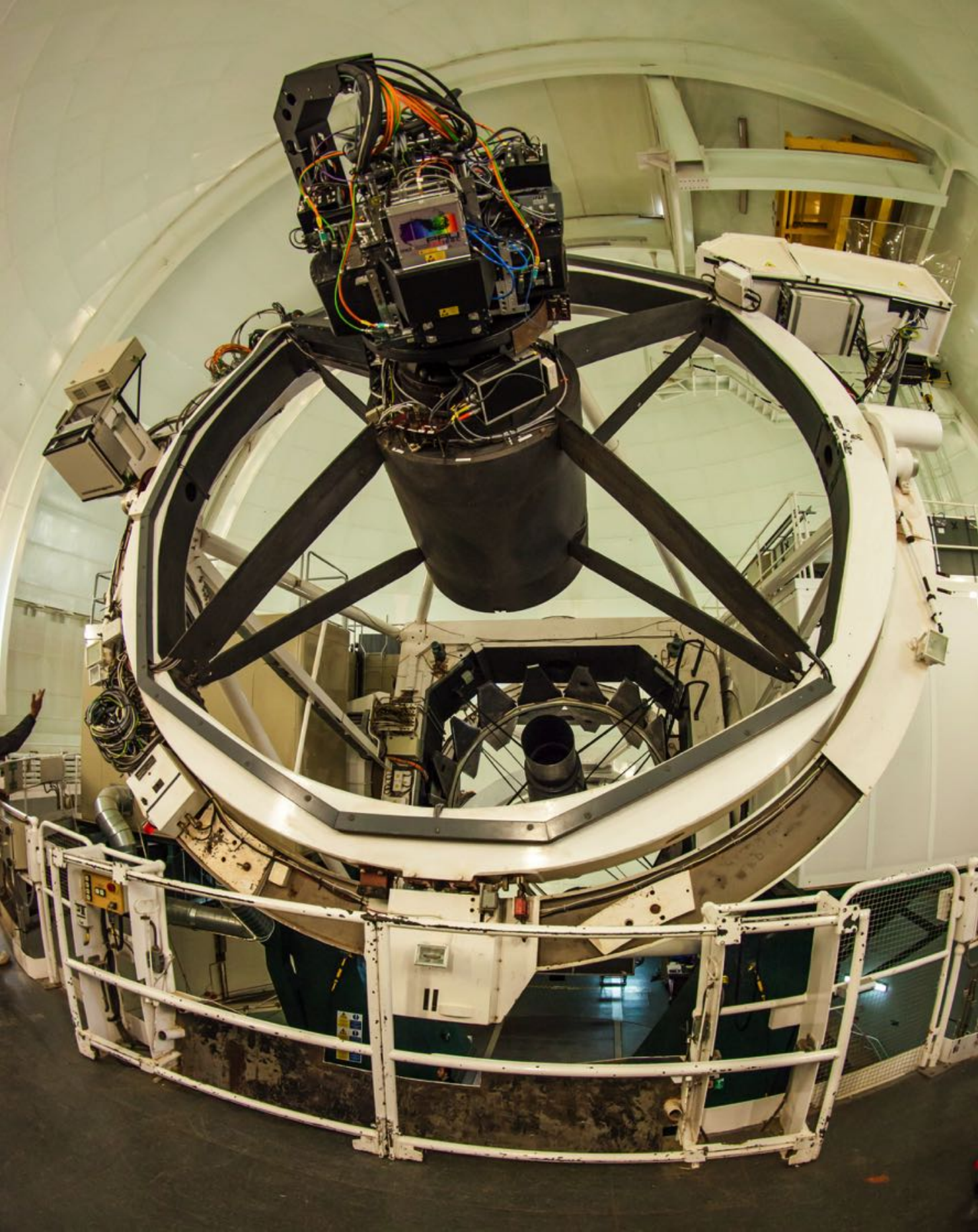}
\caption{\label{fig:PAUatWHT} 
The PAU Camera installed at the prime focus of the WHT. The carbon fiber camera vessel with the electronics mounted on top and all service cables routed through a swan neck to the telescope spider and the ring can be seen.}
\end{figure} 

Figure~\ref{fig:PAUatWHT} shows a picture of the PAU camera installed at the WHT prime focus. The overview shows the carbon fiber vessel, the front-end electronics and all services needed near the camera. A swan neck structure\footnote{The WHT is an alt-az telescope and therefore the focal plane needs to rotate with respect to the telescope structure to follow the sky. The swan neck structure provides this rotation capability.} guides all cables and services to the telescope ring and, from there to one of the Nasmyth platforms where the PAUCam control electronics and power supplies are installed.

\section{THE PAU CAMERA}
\label{sec:Camera}

Here we describe all the hardware elements of PAUCam. The description follows the path that light travels through the system.

\subsection{Optical System}


PAUCam is attached to the Prime Focus of the WHT. The main parabolic mirror of the WHT has a $4 \unit{m}$ diameter, a focal length of $10.5 \unit{m}$ and a focal ratio of f/2.5.

To correct the image in the Prime Focus an optical corrector of 4 elements\footnote{The official webpage (\url{http://www.ing.iac.es/PR/wht_info/whtoptics.html}) states that the corrector is composed of 3 elements but in reality there are 4.} plus an Atmospheric Dispersion Corrector is installed close to the Prime Focus providing an effective focal length of $ 11.79 \unit{m}$ (f/2.8) and an unvignetted field of view of $40 \unit{arcmin}$ (see Figure \ref{fig:vignetting}).

The distance between the last lens face of the corrector and the PAUCam focal plane is $118 \unit{mm}$. This is a small gap to place the optical elemnts of the camera. There are two optical elements: a) the entrance window and b) the filters. The light path transverses these elements in a converging, and tilted, beam.

Using the optical design provided by the Isaac Newton Group (ING), the company Fractal S.L.N.E.\footnote{\url{https://www.fractal-es.com/}} designed the entrance window and the filters, analyzing their effects in the final image.

\subsubsection{Entrance Window}
\begin{figure}
\epsscale{1.17}
\plotone{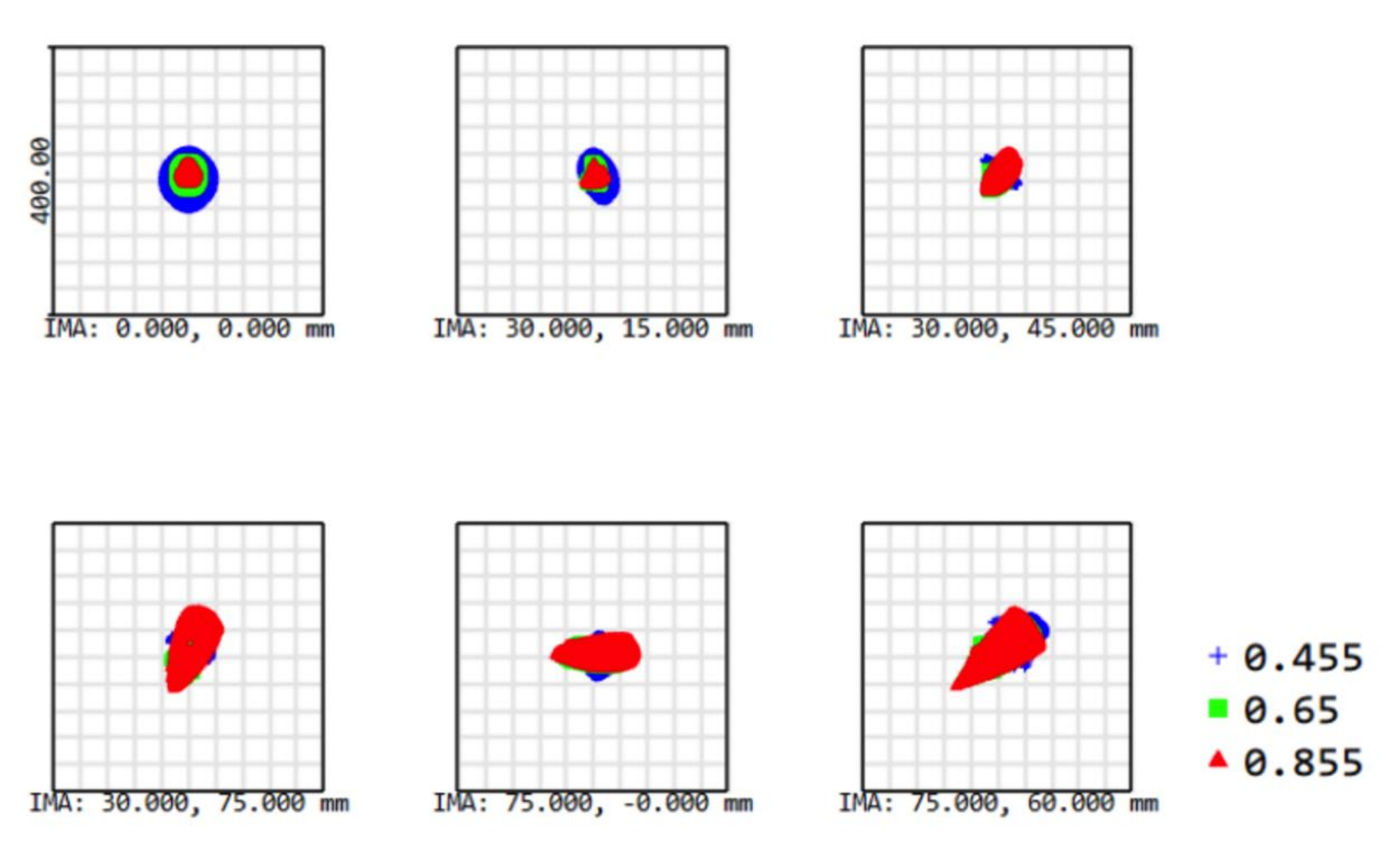}
\caption{ \label{fig:spotdiagram} Spot diagram generated with Zemax (\url{https://www.zemax.com}) using the corrector's design and the entrance window of PAUCam at different positions in the PAU Camera Focal Plane. Each diagram square is $400 \unit{\micron}$ in size. The colors correspond to the different wavelengths shown in $\micron$ in the legend.}
\end{figure}

As the camera operates in vacuum, an entrance window is needed to allow the light in while maintaining the camera vessel in vacuum. The dimensions of the focal plane demand a window diameter of $280 \unit{mm}$ to avoid vignetting. The window produces a degradation of the image quality. The spot diagram of the telescope, with the Prime Focus Corrector gives a RMS radius of $7.3 \unit{\micron}$ in the field center.  Using a flat optical window this value would increase up to $ 14.0 \unit{\micron}$. To try to minimize this value we designed a window of fused silica with two different curvature radii: $1567 \unit{mm}$ for the external face and $1588 \unit{mm}$ for the internal one, with a thickness in the center of $20 \unit{mm}$ necessary to avoid deformations due to the pressure difference between the two window faces. With this configuration the spot RMS radius in the field center is marginally improved to a value of $13.6 \unit{\micron}$. However, the improvement at larger distances from the center is larger with respect to the flat window option. The best value for the spot diagram is achieved outside the field center at the central position of the four central CCDs (see Figure~\ref{fig:spotdiagram}) with an RMS radius of $12.3 \unit{\micron}$. The outer CCDs installed at the edges of the FoV, used for guiding and calibration, were not used to optimize the spot diagram. 

The entrance window was manufactured by the Mexican Instituto Nacional de Astrof\'isica, \'Optica y Electr\'onica (INAOE)\footnote{\url{http://www.inaoep.mx/}}, with a long experience in astronomical instrumentation.  It was coated with  BBAR-580, delivering a reflection not larger than 2\% in the wavelength range of 400 to $1100 \unit{nm}$. 

\subsubsection{Filters}
\label{sec:filters}

PAUCam has two sets of filters installed in two different jukeboxes (see \ref{sec:FilterExchangeSystem}). When observing, these filters are placed within a few millimeters of the active area of the CCDs in a vacuum environment, at low temperature ($\sim 250 \unit{K}$), and in a focalized and non-telecentric tilted beam. Since the filters are under vacuum conditions, a common manufacturing requirement from all filters is that the coating does not produce any significant out-gassing that can be deposited onto the CCDs. An overview of the filter set is given in Table~\ref{tab:filterchar}. Pictures of the filters installed in the filter exchange system can be seen in section~\ref{sec:FilterExchangeSystem}.

\begin{table*}[tb]
\centering
\begin{tabular}{ccccc}
\hline
\hline
Type & Size (mm) & Width (nm) & Range (nm) & Avg. Transmission \\
\hline
External broad & $203.4\times208.8\times3.0$ & $\sim 90-150$ & $300-1080$ & $98\%$  \\
Tray Broad & $65.7\times33.8\times3.0$ & $\sim 90-150$ & $300-1080$ & $96\%$ \\
Tray Narrow & $65.7\times33.8\times3.0$ & $13$ & $450-850$ & $97\%$ \\
\hline
\end{tabular}
\caption{\label{tab:filterchar} General characteristics of the PAUCam filter sets.}
\end{table*}

\paragraph{Broad band filter set}
The broad band filter set, was manufactured by Asahi Spectra\footnote{\url{https://www.asahi-spectra.com}}. The filters cover the whole focal plane of the camera, $203.4 \unit{mm} \times 208.8 \unit{mm}$ in size. To minimize the distortion of the focalized beam, the thickness of the filters is $3.0 \unit{mm}$. This filter set provides PAUCam with a capability to enlarge its science reach and foster the interest of other potential users. Their photometric transmission bands have the same design as the filters used in DECam \citep{ref:DECAMfilters} that were manufactured by the same company. There are six broad band filters: \textit{u}, \textit{g}, \textit{r}, \textit{i}, \textit{z} and \textit{Y} and they were calibrated at the ICE-CSIC/IEEC/IFAE labs.

The setup to calibrate the filters was composed by a light source, a collimation system, a monochromator, an integration sphere and a calibrated photo-diode. To avoid dust contamination during the tests, the filters were installed inside a hermetic box in the clean room. The transmission of the window glasses were calibrated with the same system described above prior to the filter transmission measurements.

\begin{figure}[tb]
\epsscale{1.17}
\plotone{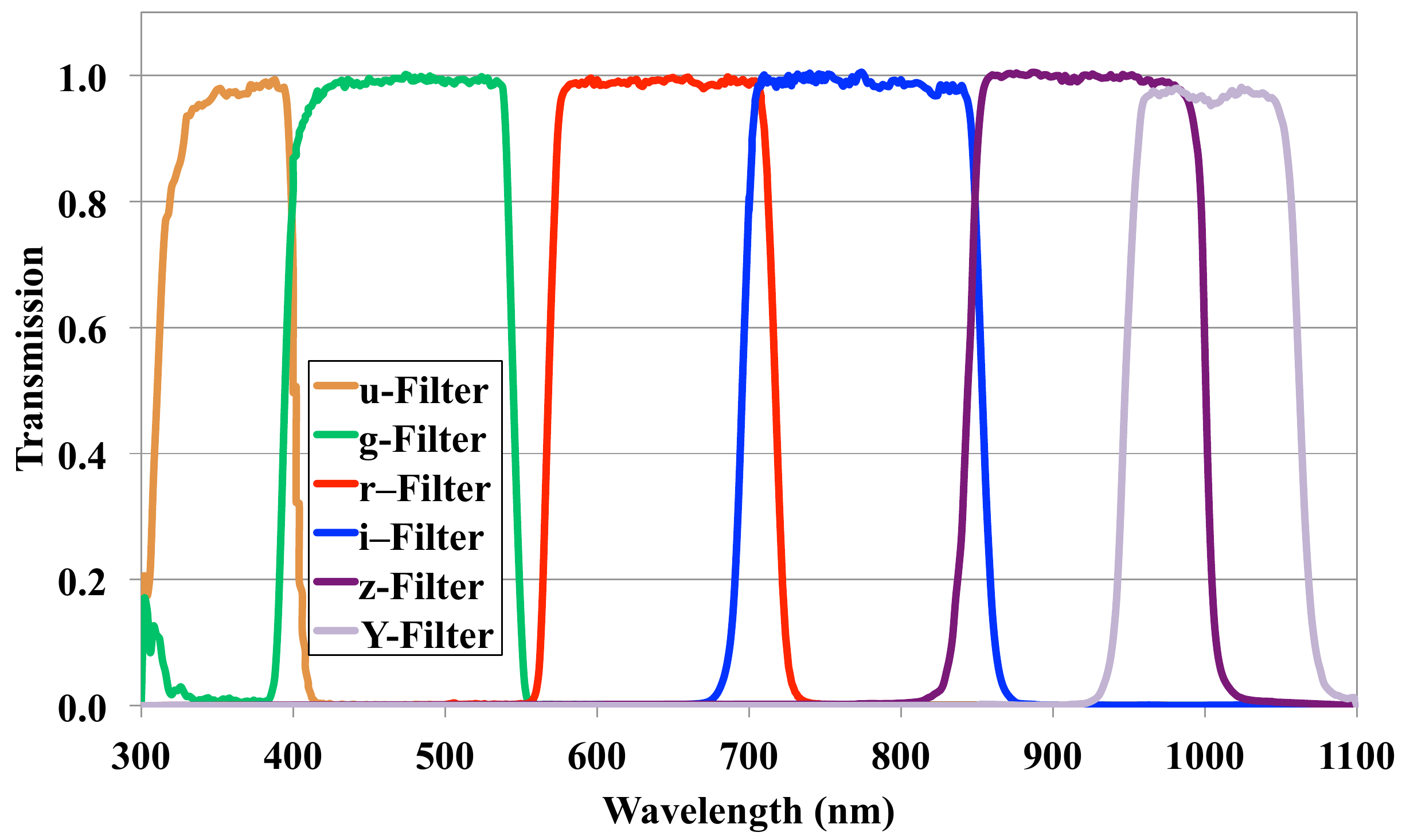}
\caption{ \label{fig:bbf_transmission} Transmission profiles for the six large broad band filters measured in the ICE/IFAE labs. The transmission wavelength gap between the g and r filters was designed to avoid the strong atmospheric [OI] emission line. From \citet{ref:Casas2016}.}
\end{figure}

Filter transmission measurements were done at different positions. These values were compared to check for a good uniformity across the filter area and averaged to obtain the transmission curves shown in Figure~\ref{fig:bbf_transmission}.

Any filter designed with multi-layers coatings changes its transmission profile depending on the incidence angle of the light beam.
The effect of the focalized and tilted beam compared to a parallel incidence beam in the final filters transmission profiles is practically negligible for these broad band filter set as the effect is much smaller than the wavelength transmission width of the filters.

\paragraph{PAU Survey filter set}

The small size filter set was manufactured by the Canadian company Iridian Spectral Technologies\footnote{\url{https://www.iridian.ca}}. The PAU Survey uses this set. It consists of 40 Narrow Band (NB) filters covering the 8 central CCDs in five trays and several Broad Band (BB) small filters covering the 10 external CCDs of each tray. Each of the eight central CCDs, for which the PAU Survey has been optimized, has five different NB filters associated to it, one for each tray.

The optical properties of the incoming beam and the environmental conditions influence the effective transmission profiles, which are more critical in the NB filters and can have an impact on the photometric redshift performance of the PAU Survey. To understand the variation of the transmission differences we modeled their transmission profiles in different parts of the filters. The focus position will also shift with wavelength, but this has been measured to be negligibly small. 


\begin{figure}[t]
\epsscale{1.15}
\plotone{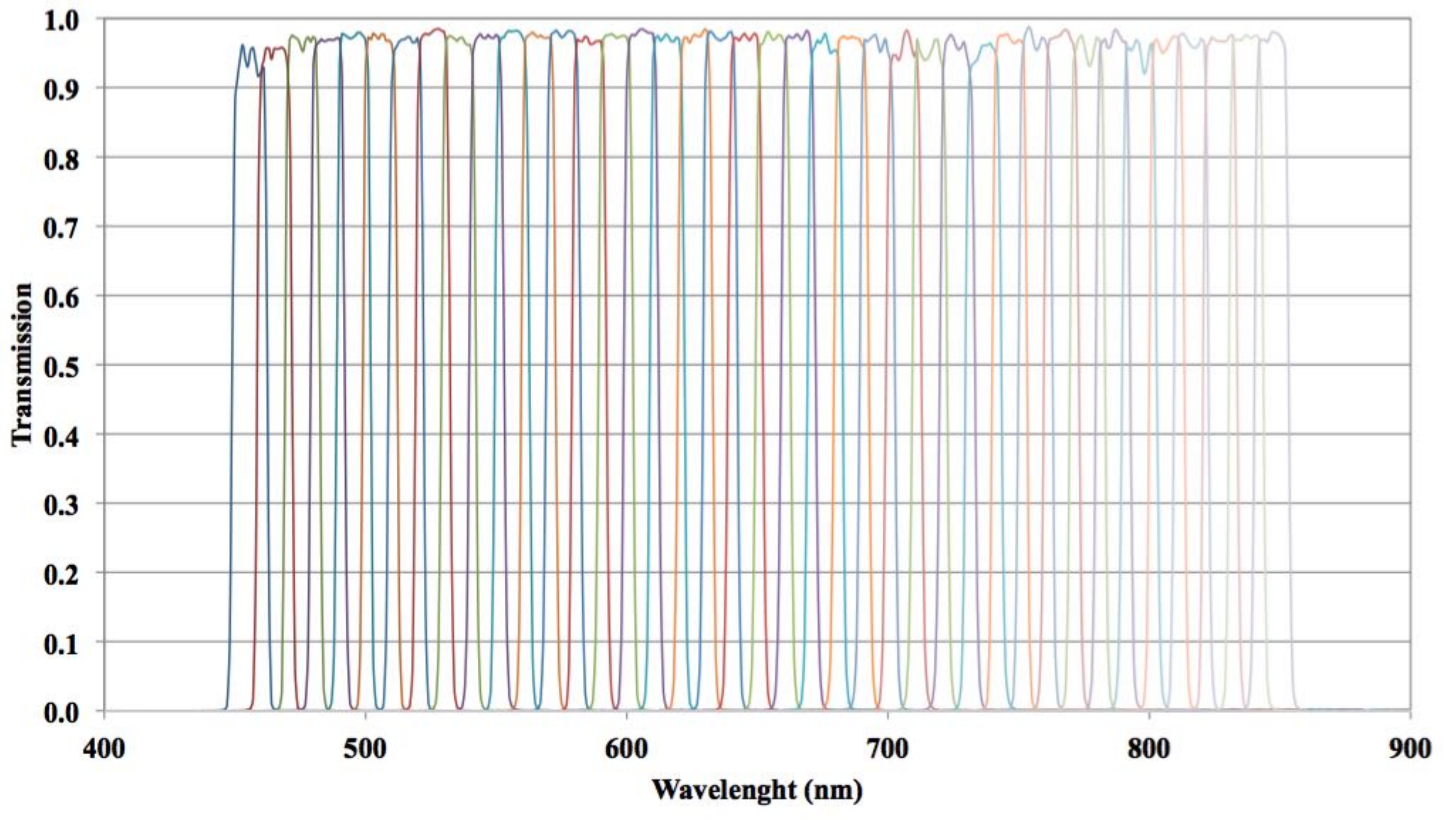}
\caption{\label{fig:nbf_transmission} Transmission profiles for the forty NB filters measured at CIEMAT. From \citet{ref:Casas2016}.}
\end{figure}

The BB set covers is composed of \textit{griz} filters that share the same characteristics as the external filter set described above, with a somewhat lower transmission for these smaller sets in the \textit{g} band, specially at bluer wavelengths (85\% vs 95\%). The NB filters display a 'top-hat' transmission curve (Figure \ref{fig:nbf_transmission}) with a nominal Full Width Half Maximum (FWHM) of $13 \unit{nm}$, and a central wavelength distance between two consecutive filters of $10 \unit{nm}$. This leads to some overlap between filters adjacent in wavelength. The central wavelength of these narrow band filters span a wavelength range from $455 \unit{nm}$ to $845 \unit{nm}$. They were measured at a calibration setup at CIEMAT \citep{ref:Casas2016}. To test the homogeneity, the transmission was measured in a matrix of 15 points obtaining a very low dispersion. The mean transmission measured for each filter was averaged and turned out to be above 95\%.

\subsection{Detector system}

\begin{figure}[t]
\epsscale{1.15}
\plotone{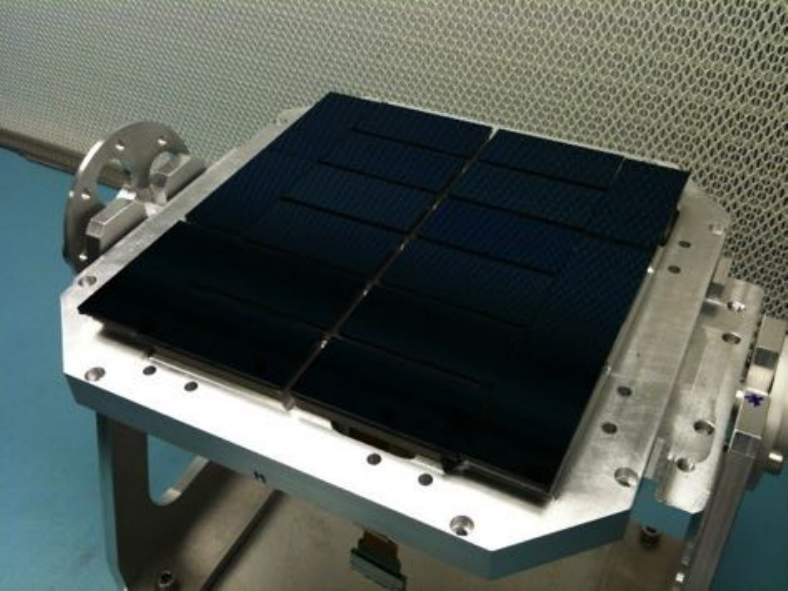}
\caption{\label{fig:paucamfpa} Hamamatsu S10892-04 CCDs covering the PAUCam focal plane.}
\end{figure}

The camera focal plane (FPA) is composed of 18 scientific-grade Hamamatsu Photonics\footnote{\url{http://www.hamamatsu.com}} CCDs type S10892-04 of $2\mathrm{k}\times 4\mathrm{k}$ pixels. The CCDs are back-illuminated and fully depleted with $15 \unit{\micron}$ size pixels (see specifications in Table~\ref{tab:ccdspecs}). In addition, these CCDs showed very low dark current  (\textless1 e$^{-}$\textbackslash pixel\textbackslash h) at the operational temperature of $\sim -100^\circ \mathrm{C}$, low readout noise (\textless5 e$^{-}$~RMS) and low number of cosmetic defects. The CCDs are distributed on the focal plane (Figure~\ref{fig:paucamfpa}) according to their performance during characterization. The CCDs that showed the poorest performance (2 units) were installed at the corners of the focal plane and used for guiding, while the best CCDs were installed populating the focal plane from the center to the borders, ensuring that the eight central CCDs are the ones with the best performance. The planarity of the assembled chips in the FPA were checked with a Stil OP10000\footnote{\url{http://www.stilsa.com}} confocal sensor with a precision of $0.9 \unit{\micron}$ mounted in a specially manufactured x-y movable table to scan the complete focal plane. The peak-to-valley planarity of the FPA was measured to be $26\unit{\micron}$.

\begin{figure} [t]
\epsscale{1.15}
\plotone{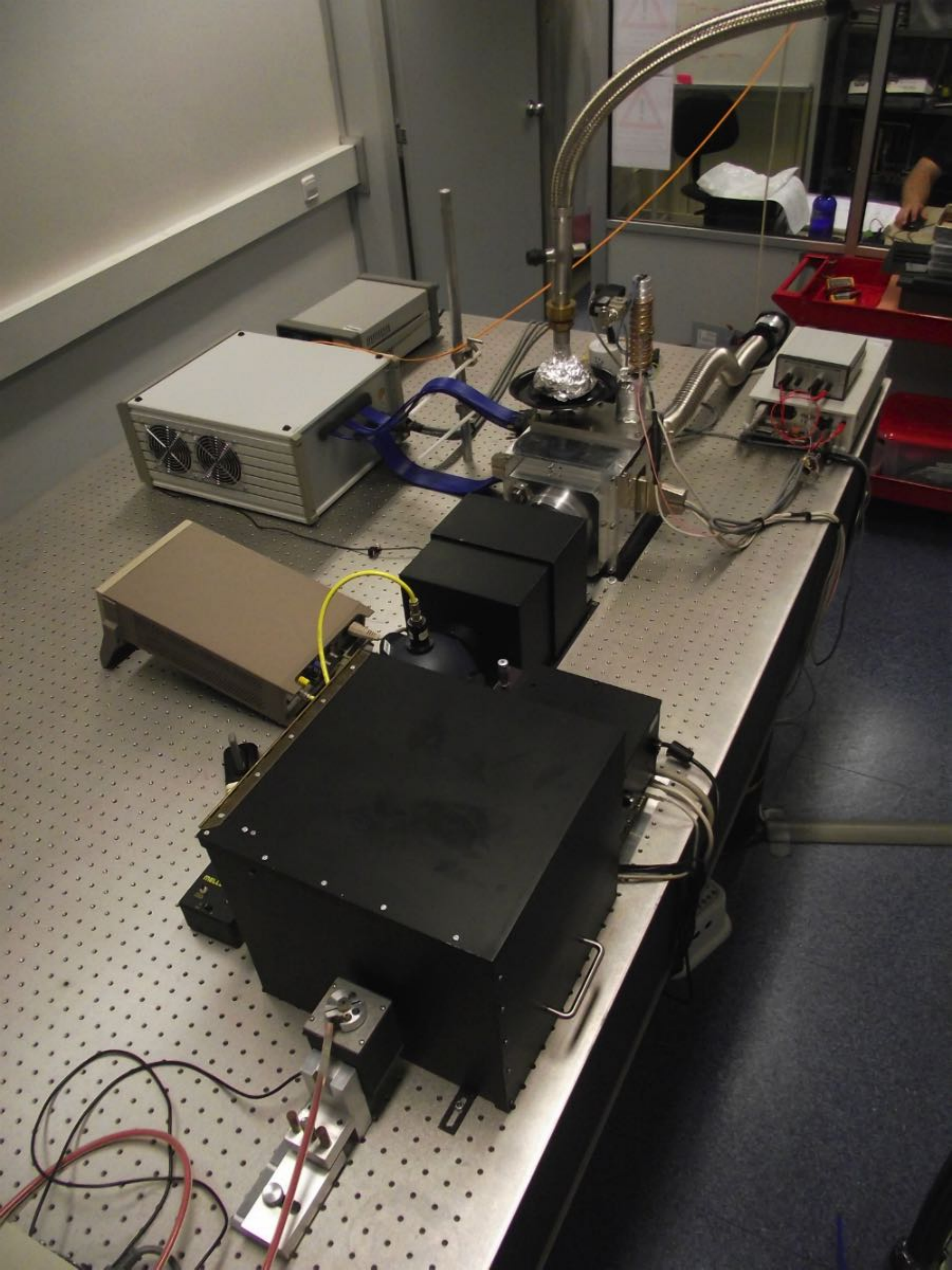}
\caption{\label{fig:ccdtestbench} Optical setup used for the CCDs characterization.}
\end{figure}

\begin{table*}[bt]
\centering
\begin{tabular}{ll}
\hline
\hline
Parameter & Specifications \\
\hline
Pixel Size & 15 $\mu$m \\
Number of Pixels & $2080\times 4224$\\
Number of Active Pixels & $2048\times 4096$\\
Active Area & $30.72\times 61.44$ mm\\
Fill Factor & $100 \%$\\
Output Amplifiers & $4$\\
Full Well (Typ) & $150$ ke$^{-}$\\
Dark Current (@$173K$) (Max) & $5$ e$^{-}$\textbackslash pixel\textbackslash h\\
Readout Noise (@$133$ kHz) (Max) & $5$ e$^{-}$ rms\\
\textbf Quantum Efficiency (Typ) & $70\%$ @ $400$ nm\\
        & $90 \%$ @ $650$ nm\\
        & $40 \%$ @ $1000$ nm\\
Charge Transfer Efficiency (Min) & $0.999995$\\
PRNU (Max) & $2.5 \%$\\  
\hline
\hline
\end{tabular}
\caption{\label{tab:ccdspecs} Hamamatsu CCD Type S10892-04 specifications.}
\end{table*}

Two test benches at IFAE/ICE-IEEC (Figure \ref{fig:ccdtestbench}) and CIEMAT were developed in order to characterize the CCDs \citep{jimenez2012}. The system is composed by a cryostat that allows to readout one CCD at a time via a MONSOON (NOAO)\footnote{\url{https://www.noao.edu/nstc/monsoon/}} controller. The CCDs are cooled down to $173 \unit{K}$ while being pumped below $1\times10^{-6} \unit{mba}$r. The test bench is composed of a closed optical path illuminated by a 150~W Xenon arc lamp. The beam coming from the lamp source, whose intensity on the focal plane can be controlled by a neutral density filter, crosses a set of lenses in order to collimate the light as much as possible before entering a filter wheel that holds 3 long-pass filters. A monochromator with a resolution of $1 \unit{nm}$ is placed after the filter wheel and the resulting monochromatic light is sent to an integrating sphere holding a monitor photo-diode characterized versus a NIST-Calibrated reference photo-diode head in absolute terms of W/cm$^{2}$. The final result is that the light coming out from the integrating sphere is a flat field monochromatic light with a known absolute light power and intensity.

\begin{figure}[tb]
\epsscale{1.17}
\plotone{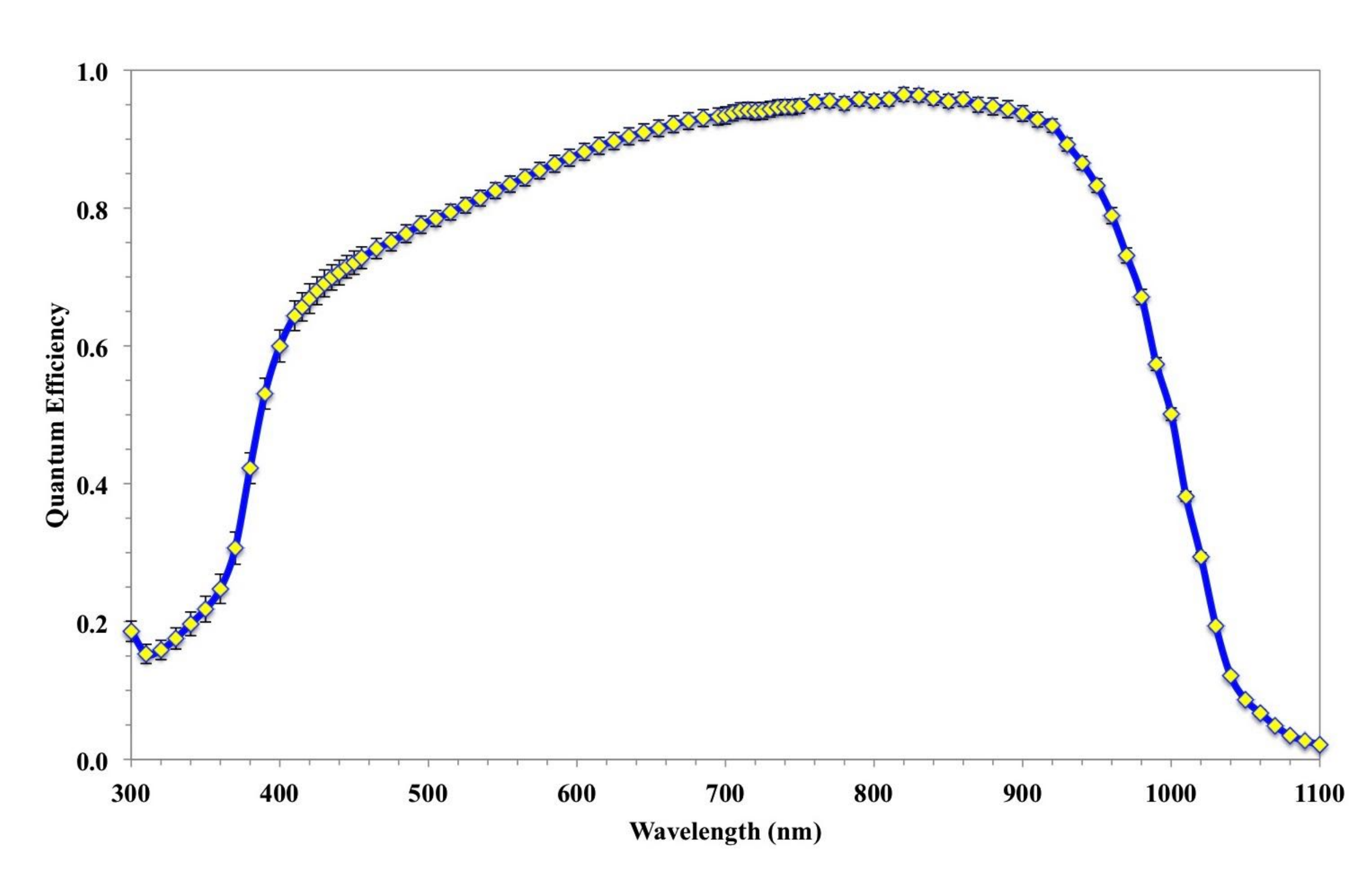}
\caption{\label{fig:CCD_QE} Quantum efficiency of the Hamamatsu CCDs used in PAUCam measured in the IFAE/ICE-IEEC labs at operational temperature. The dots are the average over the 22 tested CCDs. The errors (smaller than the dots in some regions) are the RMS over the different CCDs }
\end{figure}

Due to the importance of controlling these complex test bench instruments and synchronizing them with the CCD readout electronics (Section~\ref{sec:readout}) to get the desirable CCD characterization data, we developed several LabVIEW\footnote{\url{http://www.labview.com}} and TCL/TK applications that allow us to manage all the systems and collect and save the characterization data for their further analysis. The setup is remotely controlled in fully automatic mode by suitable software applications that can perform wavelength scans from $300 \unit{nm}$ up to $1100 \unit{nm}$. Figure~\ref{fig:CCD_QE} shows the quantum efficiency averaged for all CCDs determined at operational temperature.

\subsection{Readout Electronics System}
\label{sec:readout}
The PAUCam readout electronics \citep{ref:ROE1, ref:ROE2} is based on a MONSOON front-end and a panVIEW \citep{2010SPIE.7740E..1KH} application as the readout electronics controller. Both systems have been extensively used in several world-class instruments and there is enough know-how around this tandem to be considered as a reliable solution. panVIEW is a specific application of a Pixel Acquisition Node (PAN) developed at CTIO (La Serena, Chile)\footnote{\url{http://www.ctio.noao.edu/noao}}, which is in charge of acquiring the pixel data from the MONSOON front-end. In addition to that, it is able to control all the memory registers and variables used by the front-end electronics to perform the CCD readout and also generate suitable Flexible Image Transport System (FITS)\footnote{\url{https://fits.gsfc.nasa.gov/fits_standard.html}} headers, define and log temperature test points, voltages and many others user defined variables. In PAUCam, there are four panVIEW applications, one for each MONSOON crate, which run in an individual Linux server equipped with a Four Input Links for Atlas Readout (FILAR) board\footnote{\url{https://hsi.web.cern.ch/hsi/s-link/devices/filar}}. Finally, each panVIEW is commanded by the PAUCam Data Acquisition Interface (DAQi) (see section~\ref{sec:DAQi}) in order to startup and configure all the required settings.

Concerning the MONSOON front-end, PAUCam uses the DECam \citep{ref:DECAMfilters} physical implementation, which roughly speaking is composed of a Master Control Board (MCB) supporting a High-speed Optical Link for Atlas (HOLA) interface board\footnote{\url{http://hsi.web.cern.ch/hsi/s-link/devices/hola}} to use a S-LINK protocol for data transfer; a 12 channels Acquisition Board (ACQ) used to generate bias and capture detectors video, and a Clock Board (CB) which generates the clocks  and is able to drive up to 36 independent clocks and replicate them up to 135 times. 

The readout electronics configuration can be changed on-the-fly by software to use the whole focal plane for science images (Normal Mode) or to use the top and bottom CCDs as guiders (Guiding Mode) while the rest of them remain on science mode. The readout speed for the science mode is $150 \unit{kHz}$ and can be increased up to $300 \unit{kHz}$ while reading one Region Of Interest (ROI) for the Guiding Mode.

\begin{figure}[tb]
\epsscale{1.17}
\plotone{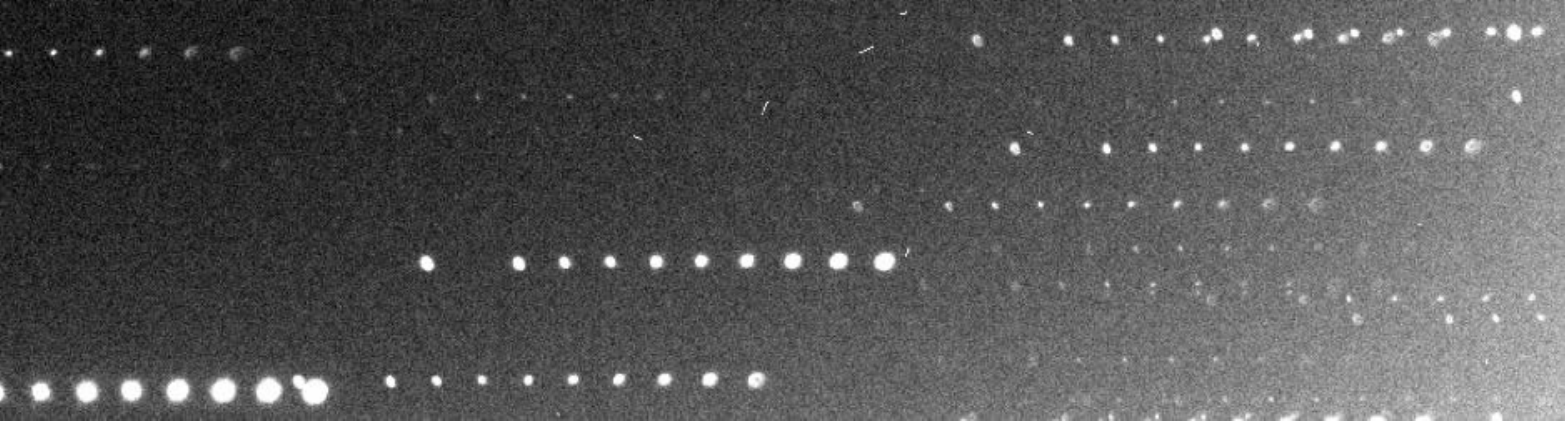}
\caption{\label{fig:focus} Focusing image resulting from a focus search iteration. For every star, several images are taken in the same file.}
\end{figure}

In order to comply with the specifications for the camera readout system, it was necessary to use four MONSOON crates: two of them reading out eight CCDs each and the other two crates reading out just one CCD per crate in order to reach different speeds and exposure times for guiding tasks. In order to get the whole readout synchronized, it was necessary to use a master-slave configuration with dependency of the readout mode selected. There are three readout modes:
\begin{itemize}
\item[--] The full mode is the default operation mode of the camera. It reads out all the CCDs at the same time. For this mode, there is just one master MONSOON which commands the others, set as slaves. 
\item[--] The guider mode is able to readout the sixteen central CCDs while using two of the CCDs at the corners for guiding purposes. This mode requires to have two MONSOONs as masters, one for the sixteen CCDs configuration and the other for the guiders CCDs.
\item[--] The focus iteration mode is used for the focusing procedure in combination with the focus of the telescope. For this configuration the same image is exposed several times. Between each exposition a number of rows are displaced and the telescope focus is changed. The number of focus iterations, focus increment between expositions and CCD rows displaced are user-defined. At the end of the configured expositions the CCDs are readout. Finally, a software algorithm (Section~\ref{focusing}) searches for the best focus value reached on every focus change. Figure~\ref{fig:focus} shows an image taken in this last focus mode.
\end{itemize}

The camera is mounted at the prime focus and due the limited weight supported by the telescope in this position, it was necessary to mount the whole readout electronics power supplies at the Nasmyth level. This limitation required more than 30 meters of cables to connect the front-end electronics  and demanded a dedicated filter stage to stabilize the power supplies and to guarantee the required low noise in the power sources.

Besides that, it was necessary to avoid any uncontrolled grounding loop around the readout electronics system since the camera is equipped with several servo-motors, pumps, sensors and cryocoolers which are all controlled by a Programmable Logic Controller (PLC). Therefore, from the very beginning, the camera design was conceptually built around a \emph{star-ground} grounding configuration with enough flexibility to be adapted to the telescope conditions.

\paragraph{PREAMP and MIX boards}

\begin{figure}[t]
\epsscale{1.17}
\plotone{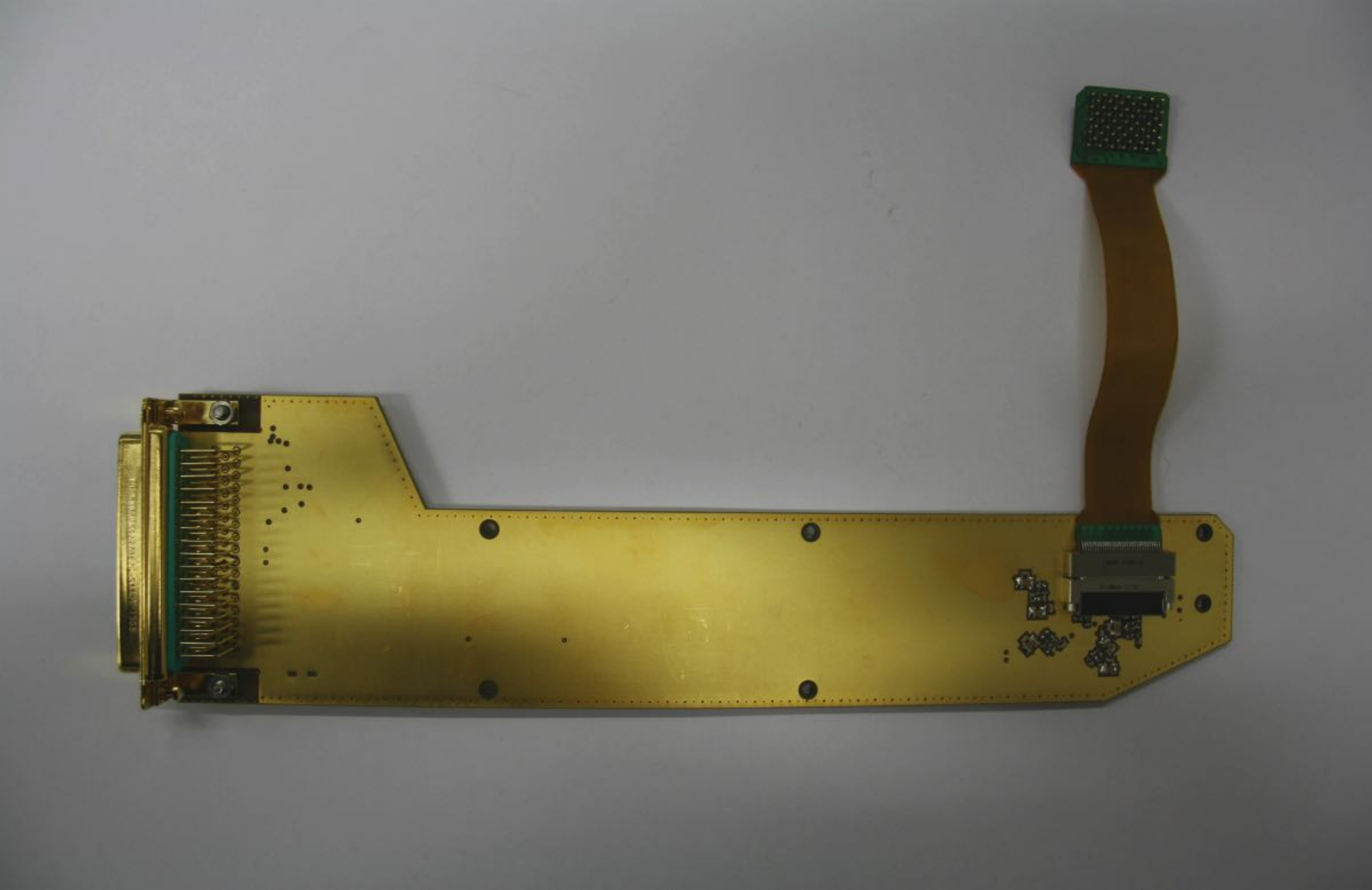}
\caption{\label{fig:PREAMPBoard} Picture of the PREAMP board with a  kapton cable that connects it to the CCD.}
\end{figure}

The PREAMP board (Figure \ref{fig:PREAMPBoard}) is located inside the cryostat (vacuum) and it is devoted to amplify the video output and connect the clocks and bias signals coming from the CCD through the D-SUB50 pins feed-through connector installed in the vessel wall towards the MIX board located outside the cryostat (atmosphere).

The final PREAMP design is based on a fully differential pre-amplification stage with twelve layers in order to separate bias, clocks and video signals with shielding layers between them to avoid electrical coupling (cross-talk). A gold-coating finishing was used in order to reduce the thermal radiation and the board outgassing. The PREAMP was designed at IFAE and CIEMAT in order to provide low readout noise (\textless$25 \unit{\mu V}$ RMS), a moderated cross-talk (\textless$1/10.000$), reduced power consumption and power dissipation ($\sim 200\unit{mW}$) and low-to-moderated outgassing and thermal radiation. A picture of the PREAMP boards as installed inside the camera is shown in Figure~\ref{fig:JukeboxMounted}.

\begin{figure}[t]
\epsscale{1.17}
\plotone{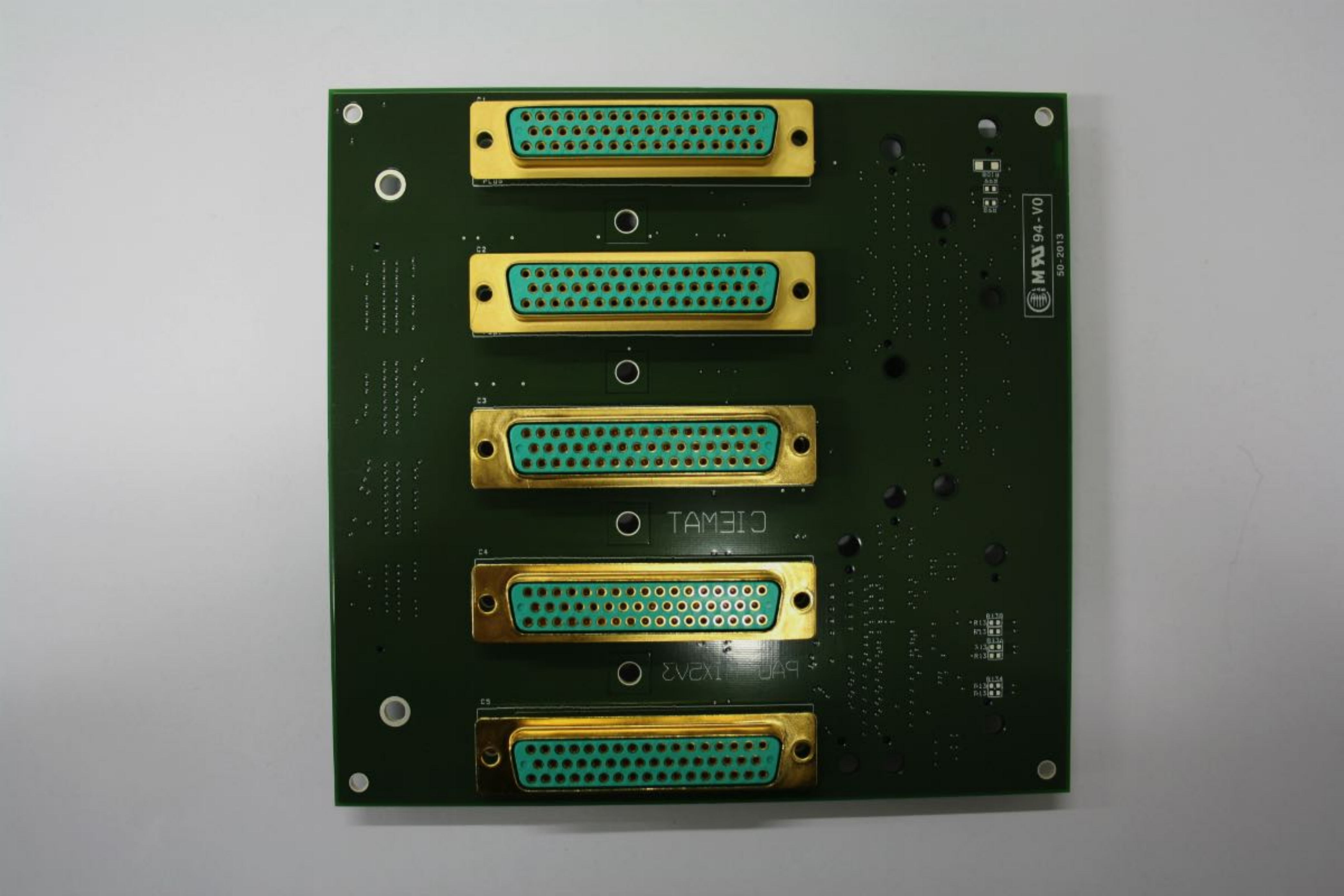}
\caption{\label{fig:MIXBoard} Picture of the MIX5 board.}
\end{figure}

The MIX board (Figure~\ref{fig:MIXBoard}) was designed at CIEMAT and it resides outside the cryostat (atmosphere) just behind the D-SUB50 pins feed-through connector. Its main aim is to fan out the CCD signals coming from the PREAMP board and route them towards the MONSOON front-ends. The MIX board collects bias, temperature and video signals and is able to drive up to five CCDs on the MIX5 version or up to three CCDs using the MIX3.

Both, PREAMP and MIX boards were manually assembled at IFAE facilities. Once ready, the MIX boards were redirected to CIEMAT for testing whereas the PREAMP boards were tested at IFAE. The PREAMP boards were tested electrically for short-cuts, open circuit, cold solders and any other defects produced during the PCB’s manufacturing and assembly processes. Once they were electrically functional, different performance tests were done:  readout noise, linearity, readout speed noise and signal stability. These tests were performed using the  MONSOON's own front-end as a pixel generator by using a clock signal with the same level as expected in the CCD (i.e.: $\sim 100 \unit{mV}$ for bias level) and synchronized with the Correlated Double Sample (CDS) of the video board. PREAMP boards with low performance were discarded.

\subsubsection{Electronics System Performance}

\paragraph{Gain and bias}

\begin{figure}[t]
\epsscale{1.17}
\plotone{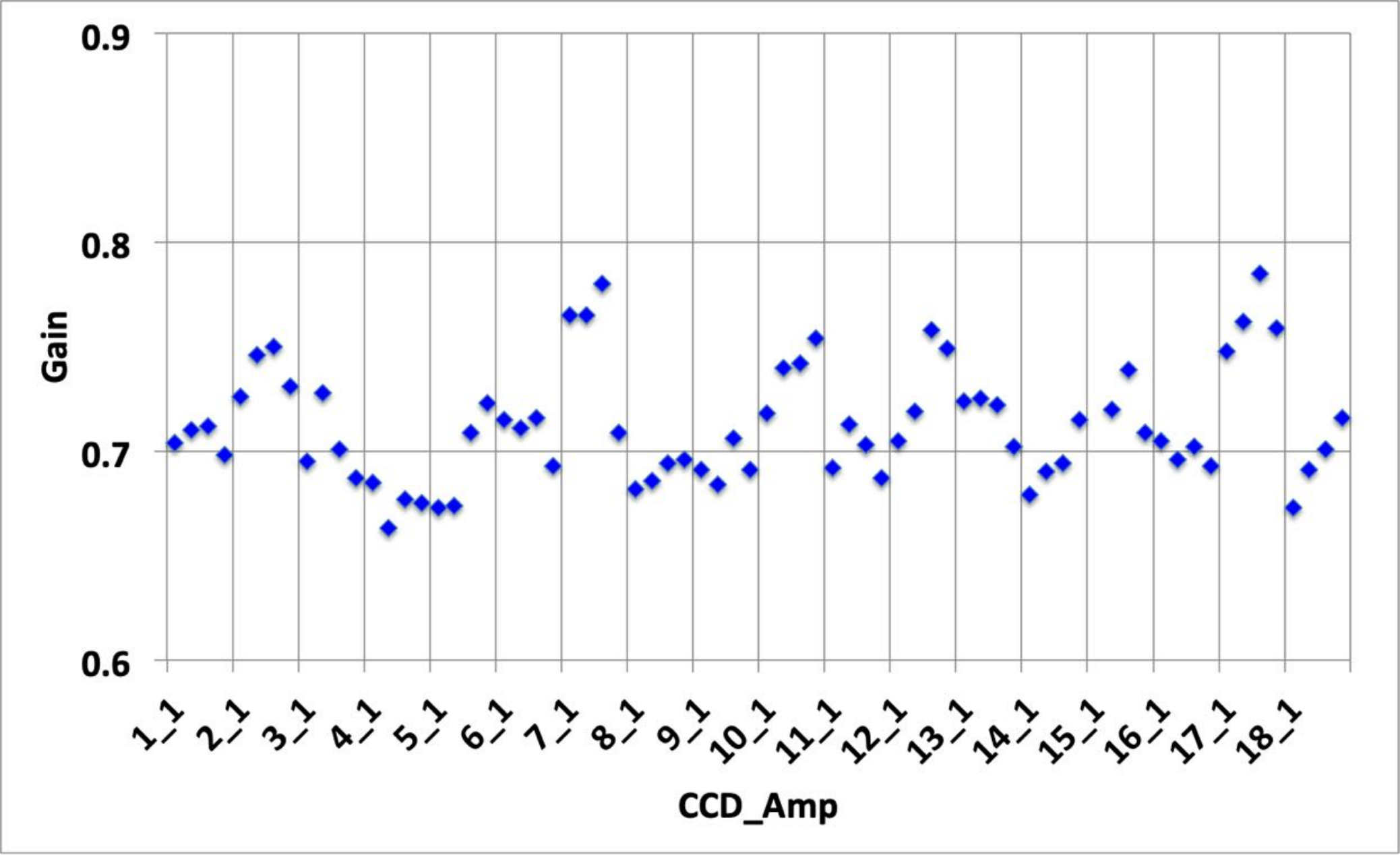}
\caption{\label{fig:PreviousGain} Gain levels plot showing the distribution for the whole set of PAUCam CCDs in a previous measurement made in the laboratories before delivering of the camera to La Palma in June 2015.}
\end{figure}

\begin{figure}[t]
\epsscale{1.17}
\plotone{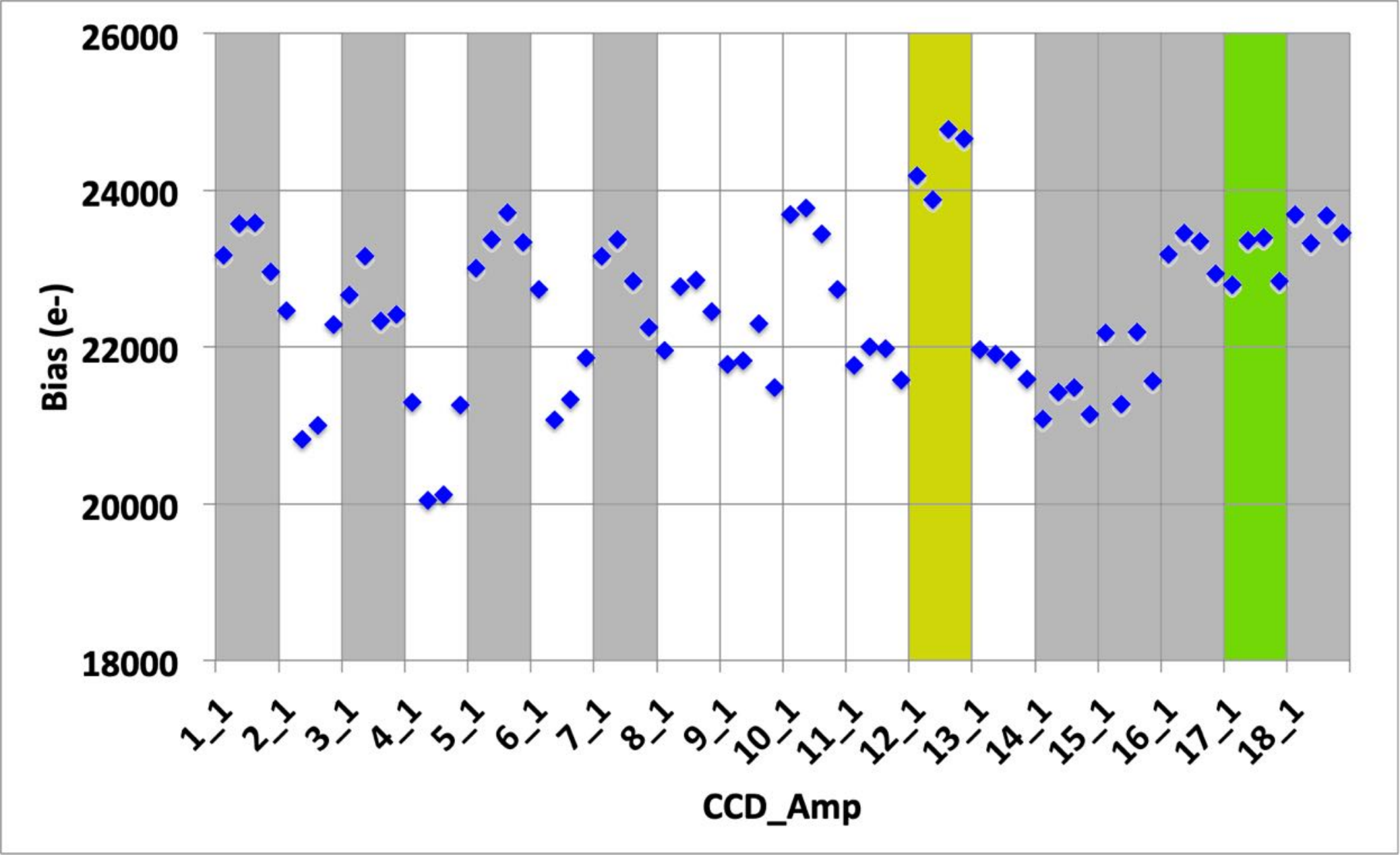}
\caption{\label{fig:PreviousBias} Bias levels plot showing the distribution for the whole set of PAUCam CCDs in the previous measurement made in the laboratory before the camera was delivered to La Palma in June 2015. The different colors for the CCD are related to the different crates in the readout. White and gray colors are related to the left and right crates, while the green and yellow (12 and 17) are the guiding CCDs read in their specific crate. See Figure~\ref{fig:TipTiltFocalPlaneCorr} for the CCD numbering scheme of the FPA.}
\end{figure}

The gain for each amplifier was measured at the laboratory just some days before delivery of the camera to the WHT, using the same electronics but with different cable and ground distribution. The mean value for the individual gains, having in  mind that each board has similar but different amplifiers and the response of each CCD/amplifier can be different, is $0.71 \unit{e^{-}/ADU}$ with a rms of $0.02 \unit{e^{-}/ADU}$.  Using this averaged gain, the mean averaged bias is of $ 22 \unit{ke^{-}}$ which is in fact, quite high for an 18 bit Analog to Digital Converter (ADC) but good enough to allocate the $150 \unit{ke^{-}}$ of the CCD full well. Figures~\ref{fig:PreviousGain} and ~\ref{fig:PreviousBias} show the distribution plots of the gain and bias levels respectively for the whole set of PAUCam CCDs.

\paragraph{Readout noise}
The readout noise is a complex issue to deal with in a CCD mosaic since several noise sources interact with each other. Even a very well planned grounding scheme can show drawbacks once the instrument is finally installed on the telescope: i.e.: the electromagnetic interference (EMI) from telescope motors and drives can induce strong noise patterns over the readout electronics despite that the mean noise is still at the lowest levels. Readout noise changes also during the night for different telescope positions.

To foresee these issues from the very first camera design, a \emph{star-ground} scheme (a single ground point) was selected. Therefore every component and electrical device was configured following this guideline and, preferably, having more than one path to reach it. Thereby, during the camera tests at IFAE’s laboratory, a readout noise level below $8 \unit{e^{-}}$ RMS was measured.

\begin{figure}[t]
\epsscale{1.17}
\plotone{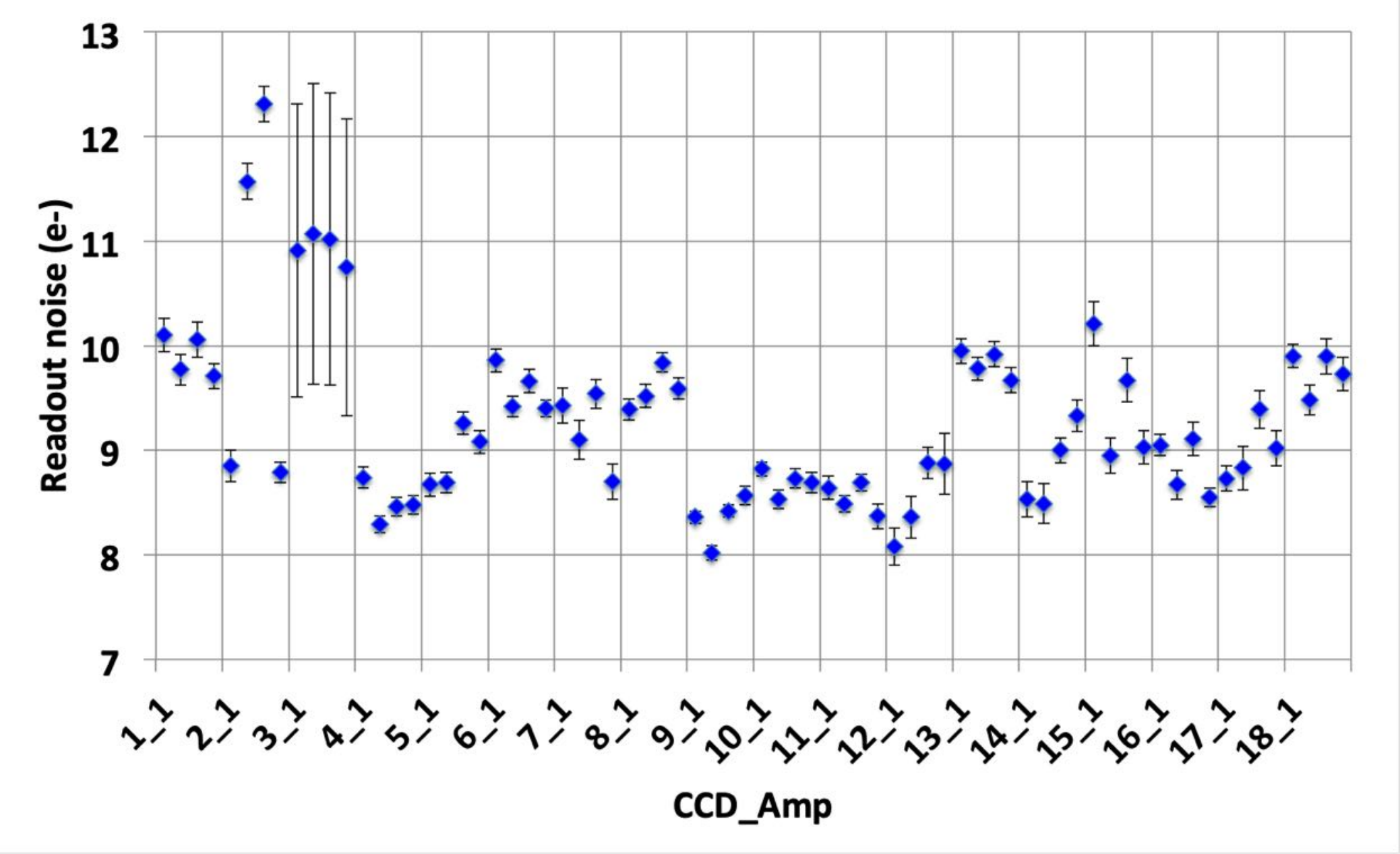}
\caption{\label{fig:NoiseDistribution} Noise distribution for a set of images from a PAUCam run. 
}
\end{figure}

Once installed at the telescope and after a few attempts to improve the camera readout noise during commissioning run (see section ~\ref{sec:commissioning}), the mean noise for the whole set of CCDs stabilized at around $9.3 \unit{e^{-}}$ RMS, although  CCD2 and CCD3 (see Figure~\ref{fig:focalplane} for the CCD numbering scheme of the FPA) showed unexpectedly high noise levels (\textless$11 - 12 \unit{e^{-}}$ RMS), but related to the readout electronics itself (see Figure~\ref{fig:NoiseDistribution}). 

\begin{figure}[tb]
\epsscale{1.17}
\plotone{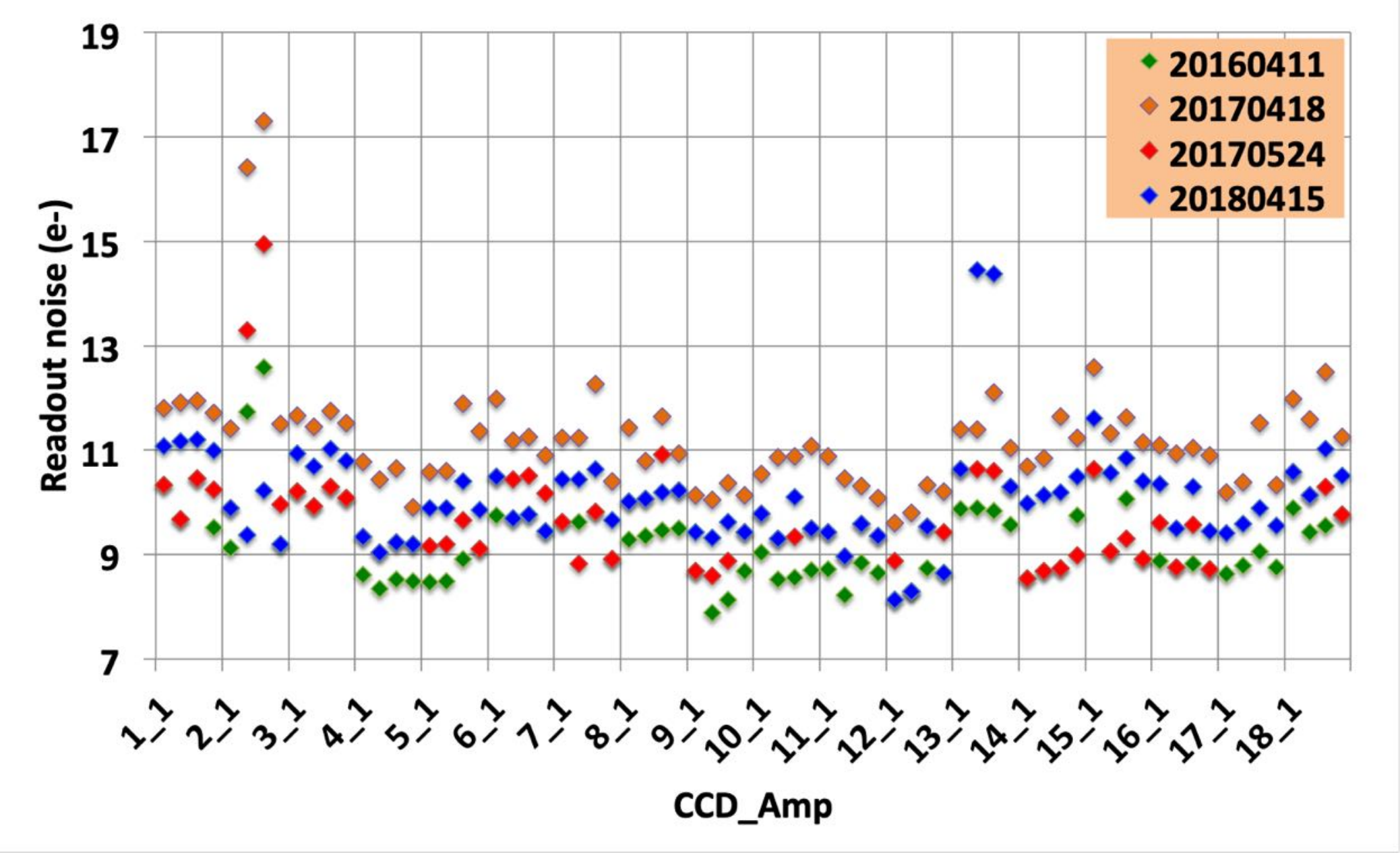}
\caption{\label{fig:readout_noise_years} Readout noise for each CCD amplifier at different times spanning the running period from 2016 to 2018. In the last period, we changed the cabling scheme and transferred the high readout noise from CCD 2 to CCD 13 in order to improve the global performance of the central CCDs.}
\end{figure}

The history of the readout noise in the different observation periods shows a change in the behavior depending of the camera installation and other instrumentation present in the telescope. Figure~\ref{fig:readout_noise_years} shows the changes during the years.

\paragraph{Cross-talk}

The readout electronics cross-talk (that is, the charge induced in a readout channel from the charge collected from a source in a different channel) was measured in the laboratory using a couple of CCDs installed on a test bench cryostat.  At that time, the worst case cross-talk was reported to be smaller than $3/10000$ in electron charge units. At the WHT, operational conditions are different since eighteen CCDs are connected and four MONSOON crates are synchronized to readout the whole FPA. Difference sources of cross-talk are being investigated in the installed system. In fact we have been able to detect cross-talk only in the presence of very large signals somewhere in the focal plane, when a substantial number of pixels reach saturation in a CCD. These effects and their correction are described in \citet{DPPaper}.

\subsection{Mechanical design and camera services}

\begin{figure}[t]
\epsscale{1.15}
\plotone{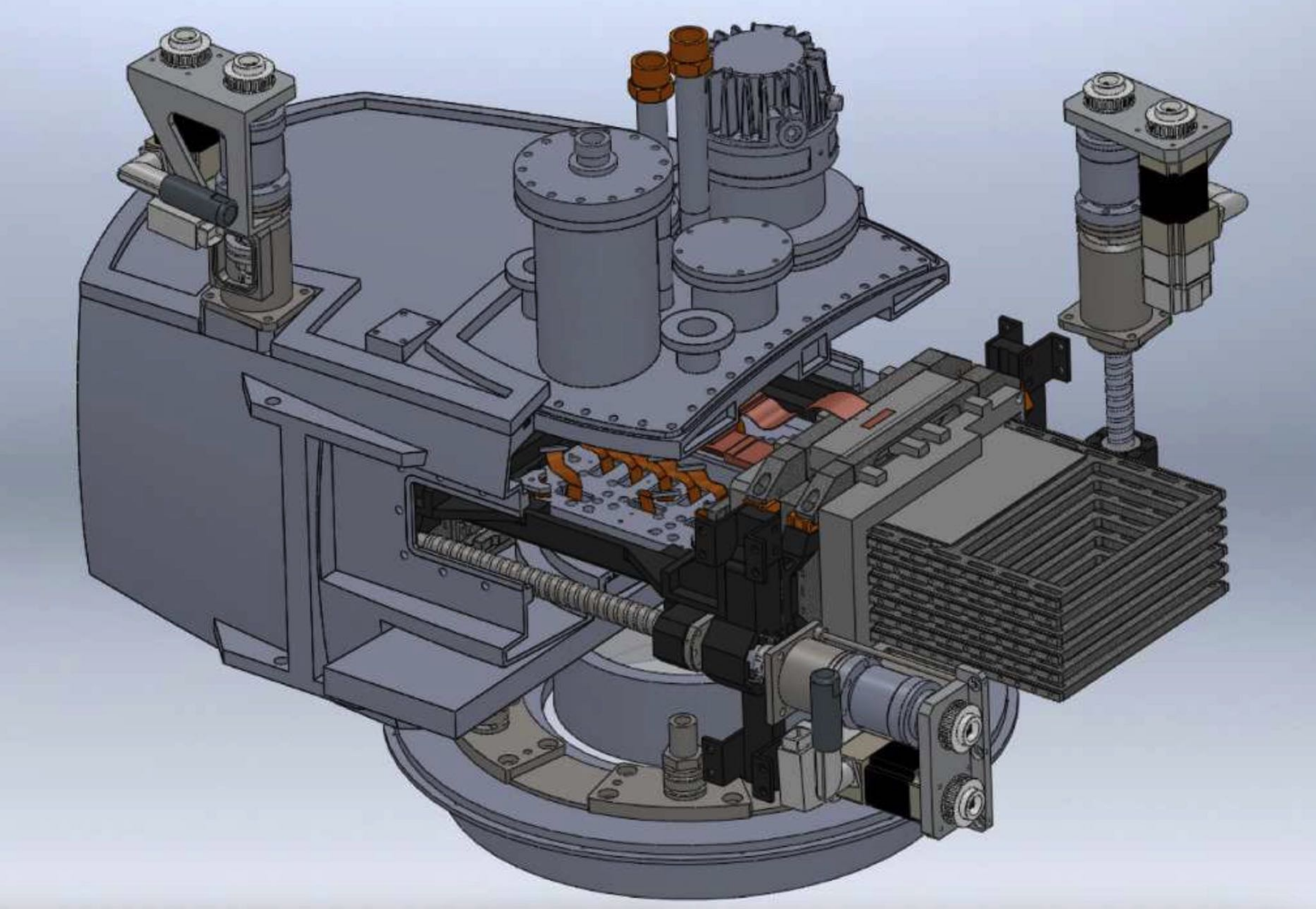}
\caption{ \label{fig:PAUCamcutout} 
Cut-out of the PAUCam design showing the filter exchange system and the overall camera vessel.}
\end{figure}

As stated in Section~\ref{sec:designspecs}, the scientific goals and operational constraints of the PAU Camera required the use of some challenging and non-standard mechanical technologies. Notably, the need to use carbon fiber whenever possible to minimize the weight, to have a filter exchange system inside the vacuum vessel and in cryogenic conditions and to design a dual cooling system to minimize the waiting time to have the camera ready for observations after its installation. The next sections describe these mechanical aspects. An overall sketch of the camera is given in Figure~\ref{fig:PAUCamcutout}. 

\subsubsection{Camera Vessel}

The camera vessel has to be able to maintain a vacuum pressure below $10^{-5} \unit{mbar}$ during operation. The instrument should work in an environment between $5 \unit{^\circ C}$ and $40 \unit{^\circ C}$ and with a humidity between 10\% and 90\%. In addition, the vessel design should include all possible ports for electronics connections, sensors, tray movement motors, vacuum pumps and cryogenic system, ensure a low vibration transmission to all subsystems, control all electrical connections to provide a star-grounding scheme, and minimize the total weight to accomplish with the limitation for instruments working at the WHT prime focus.

In order to minimize the weight of the camera, the material chosen for the vacuum vessel was carbon fiber. A similar vessel made with aluminum would have more than doubled the total camera weight. We made a detailed study to validate that the material works under vacuum conditions. The carbon fiber is a composite of fibers of carbon and non out-gassing epoxy. For our application the air tightness was made by the last layer of epoxy, with an approximate thickness of $0.1 \unit{mm}$. To validate the material, we used a Residual Gas Analyzer (RGA) in two equivalent cubic chambers of $20 \unit{cm^3}$ made of aluminum and carbon fiber. The results of the RGA showed minimal differences between the two.

\begin{figure}[t]
\epsscale{1.16}
\plotone{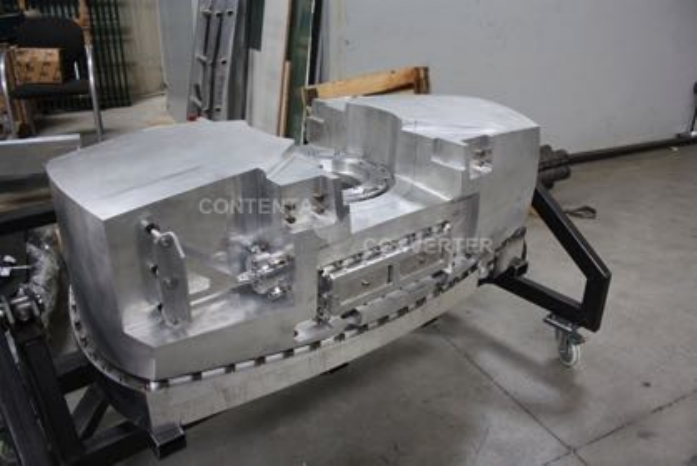}
\plotone{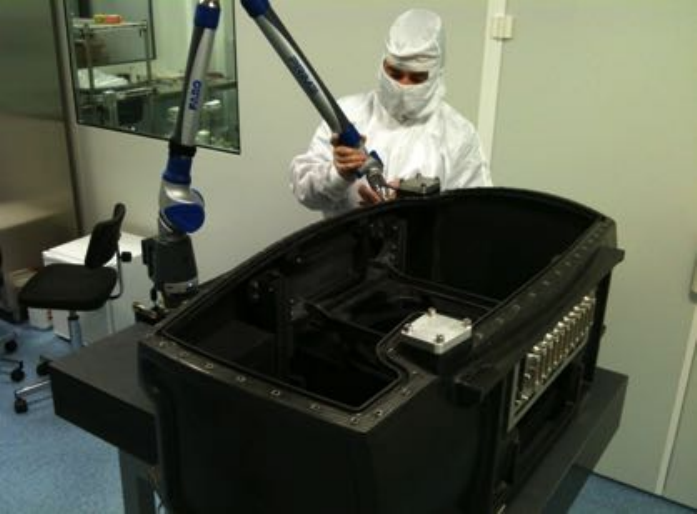}
\caption{ \label{fig:Vessel} 
Aluminum mold specially fabricated for the manufacturing of the PAUCam vessel (top) and the final manufactured empty vessel (bottom).
}
\end{figure}

In order to further minimize the vessel wall thickness, we decided to build it in an egg-shape. After thorough simulations, we chose a thickness wall value of $6 \unit{mm}$, which was sufficient to hold the vacuum pressure forces. We used Viton O-rings for all the services ports. Special molds were made to ensure a smooth surface in the joints in contact with the Viton O-rings. This was specially important for the non-flat (to follow the egg-shape) superior cover joint. Figure~\ref{fig:Vessel} shows pictures of the camera mold and the finished empty vessel. All carbon fiber parts for the PAU Camera were made at Magma Composites\footnote{\url{http://www.magmacomposites.com}}.

The empty vessel tightness was verified with a leak detector and reached a vacuum pressure of $5\times 10^{-6}\unit{mbar}$. After the complete assembly of the camera, it was baked out inside a tent at $60 \unit{^\circ C}$ for 3 days. The low baking temperature was limited because of the CCD specification requirements. Nowadays, during operation and in cold conditions, the PAU camera pressure is stable at $2-3\times 10^{-7} \unit{mbar}$.

\subsection{Vacuum and Cooling system}
\label{sec:VacuumCooling}

The vacuum in the camera vessel is obtained by means of an Agilent\footnote{\url{https://www.agilent.com}} TV-301 turbo-molecular pump installed in the cryostat lid. An additional Agilent IPD-15 scroll pump is placed in the WHT ring. Both pumps are connected through a corrugated tube. For safety reasons the camera is also equipped with a Saes Getters\footnote{\url{https://www.saesgetters.com/}} GP 500 MK5 NEG getter pump which is regenerated at the start of every operation period. 

\begin{figure}[t]
\epsscale{1.15}
\plotone{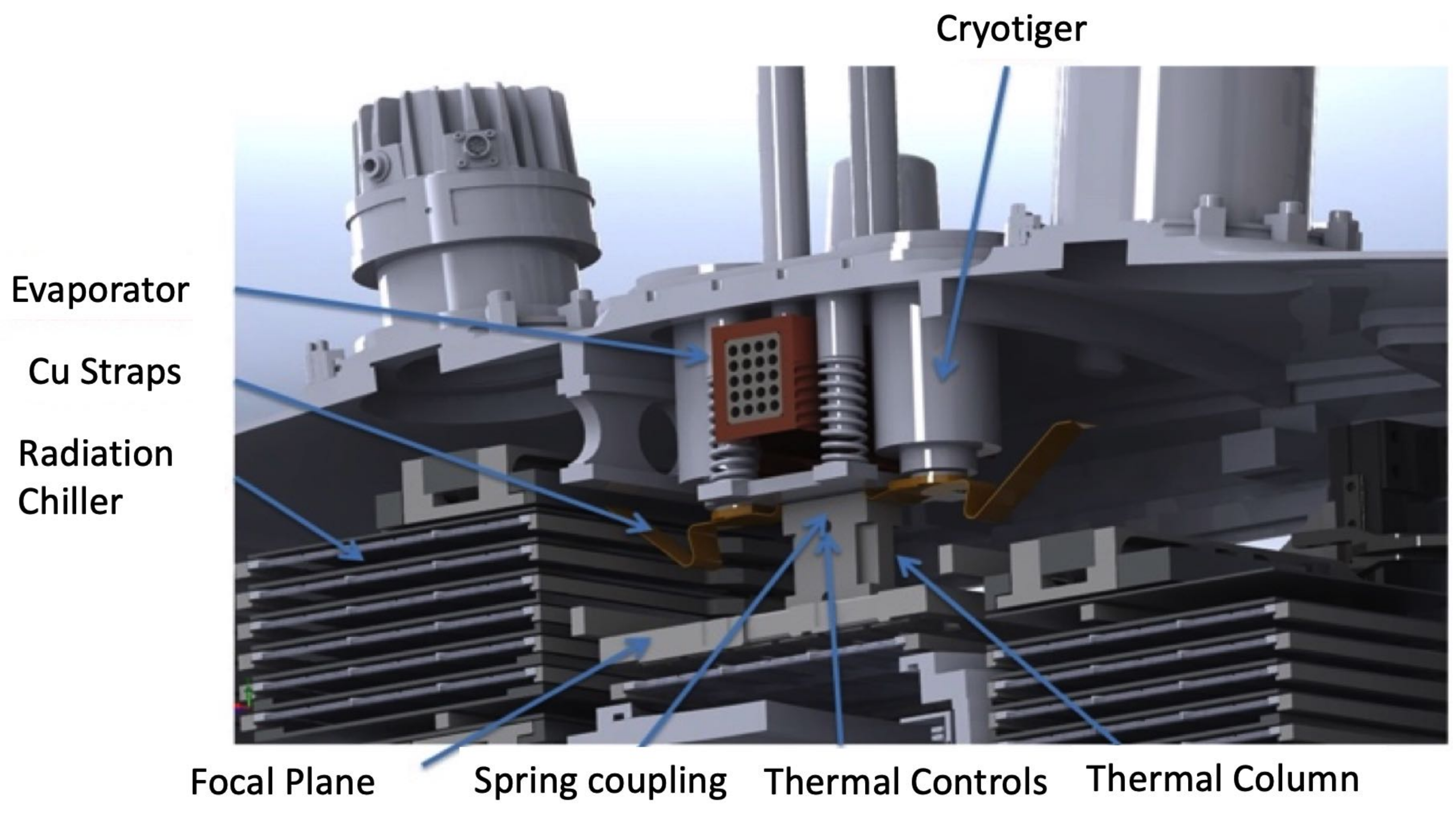}
\caption{ \label{fig:Coolingsytem} 
Scheme of the PAUCam cooling system. Two cryo-tiger heads and the evaporator are connected to straps that cool down the focal plane and the NB filter tray. The evaporator has two tubes (inlet and outlet) for the nitrogen cooling.}
\end{figure}

The camera is equipped with a hybrid cooling system. During regular operations, a set of two Polycold\footnote{\url{https://www.mandtsystems.com/index.php?page=about}} PCC PT-30 cryo-tigers with PT-30 gas maintain the temperature of the cold plate absorbing the power output of the CCDs and preamplifiers mounted inside the cryostat. The system also maintains the NB filter trays (all installed in the same jukebox) close to the equilibrium temperature when extracted and positioned next to the focal plane. This avoids systematic effects arising from the variation of the filter spectral transmission with temperature for these NB filters. After the installation of the camera during the instrument change period at the WHT, the cryo-tigers would take too long (around 12 hours) to cool down the complete system to the operational temperature. Therefore, an additional liquid nitrogen system is installed to allow this process to be performed in less than 4 hours and ensure the camera is operational during the first night after installation. Figure~\ref{fig:Coolingsytem} shows the scheme of the cooling system inside PAUCam, with the two cryo-tiger heads and the two nitrogen cooling tubes connected to the evaporator inside

The heat transfer to the CCD focal plane and the active NB filter tray cooling is made by thermal conduction, connected to the cooling system through flexible copper straps in order to absorb the thermal contractions.

The Focal Plane Assembly (FPA) is made of polished aluminum, in order to avoid thermal losses by radiation with the rest of the camera body, which is black with a radiation emissivity coefficient of 0.6. The CCDs are fixed to the focal plane with stainless steel screws and their aluminum nitrate body is thermally connected to the FPA. In order to avoid stresses in the detectors due to differential contractions, the FPA alignment pins have been thoroughly designed. The FPA is mounted in the vessel structure with flexible brackets to reduce deformations and maintain flatness, and it is electrically isolated from the carbon fiber vessel in order to ensure a controlled grounding scheme. 

Since the filters cannot be touched, their heat transfer works by radiation. We use an aluminum chiller with several plates face to face to the filters trying to keep the view factor as close as possible to 1 and with a black matte anodized surface of $10 \unit{\micron}$ thickness. The temperature control of the whole system is done through a thermal resistance on the top of the chiller. The aluminum plates are connected between them with metal plates of different thickness to ensure an homogeneous temperature among all the filter trays.  A special filter tray with temperature sensors connected to dummy filters was installed. It allows calibrating the system by analyzing the changes in temperature when the filter tray was extracted from its jukebox and checking the temperature homogeneity in all filters in a tray. 

The FPA and the NB cooling system have active heaters to stabilize the temperature. This operation is made by the slow control system (see Section~\ref{sec:slowcontrol}). During operation, the temperature of the CCDs is set at $173 \unit{K}$ and the filters at around $246 \unit{K}$ with fluctuations of the order of $\pm 0.2 \unit{K}$.

\subsubsection{Filter Exchange System}
\label{sec:FilterExchangeSystem}

\begin{figure} [t]
\epsscale{1.17}
\plotone{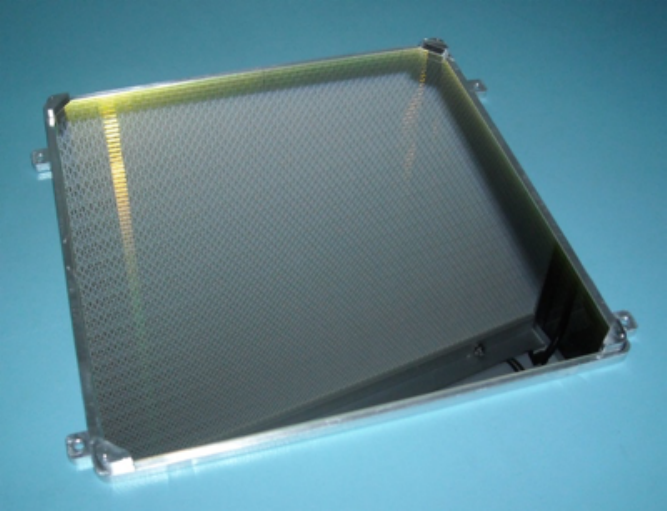}
\plotone{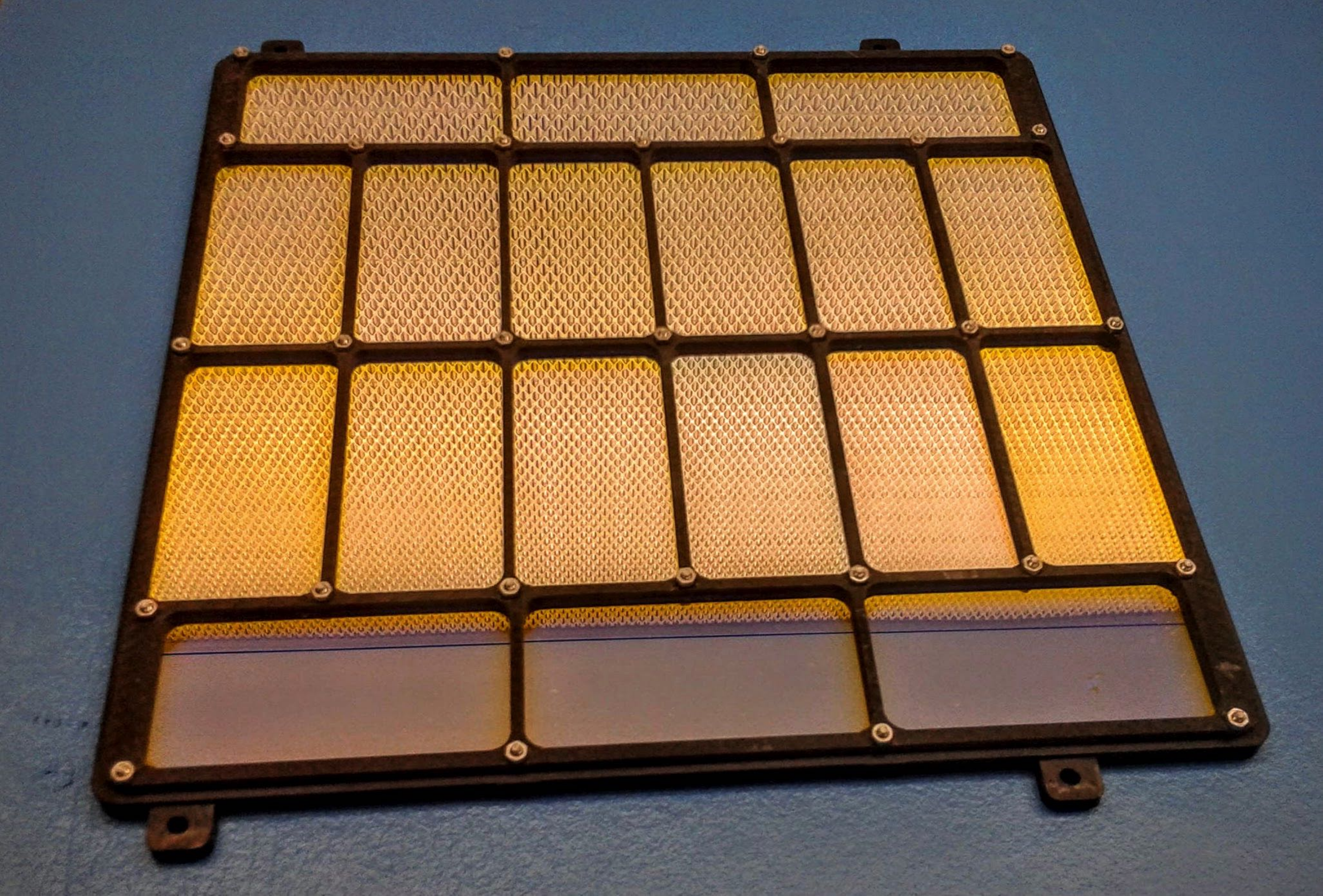}
\caption{ \label{fig:Filters} 
Broad Band filter with its support (top) and Narrow Band filter mounted tray (bottom).}
\end{figure} 

As specified in section~\ref{sec:designspecs}, we decided to place the filter exchange system inside the camera vessel. It consists of two jukeboxes installed on the sides of the camera, each containing 7 tray slots. One contains the 6 BB filter trays and the other (which is actively cooled) the 5 NB filter trays. Two other slots are occupied with the dummy filter tray for temperature monitoring described in Section~\ref{sec:VacuumCooling} and with a shielding. 

The majority of the jukebox and filter tray materials are made of carbon fiber. BB filters cover the whole focal plane and are attached to an aluminum frame interface  mounted in the filter tray structure. The NB filter trays have a filter per CCD and all the structure is made of carbon fiber. Figure~\ref{fig:Filters} shows a picture of both kinds of filter trays.

The up and down (vertical) movements of the trays are guided by a lead screw coupled with a servomotor located outside the vessel. The servomotors are controlled by the slow control system (Section~\ref{sec:slowcontrol}). With these lead screws one selects the tray to be inserted close to the FPA. The range of movement is $120 \unit{mm}$. Once the tray is positioned vertically, a magnetic coupling takes it and a horizontal motor moves it $280 \unit{mm}$, to place it in front of the FPA. The positioning tolerance was verified to be below $0.1 \unit{mm}$.

\begin{figure}[t]
\epsscale{1.17}
\plotone{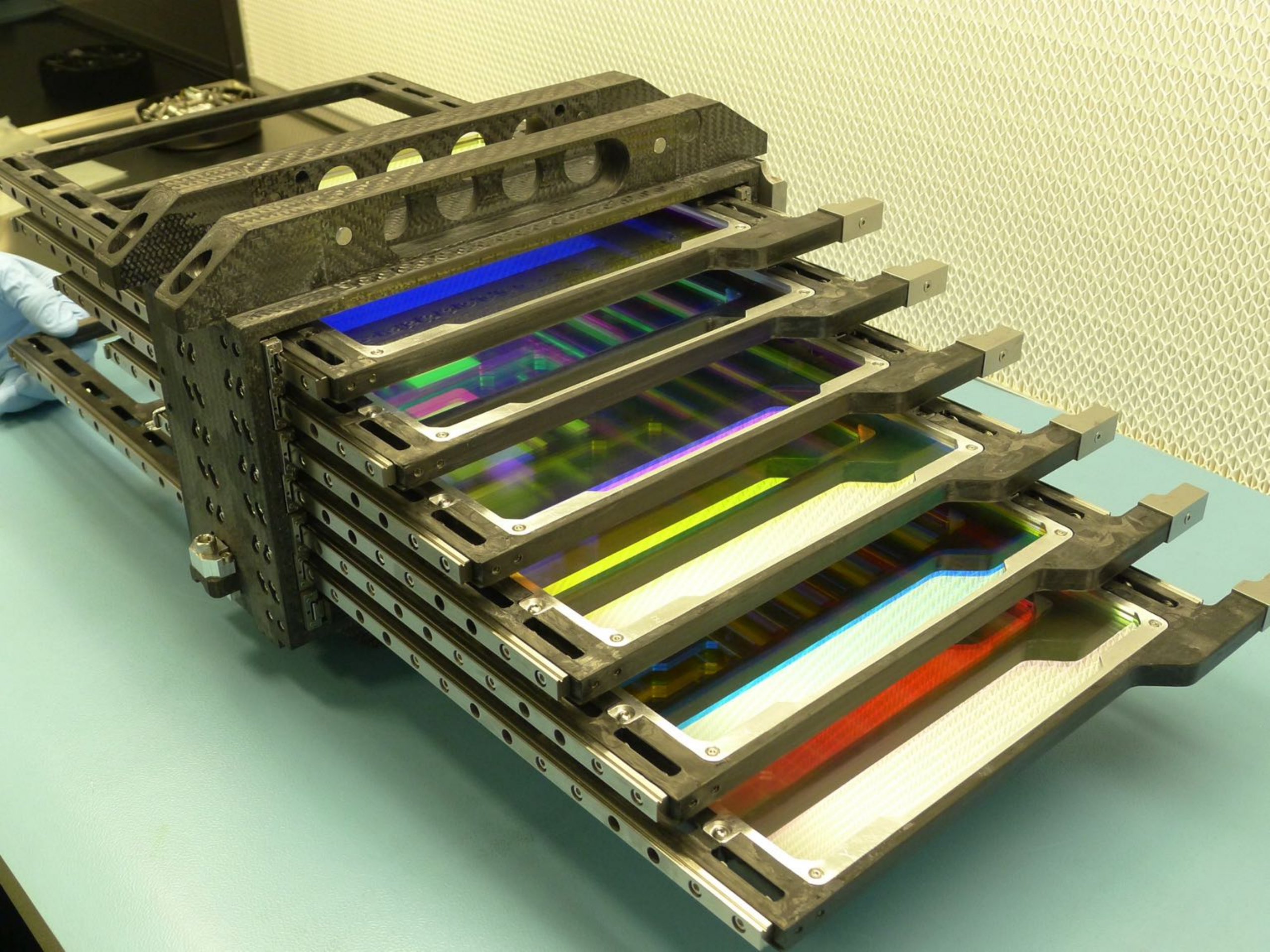}
\caption{ \label{fig:Jukebox} 
A complete jukebox of the Broad Band filters. }
\end{figure}

\begin{figure}[t]
\epsscale{1.17}
\plotone{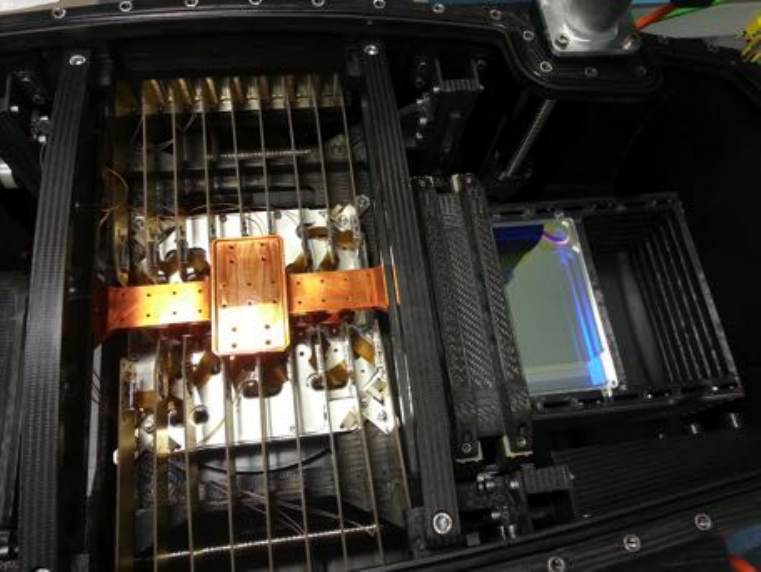}
\caption{ \label{fig:JukeboxMounted} 
Picture of the jukebox inside the camera. One can see the trays in its jukebox in one side of the camera. They will move the center underneath the focal plane which is in the left of the picture underneath the PREAMP boards. }
\end{figure}

Figure~\ref{fig:Jukebox} shows one of these jukeboxes already mounted  with all the BB filter trays. Figure~\ref{fig:JukeboxMounted} shows the jukebox inside the camera with the Focal plane and the preamplifier boards on top of the FPA. During observations, the filters are placed $4 \unit{mm}$ above the detectors surface, sandwiched between the cryostat window and the detectors.

\subsubsection{Shutter}
\begin{figure} [t]
\epsscale{1.17}
\plotone{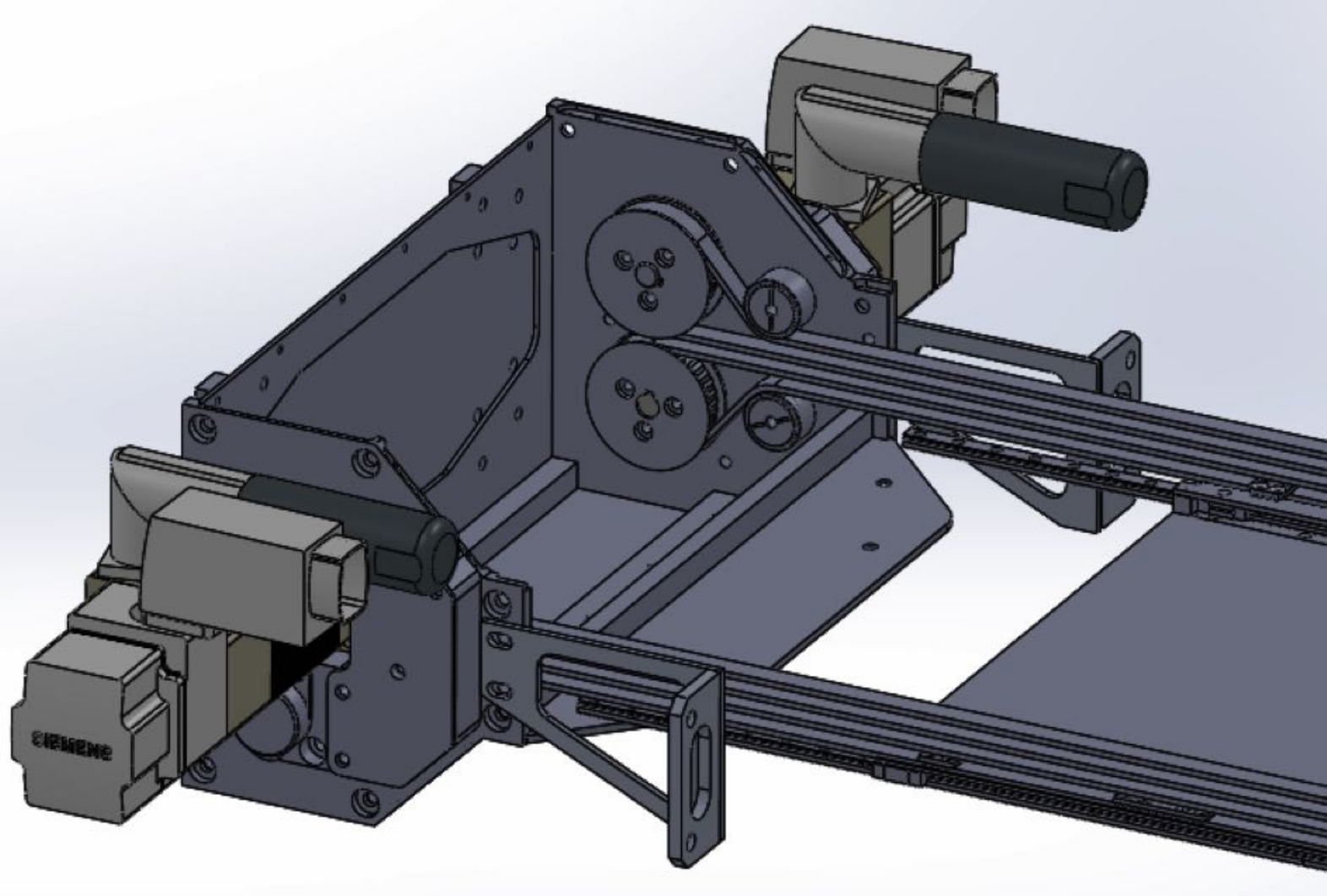}
\caption{ \label{fig:shuttertransmission} 
Detail of the shutter transmission. A Siemens motor acts on a pulleys that move the blades (one per each side of the shutter) through a transmission chain.}
\end{figure}

The shutter of the camera is installed outside the cryostat. It contains two blades sliding in and out of the optical beam. The hardware and control software used is the same as for the filter exchange system (Section~\ref{sec:shuttercontrol}), with a lighter transmission system based on pulleys and traction chains (Figure~\ref{fig:shuttertransmission}). It has an opening aperture of $220\times 220 \unit{mm}$ and a minimum opening time of $0.5 \unit{s}$.

The shutter is made of aluminum, covered with a matt black zinc laminar coating made at Asacoat\footnote{\url{http://asacoat.com}}. This process was chosen instead of the anodized aluminum because of its better anti-reflective properties. In order to avoid water condensation on the lens, a dry air blowing system was integrated into the shutter. It was thoroughly designed to distribute the air in a laminar flow in order to avoid turbulences in the light path.

\begin{figure}[t]
\epsscale{1.17}
\plotone{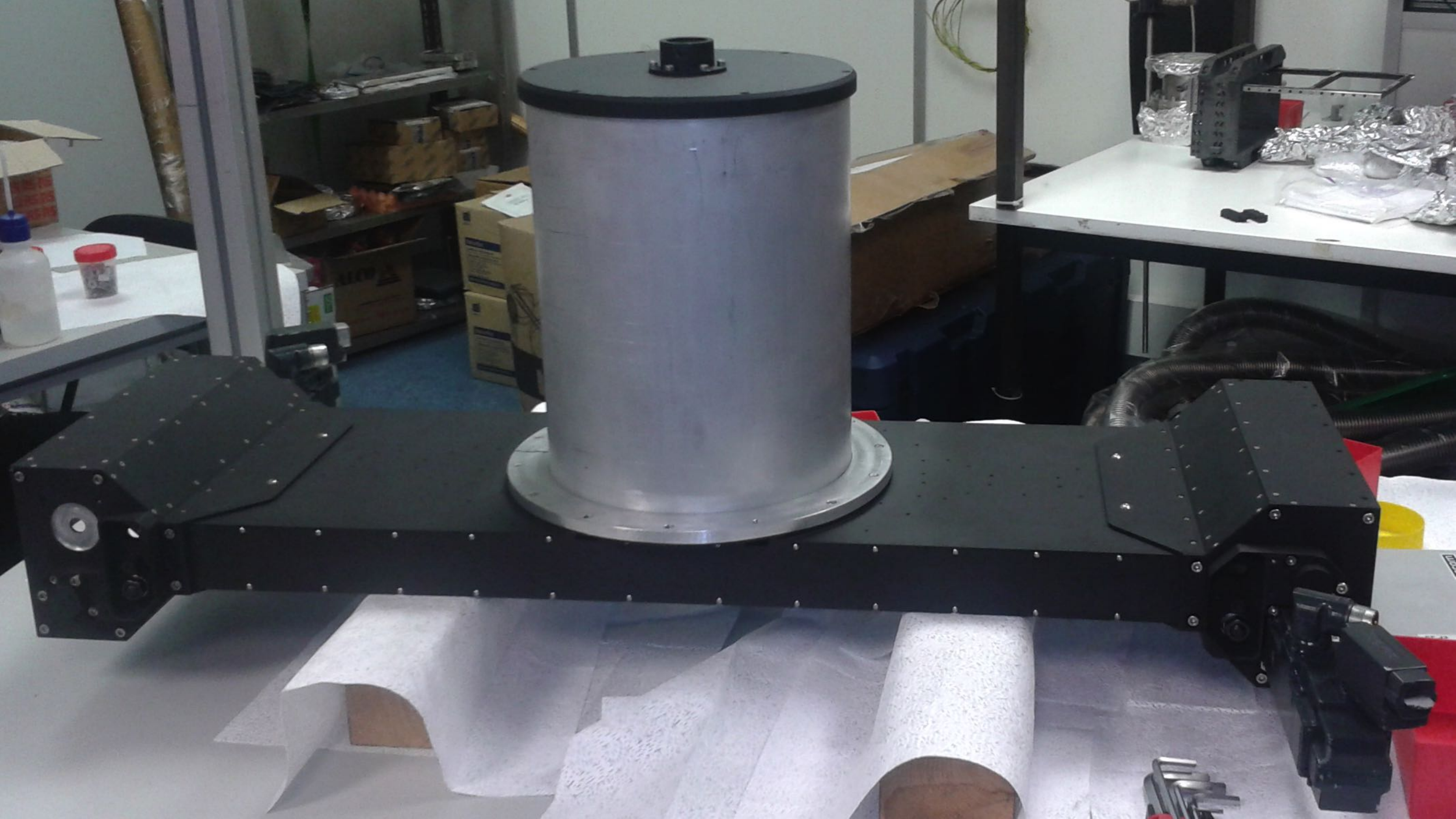}
\caption{ \label{fig:ShutterDarkTest} 
Picture of the shutter during the light tightness tests prior to be mounted in the camera. The top cylinder contains a light sensor to test the light tightness of the shutter. }
\end{figure}

Extensive mechanical stress and light tightness tests were done prior to the shutter installation with satisfactory results. Figure~\ref{fig:ShutterDarkTest} shows a picture of the finished shutter during the light tightness tests prior to be mounted in the camera. 

\subsubsection{Camera Services and Swan Neck}

Several services and electronics are mounted on top of the camera vessel: the two MONSOON electronics racks with all the readout boards; the vacuum pumps; several vacuum pressure sensors; vacuum electro valves and two boxes with the needed actuators and power supplies for the control system. The first scroll vacuum pump is at the telescope ring and connected to the turbo-molecular pump with a $10 \unit{m}$ corrugated tube. The rest of the services are outside the ring: the electronics power supplies, the cooling compressors and the slow control system electronics, are mounted in one of the Nasmyth WHT rooms and the computing systems are in the WHT control room and connected to the camera through optical fibers.

\begin{figure} [t]
\epsscale{1.17}
\plotone{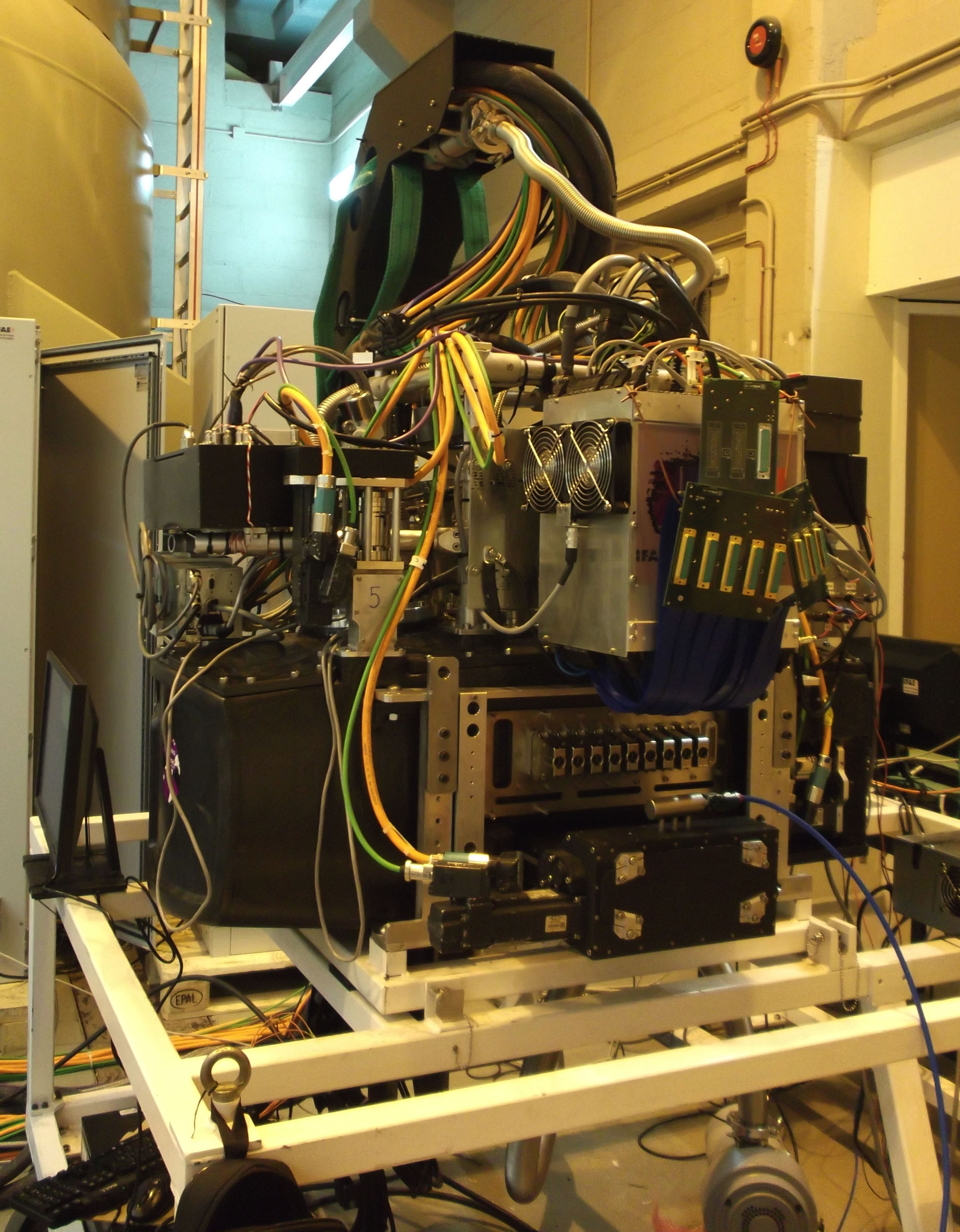}
\caption{ \label{fig:OverviewServices} 
Picture of the camera installed in the testing station at the WHT during its tests prior to the first installation at the prime focus. Detail of the periphery electronics: MONSOON crates, control electronics (inside the black boxes), shutter, and the swan neck with all the needed cabling.}
\end{figure}

Fixed service cables running from a patch panel at the ring to the Nasmyth room are permanently installed in the telescope structure. They include: motor control, cooling pipes, power supply cables, optical fibers and the Profibus\footnote{https://www.profibus.com} link for the slow control system. From the ring patch panel to the camera, all these connections are routed through the telescope spider thanks to a dedicated swan neck with all cables pre-installed. The swan neck is made of black-painted aluminum and helps to speed up the installation of the camera. Figure~\ref{fig:OverviewServices} shows the PAU Camera mounted in its testing station at the WHT prior to its first installation at the prime focus of the WHT. All the services and cables described above can be seen in this picture.

\subsection{Mechanical Verification}
\label{sec:mechanicalver}

\begin{figure} [t]
\epsscale{1.17}
\plotone{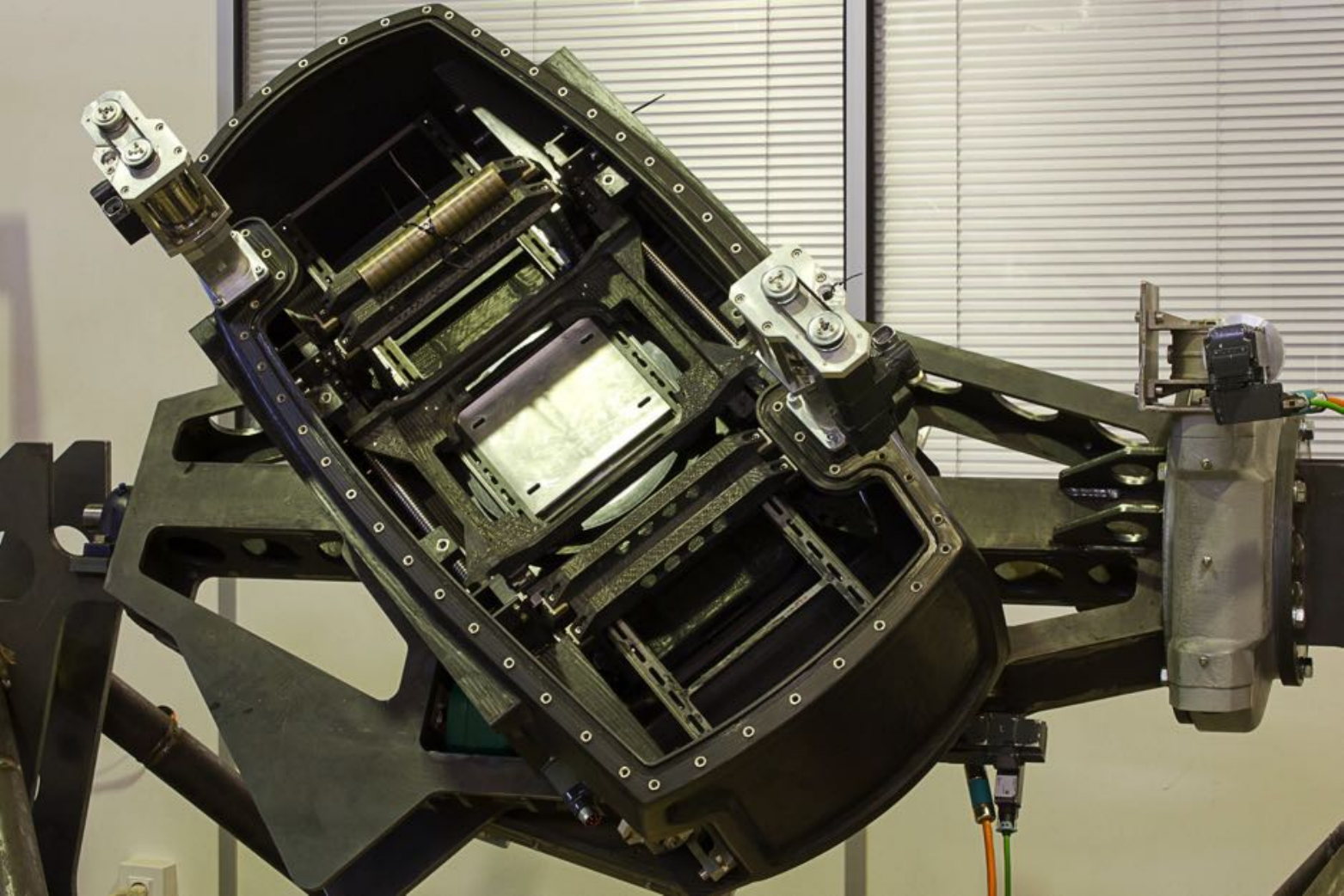}
\caption{ \label{fig:WHTSimulator} 
Picture of the Camera in its first tests mounted in the WHT simulator.}
\end{figure} 
The final verification of the system was done at IFAE. A simulator (see Figure~\ref{fig:WHTSimulator}) of the telescope was built, reproducing the interfaces of the final assembly. It can move the complete camera in every position (elevation and rotation) to simulate real conditions and verify that the tray movement system works properly in real observing conditions.

In order to align the Focal plane with the WHT interface, the Stil OP10000 confocal sensor used to check the FPA planarity, together with the x-y table, was mounted in the simulator and modification on the camera-telescope mounting interfaces where done to minimize the tip-tilt corrections needed after the installation at the WHT.

\section{SLOW CONTROL SYSTEM}
\label{sec:slowcontrol}

\begin{figure*} [tb]
\epsscale{1.1}
\plotone{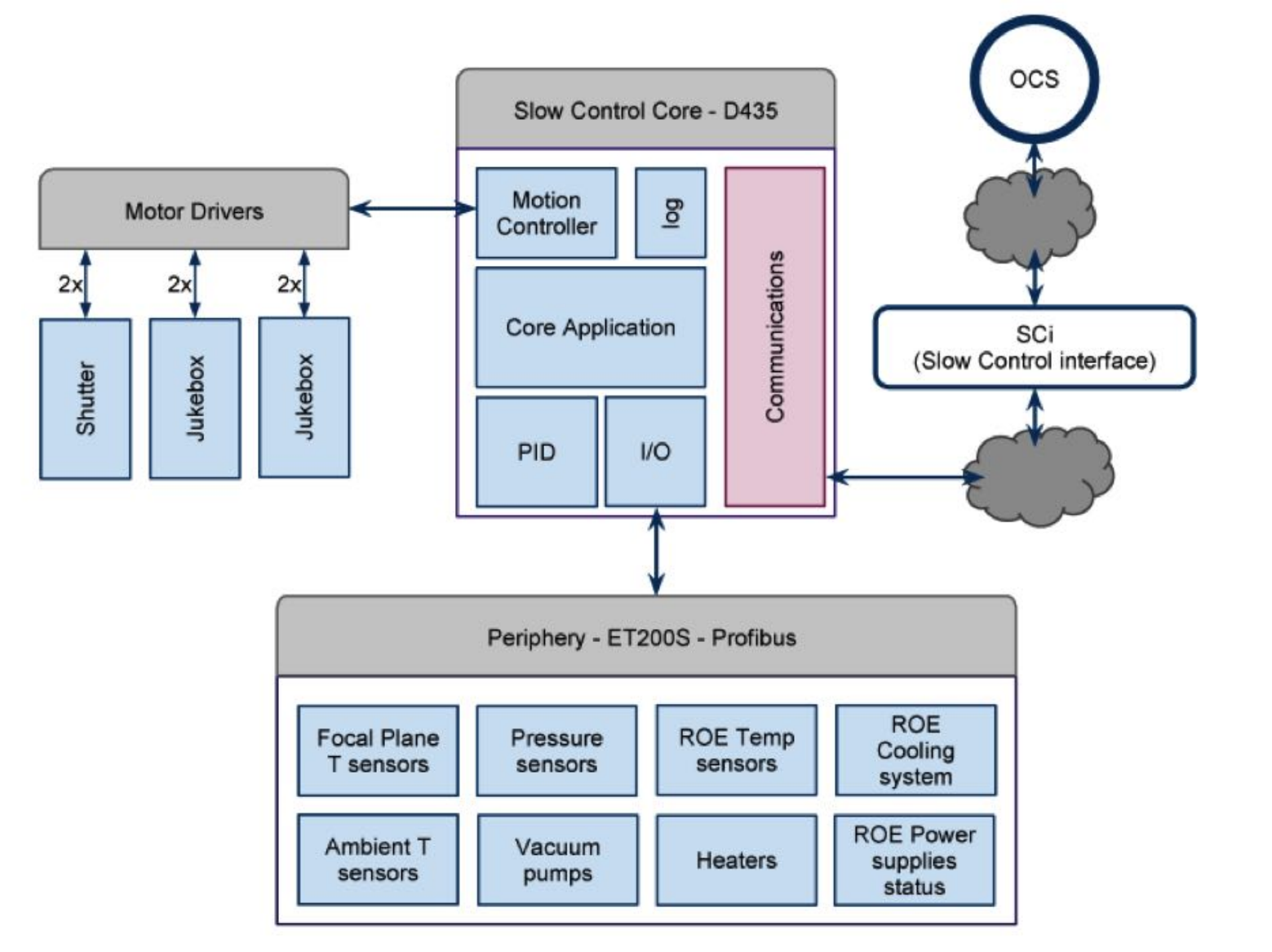}
\caption{ \label{fig:SlowControl} 
   Scheme of the Slow Control System. The SC interfaces with the PAUCam Control system. A Slow Control Core deals the all central communications and controls. It interfaces with the motor drivers and, using a Periphery Profibus, with the rest of the sensors and actuators needed to operate the camera. 
}
\end{figure*} 

The PAUCam Slow Control System (SC) is the software package of PAUCam in charge of implementing all the  motion control, sensors monitoring, actuators control and first safety reaction. The main tasks of the PAUCam SC are: focal plane temperature control, filter jukebox motors monitoring and movement, shutter control, vacuum sensor and pump monitoring, ambient temperature monitoring, and the slow control or the power supplies.

The SC is implemented using a Siemens Simotion D435 Motion Controller\footnote{\url{https://www.siemens.com/global/en/home/products/automation/systems/motion-control/simotion-hardware/simotion-d.html}} which controls all the motors and the connected Profibus periphery. 
The system is very flexible and allows connecting almost any subsystem, controlled through a single interface. Although it is a motion controller, the Simotion D435 can be thought of as a Programmable Logic Controller (PLC). It can manage the position and velocity of up to 16 motors, implement temperature controllers (PID), connect sensors along its periphery, generate as many digital and analog outputs as required and communicate through ethernet, profibus, etc. Figure~\ref{fig:SlowControl} shows the scheme of this implementation and the interface with the PAUCam Control system (see Section~\ref{sec:PAUControlSystem}). A SC Core deals with all central communications and controls. It interfaces with the motor drivers and, using a periphery Profibus, with the rest of the sensors and actuators needed to operate the camera.

A communication protocol to access the registers of the programs running on the Simotion Controller was implemented in Python. It is built on top of the TCP protocol and is based on specially created packages that are sent and received by the different processes. The protocol is multi-client and implements all the features to monitor and log the registers in the Motion Controller, either via a request or asynchronously and to send commands to the different clients to take the needed actions in the camera.

The next sections describe the three main controls that are needed during the PAUCam operation.

\subsection{Motion Control}

The main requirements of PAUCam SC are related to motion control, as filter trays are inside the vessel and they are inserted only a few millimeters away from the CCDs. Therefore, extreme care has to be taken managing the acceleration and speed of the trays, and the torque to be applied to the motors. The monitored torque is used as a main indicator of the correct behavior of the moving parts inside the vessel. The shutter motion control also needs to ensure the exposure time of the camera is consistent.

\subsubsection{Slow Control of the Filter Tray Jukeboxes}

\begin{figure} [t]
\epsscale{1.17}
\plotone{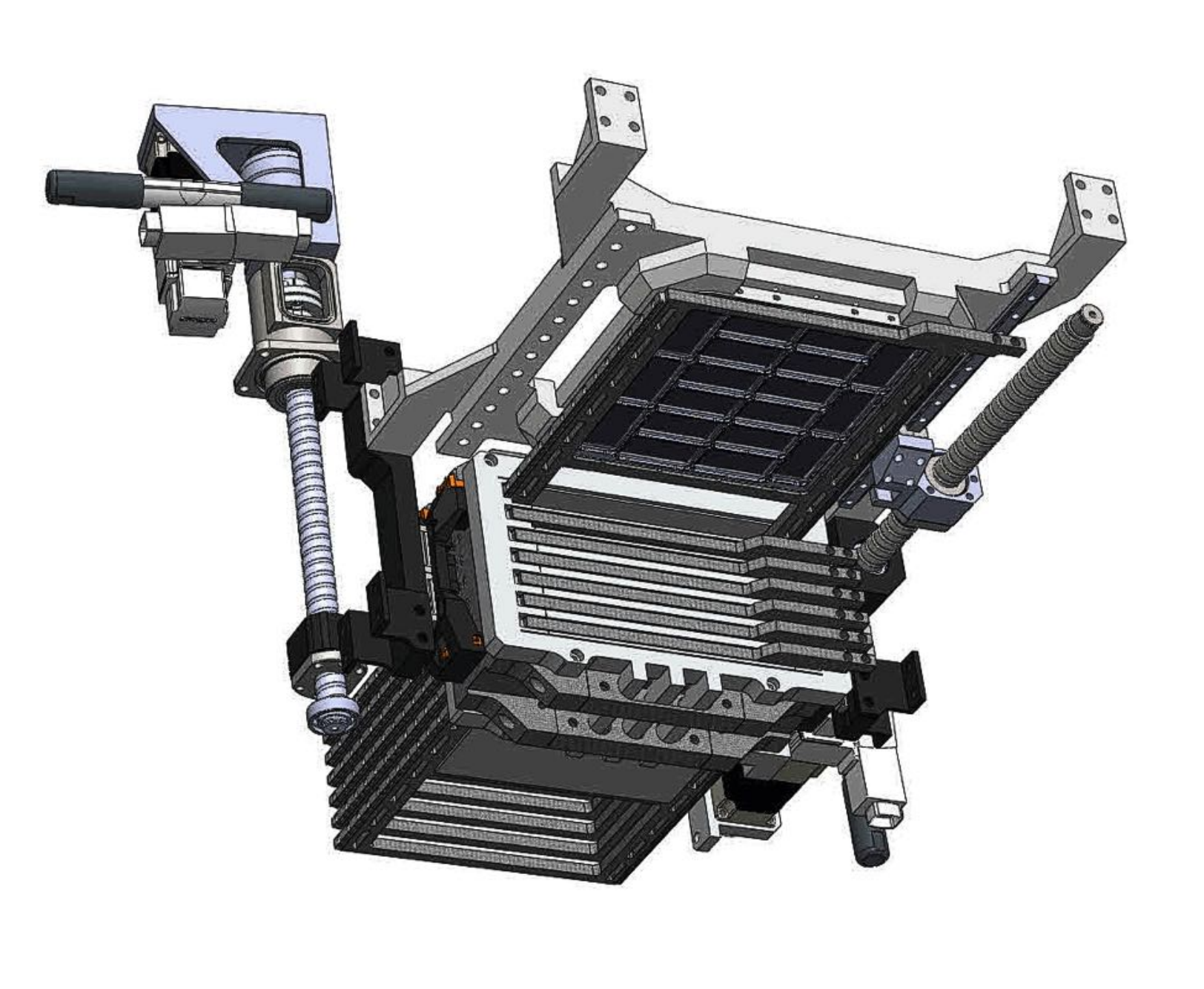}
\caption{ \label{fig:JukeboxFilter} 
 Drawing of one of the PAUCam tray jukeboxes. One of the trays is positioned in front of the Focal Plane. The two lead screws (vertical and horizontal) are controlled by two motors positioned outside the camera.}
\end{figure}

Each of the two tray jukeboxes inside the vessel uses two motors. One motor is used to select the filter tray (vertical motor) and the other one to slide in position the filter tray in front of the focal plane (see Figure~\ref{fig:JukeboxFilter}).

All movements inside the vessel have their torque limited by hardware in order to minimize the creation of particles in case of friction or collision. 

Vertical movements are allowed only to discrete positions. Only at these positions can the filter tray slide out of the jukebox. They are calibrated for each of the filter trays and their values stored into non volatile memory. As there are no hardware limits to the movements inside the vessel, the calibration of the system is a critical step of the development.

\subsubsection{Shutter Control}
\label{sec:shuttercontrol}

\begin{figure} [t]
\epsscale{1.17}
\plotone{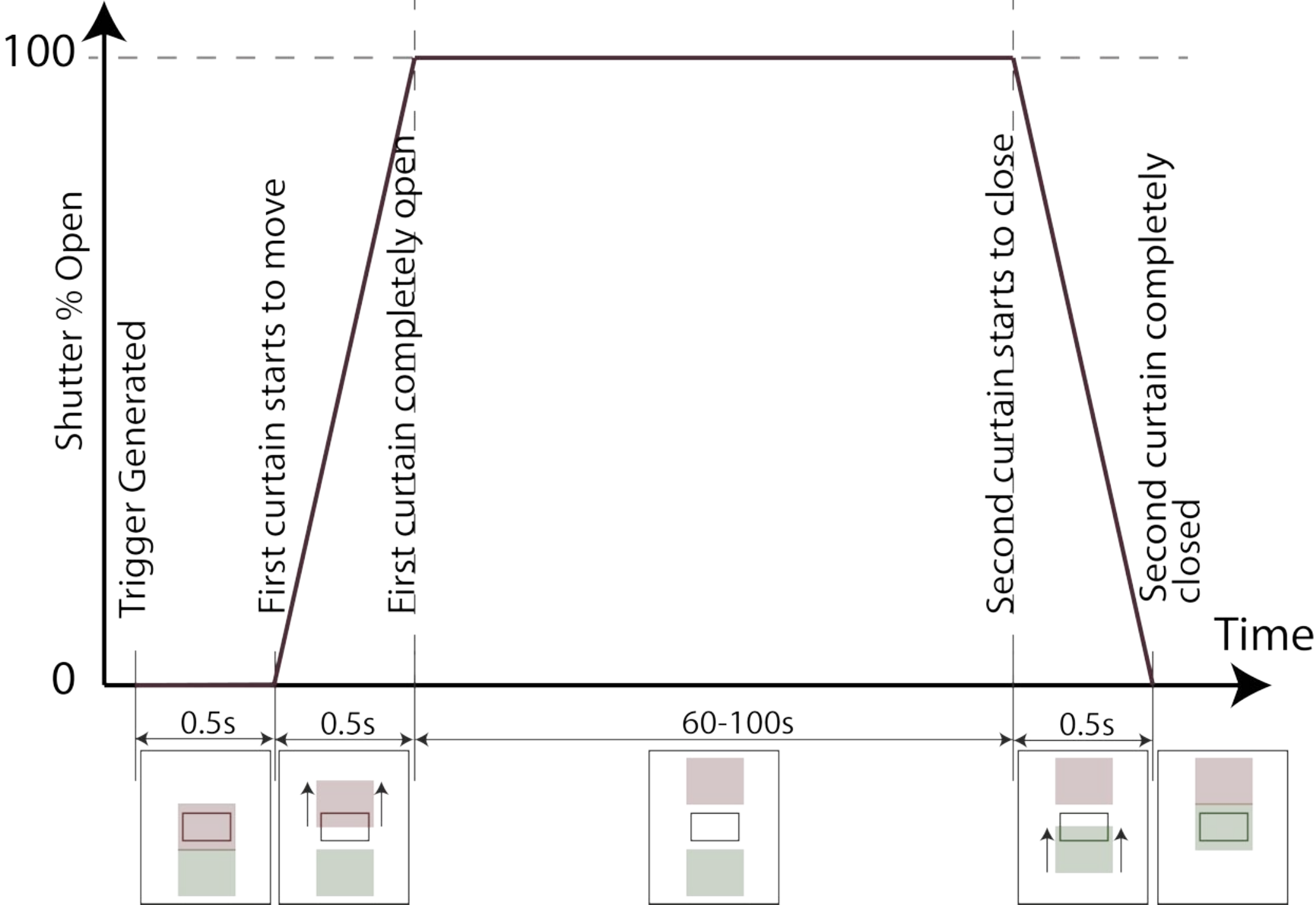}
\caption{ \label{fig:ShutterOpenClose} 
 Shutter profile. In $0.5 \unit{s}$ the first blade is open (see the schematic view of the two blades below the graph), then, the shutter is open for the defined calibrated time and, finally, the other blade closes with the same profile (in $0.5 \unit{s}$). }
\end{figure}

The PAUCam SC is in charge of opening and closing the shutter in order to take exposures. The PAUCam Data Acquisition System (DAQ) (see section~\ref{sec:DAQi}) provides a TTL signal to open and close the shutter for a given time. 
All shutter movements follow a configured profile which tries to minimize jerking movements and transit time. Two motors control each one of the shutter blades. With the usual exposure times of PAUCam, first a blade is open and, at the end of the exposure time, the other blade is closed. For shorter times, the first blade slightly opens and the second one starts closing shortly after, without waiting until the first one is completely closed. In all cases, special care must be taken to equalize the profiles of both blades and that the open and close operations are equal. Figure~\ref{fig:ShutterOpenClose} shows this profile, calibrated during the PAUCam assembly process. 
A video camera at 50 frames per second was used to estimate the exposure time error at $500 \unit{ms}$. It resulted in an error below 1\%.

\subsection{Sensor Monitoring}

PAUCam has several sensors to monitor its status. Based on sensor values and the camera state, the SC can take safety measures to protect the CCDs, starting the relevant protocols to put the camera in a safe mode. Temperature sensors are placed in the cold heads, focal plane, filter tray jukeboxes, etc. They allow the SC to monitor the current temperature and its change rate, and decide any needed action. There are two redundant  vacuum sensors in the camera vessel and additional ones in between the turbo-molecular and the scroll pumps.

To allow for several responses of the system, alarms and security routines depend on the camera state (parking, cooling, operation), which are set by the PAUCam Camera Control System (see Section~\ref{sec:PAUControlSystem}). For example, at any given time, the SC might have to heat the CCDs cold plate if the vacuum level surpasses a configured limit.

\subsection{Actuators}

The SC has limited access to non motors actuators. Basic actuators on the camera system are vacuum pumps, cooling compressors and heaters.

The turbo-molecular vacuum pump is permanently enabled and the SC is used for its activation only. 

The Cooling compressors ensure the Focal Plane and NB filter trays are at the required temperature. The compressors are monitored, activated and disabled by the SC. To correct the temperature on the focal plane and to allow the warm up the CCDs following a controlled profile, the heaters inside the camera are used. Heaters are the only active control over the temperature of the system.

\section{CONTROL SYSTEM}
\label{sec:PAUControlSystem}

\begin{figure} [t]
\epsscale{1.1}
\plotone{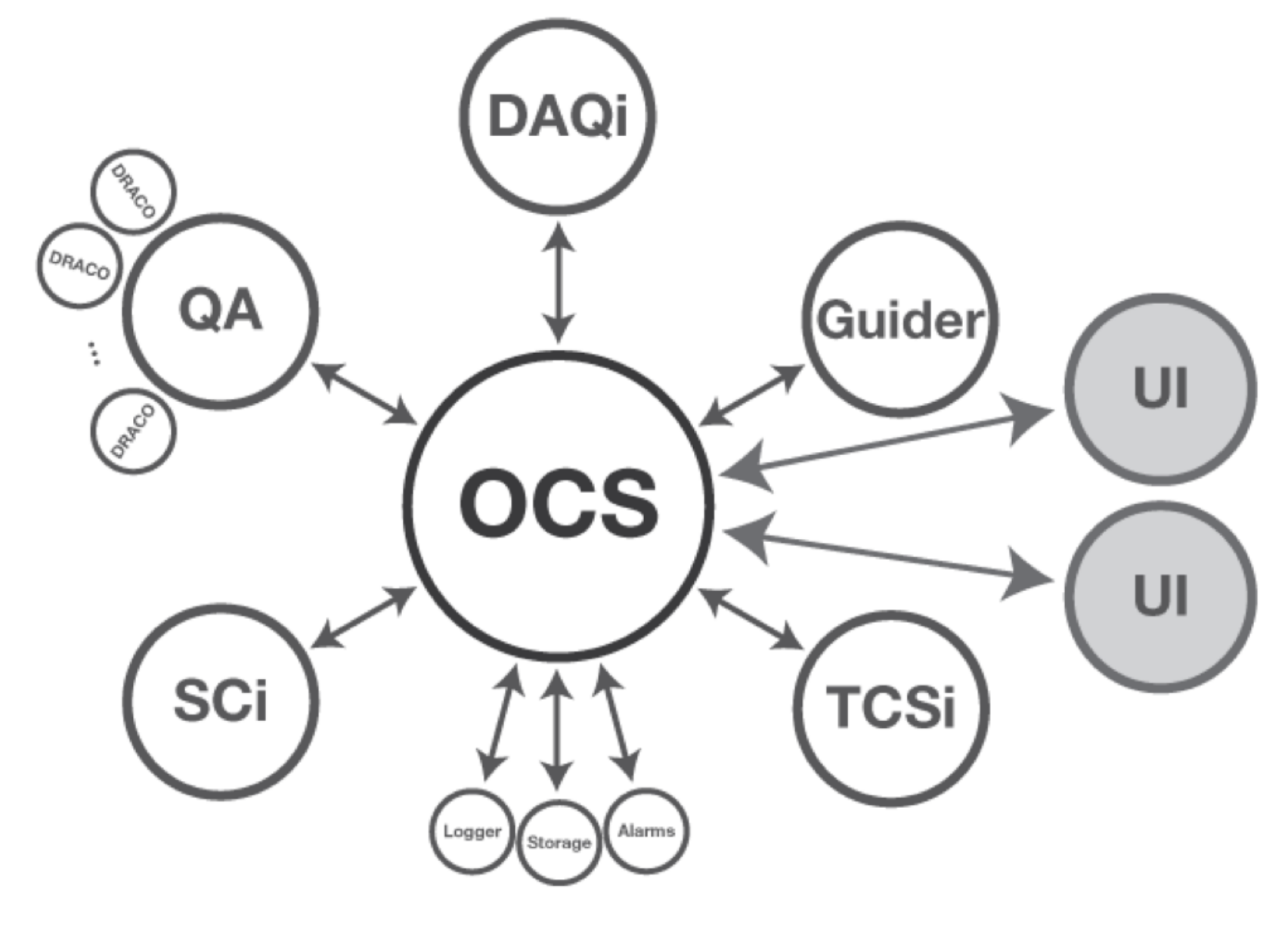}
\caption{ \label{fig:CCSScehem} 
The PAUCam control system scheme. The OCS is the central brain of the system which coordinates the rest of services, such as the interface to the telescope (TCSi), the Quality analysis (QA), the User Interface (UI), etc.}
\end{figure}

The PAU Camera Control System (CCS) is in charge of controlling all PAUCam  subsystems. It is organized as a central node, the Observation Coordination System (OCS), and several satellite nodes. The satellites implement interfaces to other subsystems and the telescope. The main task of the OCS is the coordination of all subsystems to acquire an exposure. The list of satellites includes: the Slow Control interface (SCi), the Data Acquisition interface (DAQi), the Online Quality analysis, the Guider, the Telescope Control System interface (TCSi), and Storage and Alarms. Figure~\ref{fig:CCSScehem} shows the scheme of the CCS.

The PAU CCS is designed as a distributed software system as some of its processes must run in separate computers and because certain subsystems consume large amounts of resources. Satellites can be installed on any computer on the network, and are launched by the OCS. These features require the use of a communication system to send messages between subsystems. The exchange of information is done through the Advanced Messaging Queue Protocol (AMQP)\footnote{\url{http://www.amqp.org/}}
which is an open standard for passing messages between applications. The AMQP follows a client-server based architecture. Each client sends to a central server the message to be transmitted, including tags to enable server routing. The server receives every message and redirects it to a different queue for every client subscribed to that message tag. These queues are handled by each client which take care of the messages one at a time, as the queue is emptied. The chosen server implementation of AMQP is the open source RabbitMQ\footnote{\url{http://www.rabbitmq.com/}} which is an open source broker written in erlang\footnote{\url{https://www.erlang.org}}. The main benefits of using this server are the ability to run on all major operating systems, its support of a huge number of developer platforms and its inclusion of an http server extension to check messaging server status.

 As huge files (FITS) have to be passed between clients, a file transfer library has been developed too. This alleviates the load on the AMQP server making execution and variable passing much more reliable.

Both utilities can be accessed by every client through a framework, which offers to every client all the functionality: exporting functions, subscribing and accessing variables and sending/receiving files.

The next sections detail the main tasks of the PAU CCS.

\subsection{Observation Coordination System (OCS)}
\label{sec:OCS}

The OCS is the core of the PAUCam control system. It is in charge of coordinating all other subsystems to acquire exposures. It has two types of interfaces, one to connect to the satellite nodes and another for the user interfaces. It is in charge of building the system when a new configuration is loaded, launching each of the satellites defined in the configuration at the specified IP address and executing and coordinating every observation during the night. In order to perform this task, different levels of executions are defined (see Figure~\ref{fig:OCSExecution}):

\begin{figure} [t]
\epsscale{1.17}
\plotone{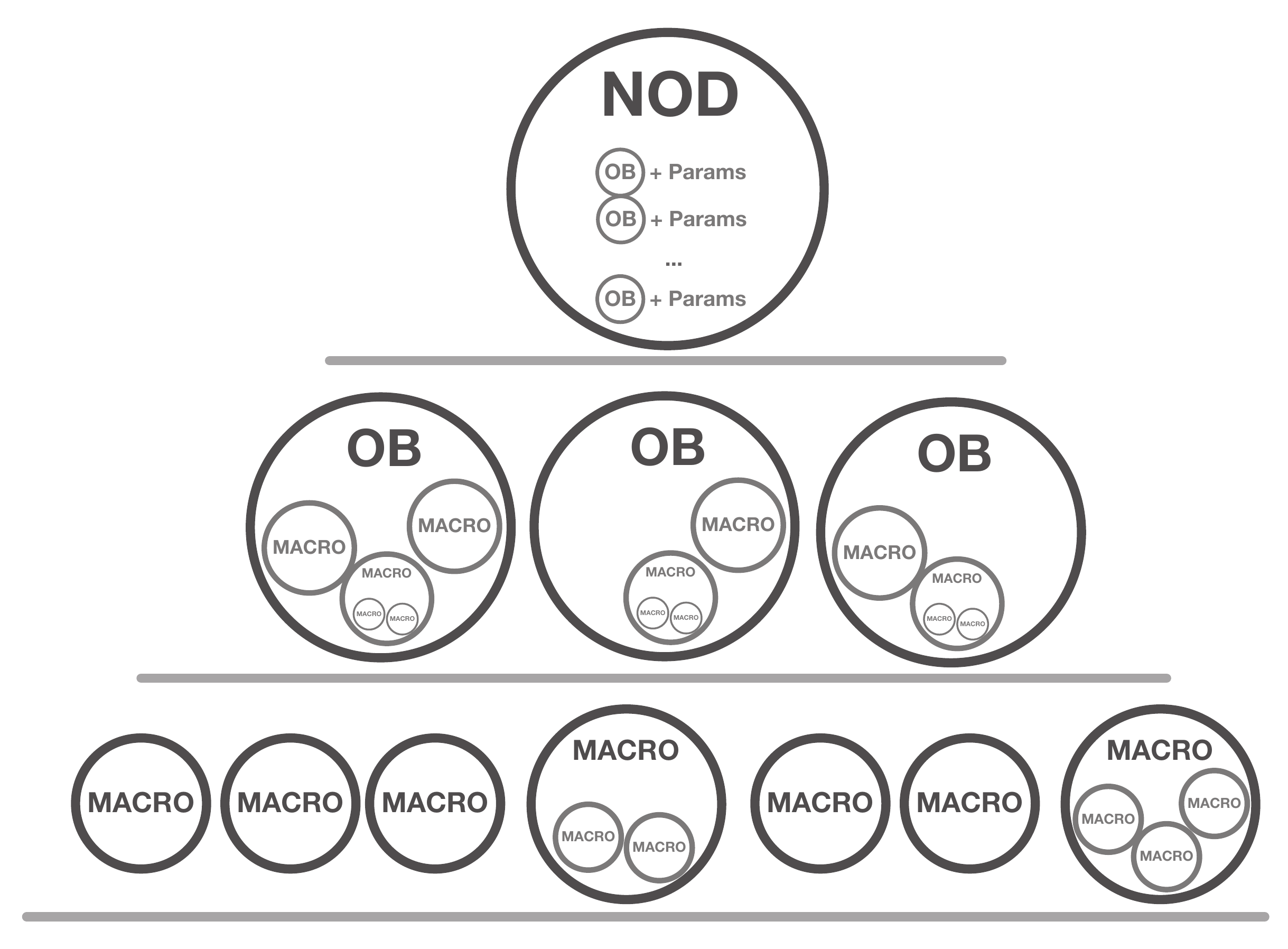}
\caption{ \label{fig:OCSExecution} 
Levels of OCS execution: macros (and recursive macros), Observation Blocks and Night Observation Definition.}
\end{figure} 

\paragraph{Macros}
The OCS loads macros defined at specific locations. Macros are the base commands made available to the user, who has no direct access to satellite functionality. Macros can call other macros. The OCS, when loading macros, checks that there is not an infinite recursive call.


\paragraph{Observations Blocks (OB)}
The OCS can define several types of exposures (bias, flats, science and others) which are also loaded dynamically at boot. Each exposure is defined in terms of high level macros and can be considered itself a macro. When a macro defines an exposure, it is called an observation block.

\paragraph{Night Observation Definition (NOD)}
An observer can define a night observation plan in advance, setting which calibration exposures and which targets wants to acquire. A NOD is composed by a set of parameter templates and a list of Observation Blocks. Once loaded, PAUCam can take all exposures automatically.

\subsection{Data Acquisition interface (DAQi)}
\label{sec:DAQi}

The system used to read out the PAUCam CCDs is based on MONSOON+PanVIEW (see Section~\ref{sec:readout} for details). The function of the DAQi is to provide the PAU CCS with an interface to several PanVIEW instances to be able to use it from the OCS as a block.

The PanVIEW system  is in charge of getting pixels and header information out of a detector system through a digital link to the MONSOON crates installed in the camera. 

The DAQi provides functionality to configure the readout electronics for several read out modes. It implements the control for regular and guiding exposures, depending on the exposition mode used. As several PanVIEW instances are used, each one generating a FITS file, the DAQi is in charge of creating a single FITS file once all parts are received. When a new FITS file is ready, it publishes its location, so any other satellites requesting it can use it.

\subsection{Telescope Control System interface (TCSi)}

PAUCam is a visitor instrument of the WHT. The TCSi is in charge of transmitting commands to the WHT Telescope Control System (TCS) and accessing the WHT Meteorological database. The TCSi has several interfaces with the WHT. 

\begin{figure} [t]
\epsscale{1.17}
\plotone{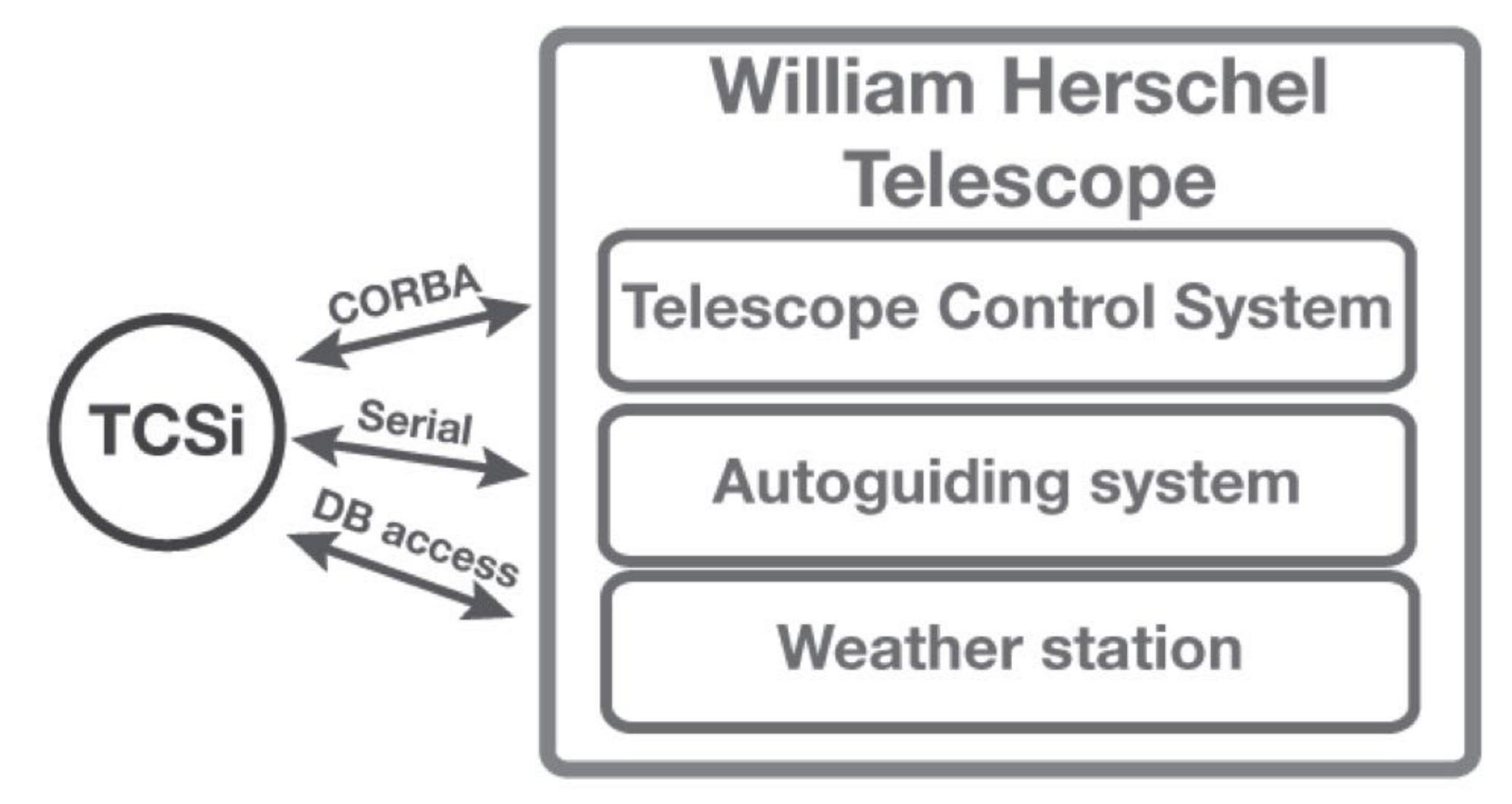}
\caption{\label{fig:TCSi} 
Scheme of the TCSi and its links to the WHT control and information.}
\end{figure} 

As can be seen in Figure~\ref{fig:TCSi}, the WHT offers a CORBA\footnote{\url{http://www.corba.org/index.htm}} based software interface to operate with its TCS. The WHT TCS enables the PAU CCS to point the telescope to a target, set the focus, enable or disable auto-guiding and send the corrections computed by the guider process (see details in Section~\ref{sec:autoguiding}).


The WHT has a local weather station that stores its data to a postgreSQL\footnote{\url{https://www.postgresql.org}} database. The TCSi accesses this database periodically and publishes its latest results to the PAU CCS system, so its data can be integrated on resulting FITS files.



\subsection{Quality Analysis (QA)}
\label{sssec:draco}


Quality Analysis is the satellite in charge of calculating the quality of the images acquired and give the possibility to generate a visual check of the acquired FITS for the observer.  The QA starts as many DRaw, Analyze and Check Observations (DRACO) instances as required either in local host or in remote machines and it is in charge of distributing jobs to them. Each DRACO then writes its results into a database. The different analysis performed with DRACO are detailed in Section~\ref{sec:Monitoring}.

\subsection{Slow Control interface (SCi)}

The PAUCam SCi is the satellite in charge of sending commands to the SC (section~\ref{sec:slowcontrol}) and publishing its sensors values. 


\subsection{Alarms and monitoring}

The PAUCam SC has a main state machine and several sensors that determine the status of the camera. On the main server of the system, there is a service that stores most of the sensor data continuously and organizes it to be able to retrieve time series of it.

A web page has been created to follow the evolution of the camera and its status throughout its complete operation history. There is also a database that stores the observation blocks executed by the OCS and its parameters as well as any exception that has occurred during the night execution. Finally, under certain circumstances, a service sends emails to a list of trained personnel when the camera is in an emergency state.

\subsection{User Interface (UI)}

In addition to the interface used for all satellites described above, the OCS has a socket interface to enable user interfaces to send commands and receive its return values. The PAU CCS has two types of user interfaces, a terminal command and a graphical user interface (GUI).

The GUI is the main interface between the operator and the system. It contains all the system information and keeps a history of the systems use. The GUI is organized hierarchically to allow the user to have the control needed for each operation (execute different exposure types, control the telescope, NODs, etc.). The GUI also  integrates displays that show the alarms, the detailed logging, the OBs already executed, etc.



\subsection{Logbook}
\label{sec:logbook}

\begin{figure}[t]
\epsscale{1.06}
\plotone{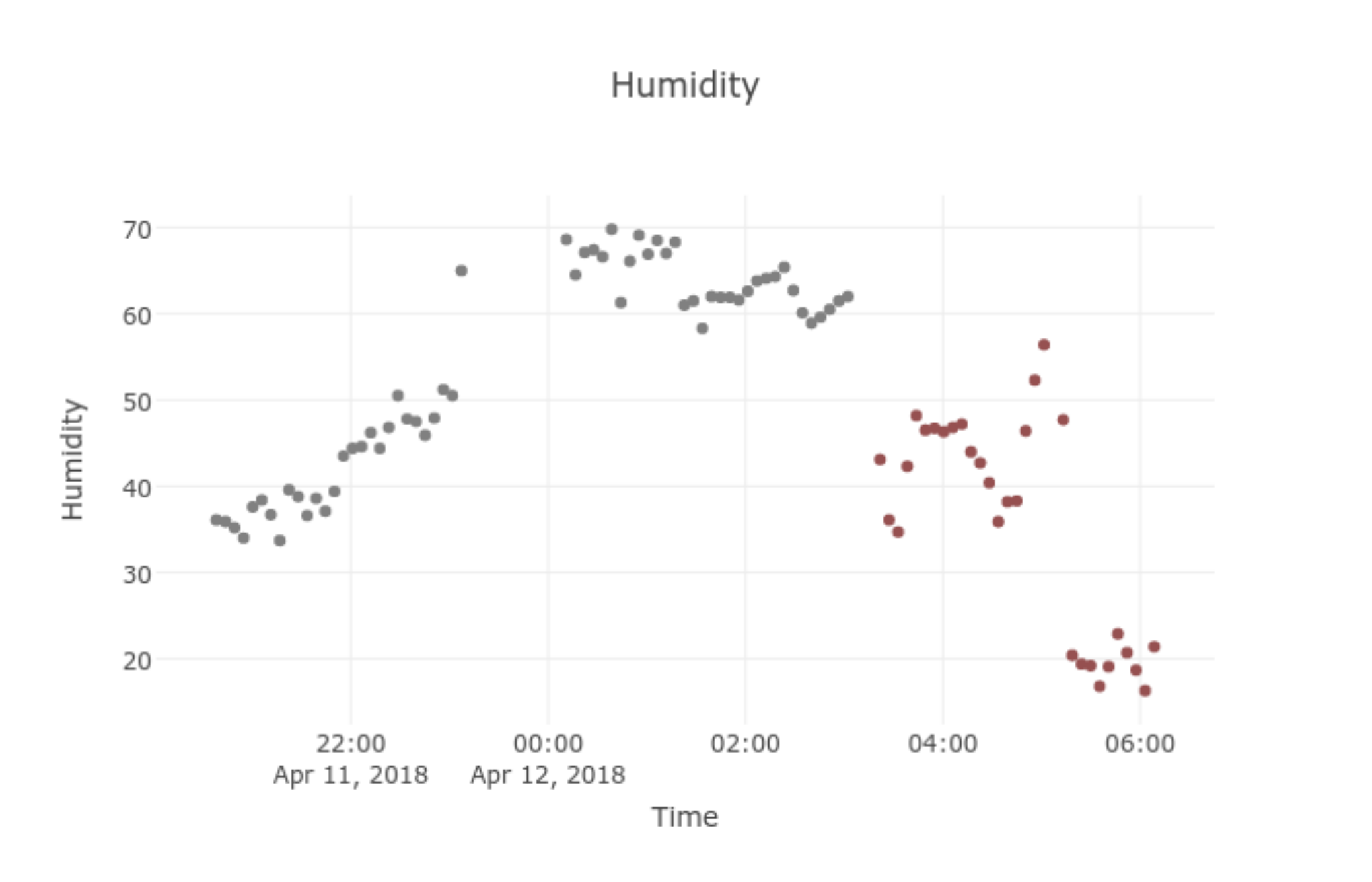}
\plotone{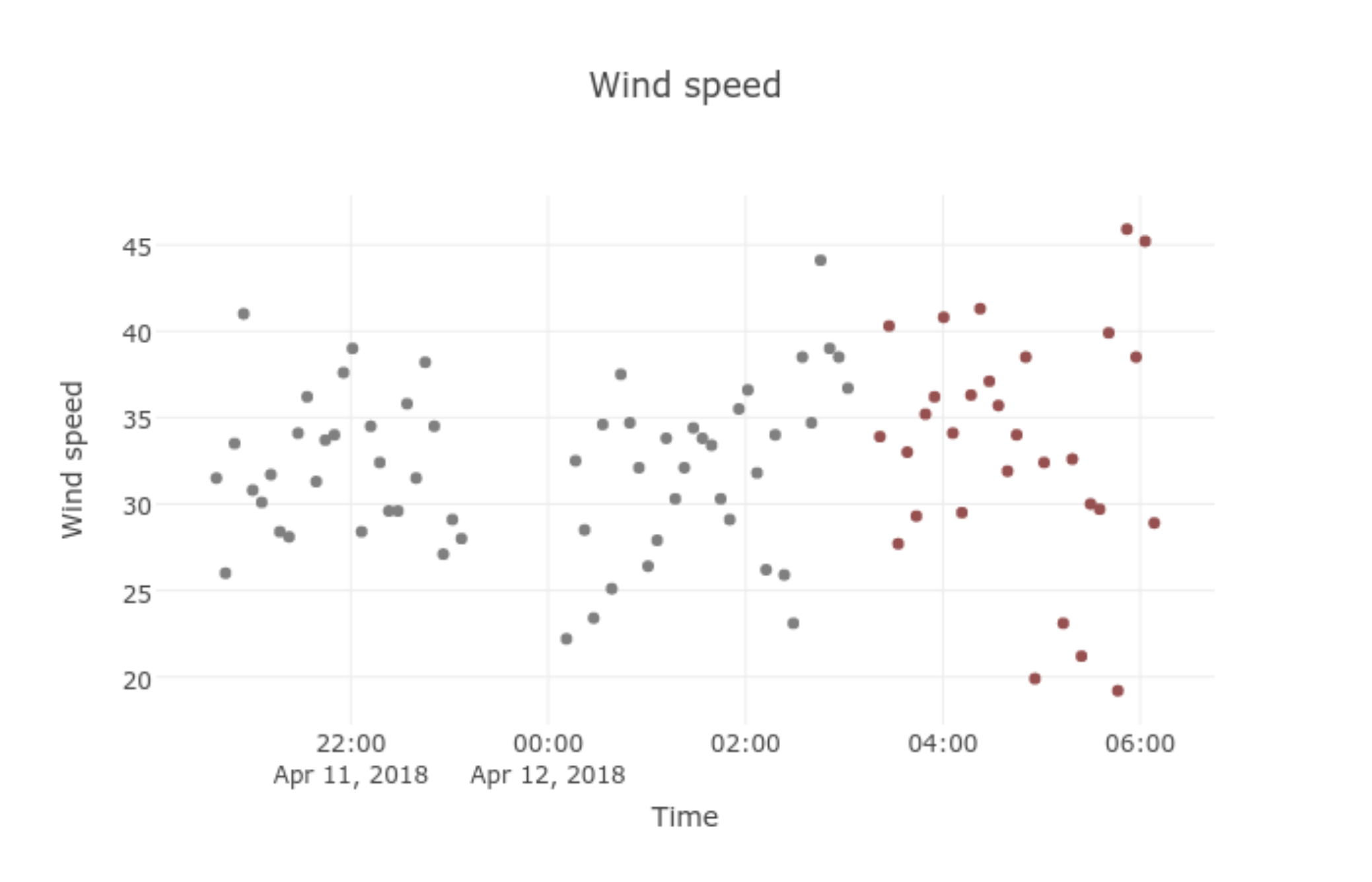}  
\plotone{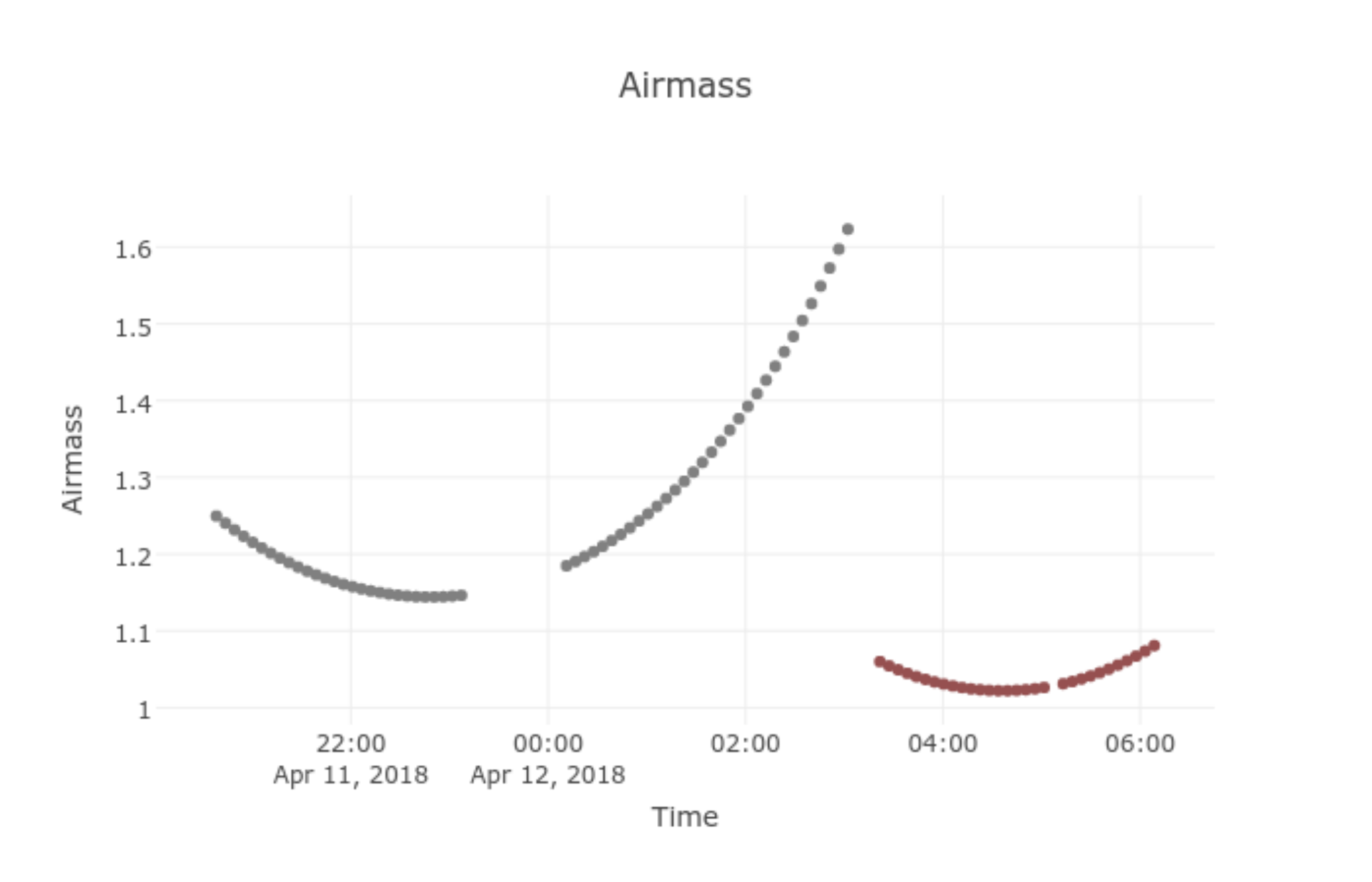}
\plotone{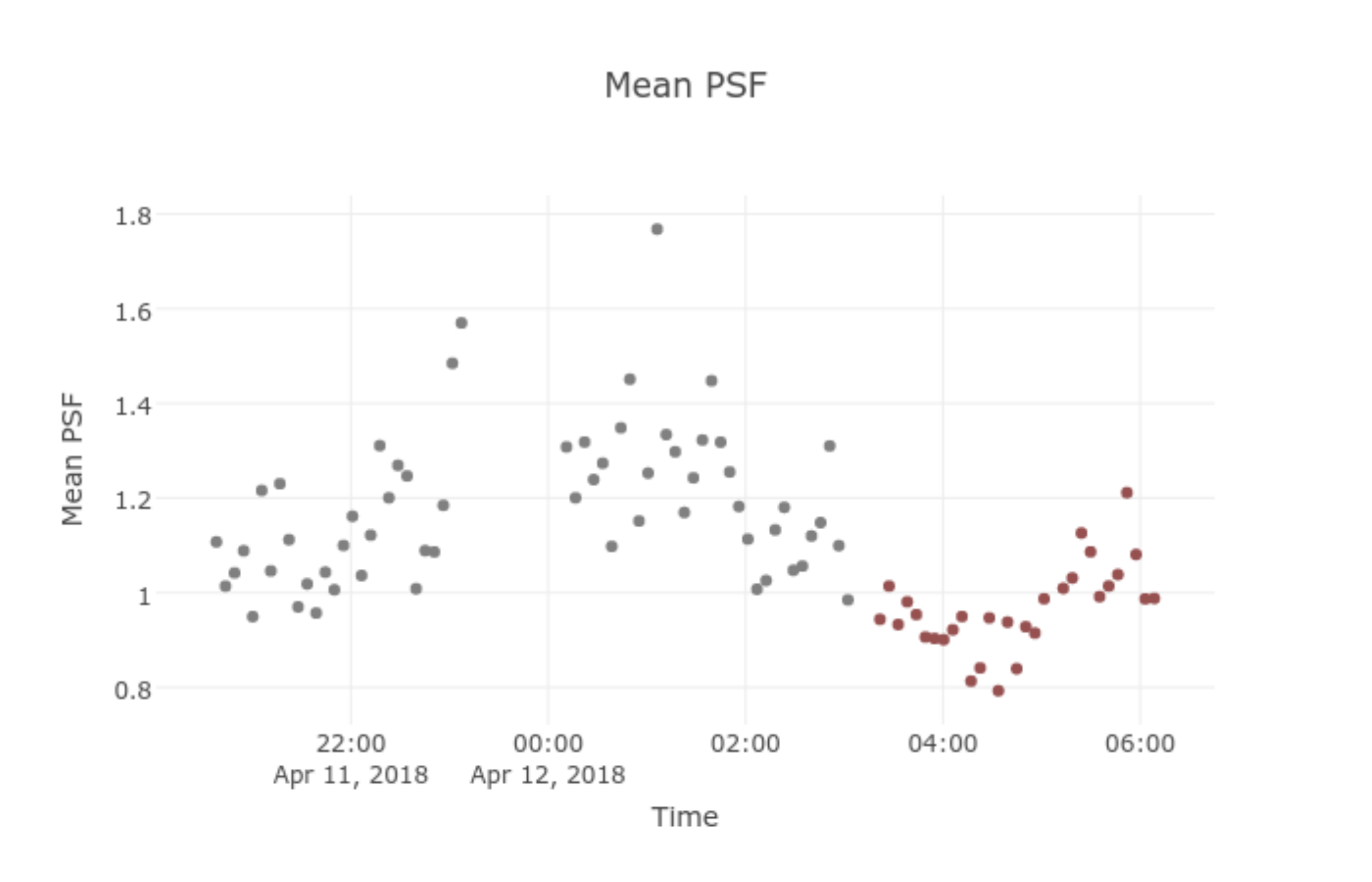} 
\caption{ \label{fig:logbook} 
   Night history of the variables as shown by the PAU Logbook. From top to bottom, the humidity (in percentage) and wind Speed (in km/h) measured at the WHT, the airmass and mean PSF FWHM measured by the camera exposures (in arc seconds). }
\end{figure} 

Integrated into the execution system of PAUCam there is an event database. These events are stored on a PostgreSQL database during execution time and can come from several sources, the most important being the macro execution system. The most typical events registered are alarms, telescope pointing status, and observation executions, but any new event can be defined dynamically.

The second part of the system is the PAUCam logbook which is the front-end of the event database. The logbook displays events versus time with optional filters by event type or date. It also lets the operator introduce log messages into the database. Another feature of the logbook is that it can make online plots of the evolution of the live QA results from exposures taken during the night, together with the environmental conditions at the observatory. An example of this information for a complete night can be seen in Figure~\ref{fig:logbook}. The PAUCam logbook is programmed in python and uses the well-known pyramid framework\footnote{\url{https://docs.pylonsproject.org/projects/pyramid/en/latest/}} as a back-end, whereas the front-end uses the javascript framework Angular\footnote{\url{https://angularjs.org}}. Plots are produced using the plotly library\footnote{\url{https://plot.ly}}.

\section{ONLINE ANALYSIS SOFTWARE}
\label{sec:OnlineAnalysisSw}

In addition to storing the FITS files from the various observations, the PAUCam online software includes several online analysis packages to help observers do some operations faster and to monitor the images being taken during the night. This section describes these tools.

\subsection{Focusing System}
\label{focusing}

In order to set the camera in focus, we have designed a specific readout and analysis system that provides accurate measurements of the focus position at each detector. As described in Section~\ref{sec:readout}, to speed up the acquisition of the image at different focus positions, we produce a single exposure image, where charges are shifted between the different steps, producing an image where the same source (star) is exposed several times, each at a different focus position. Figure~\ref{fig:focus} shows an example of such an image, where the same star is seen in different places of the CCD separated by a fixed number of pixels (a jump). The first focus is identified because the distance to the second one is doubled (a double jump).

\begin{figure*} [t]
\epsscale{1.08}
\plotone{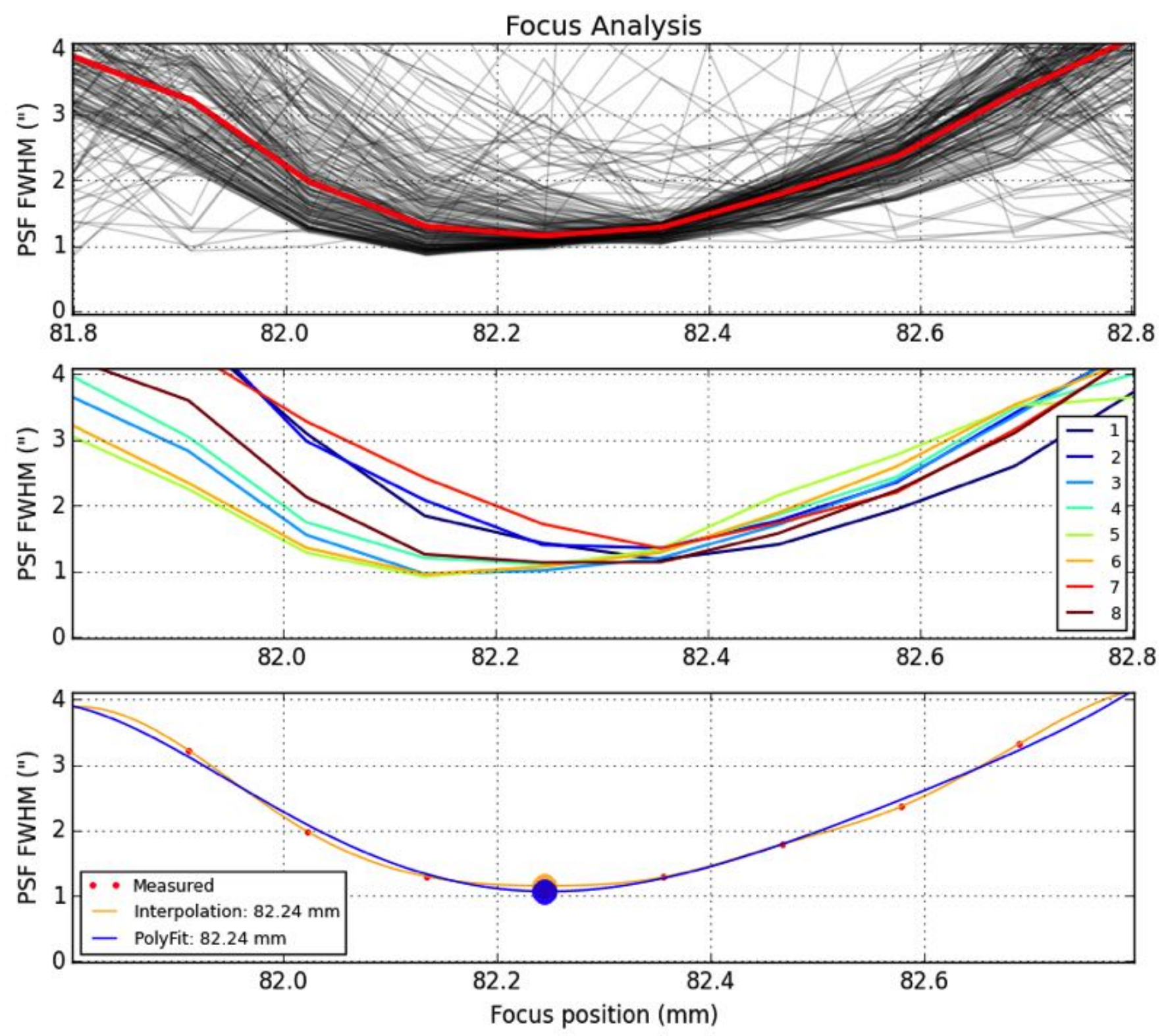}
\caption{\label{fig:FocusAnalysis} 
Focus analysis: Top, with all stars, middle with the average results per CCD and bottom with the fitted result and the preferred focus value. }
\end{figure*}

Once the image is read and built, the control system automatically triggers the analysis of this `stacked' focus exposure. In the first place the raw image is quickly reduced applying a gain and overscan correction. Over the reduced focus image, we run SExtractor \citep{1996A&AS..117..393B} to obtain a list of detected sources. As we need to identify the pattern of one double readout jump, plus N jumps, we create an R-Tree \citep{Rtree} of all detected stars, which allows making a very fast spatial matching, keeping the sequences that match all the jumps. The measured FWHM of each star in all the focus sequences identified for each CCD detector are median combined to produce a master curve. This curve is fitted to a 6th-degree polynomial that allows to automatically detect the focus position that minimizes the FWHM of the PSF. Figure~\ref{fig:FocusAnalysis} shows an example of a focus sequence analysis, automatically performed by the PAUCam Control System. The top plot shows the FWHM as a function of the focus position for all analyzed stars. Some of the stars have low signal or they are in a more distorted area of the focal plane. The red curve is the mean of all measurements. The middle plot shows the average per CCD. One can see small variations because of the tolerances in the CCD fabrication and installation process. The bottom plot shows the results with the fitted polynomial and the preferred focus value.

The algorithm runs automatically in a special DRACO instance (Section~\ref{sssec:draco})) and the results are presented in the GUI application to allow the observer to perform any needed focus correction.

\subsection{Auto guiding system}
\label{sec:autoguiding}

PAUCam has a built-in auto-guiding system to allow precise tracking of long exposures with minimal PSF size and optimal image quality. 

When the guiding system is enabled, two of the outermost corner detectors of the focal plane (G12 and G17) are devoted to guiding, leaving the science exposures to be done with the remaining 16 CCDs, including the most illuminated detectors in the focal plane.

When the shutter opens at the beginning of the exposure, the guiding detectors continuously take images with exposure times between 1 and 5 seconds, depending on the sky conditions and stellar density. The guider software then analyzes the images from these two guiding detectors and, using the stars available in the field, identifies the telescope tracking error and sends the correction signal to the TCSi.

To minimize the latency of the correction signal sent to the TCS, only a subset of the detector is read out. This subsection is called Region Of Interest (ROI), which is typically adjusted to $1000 \times 2000$ pixel subset in the most illuminated side of the detector. In this area, the image is quite distorted and partly vignetted.


Depending on the stellar density of the field, a number of guide stars from 3 to 15 are selected to measure the centroid correction. In the first place the guiding algorithm will quickly reduce the image, subtracting the bias signal and correcting the gain. The image is then analyzed by SExtractor, providing a list of source detections. These detections are filtered to ensure that stars are neither blended nor very close to each other, nor in the border, unaffected by cosmic rays, etc. The remaining list of candidate guide stars are selected and will be searched in the next iterations when the new guiding images arrive. Stars that are not found in new iterations will be removed from the list, assuming that might be moving objects, badly selected cosmic rays in the first place or too low signal to noise stars that went below the detection threshold. Having multiple sources as guide stars provides more accurate guiding, even under this area with more distortion of the focal plane. The individual tracking errors found in each star are combined using a weighted average by each star S/N ratio. After integration, readout and analysis, the error signal is sent to the TCSi every 6 to 8 seconds. 

In any case the WHT's tracking performance is very good, even in open loop (without any auto-guiding), with an error of less than $\sim 0.1 \arcsec$ drift per minute. Therefore for exposures shorter than 3 minutes, auto-guiding might not be necessary and the two guiding detectors will act as the rest of the scientific CCDs, allowing PAUCam to observe with the full 18 detector set.

\begin{figure}[t]
\epsscale{1.17}
\plotone{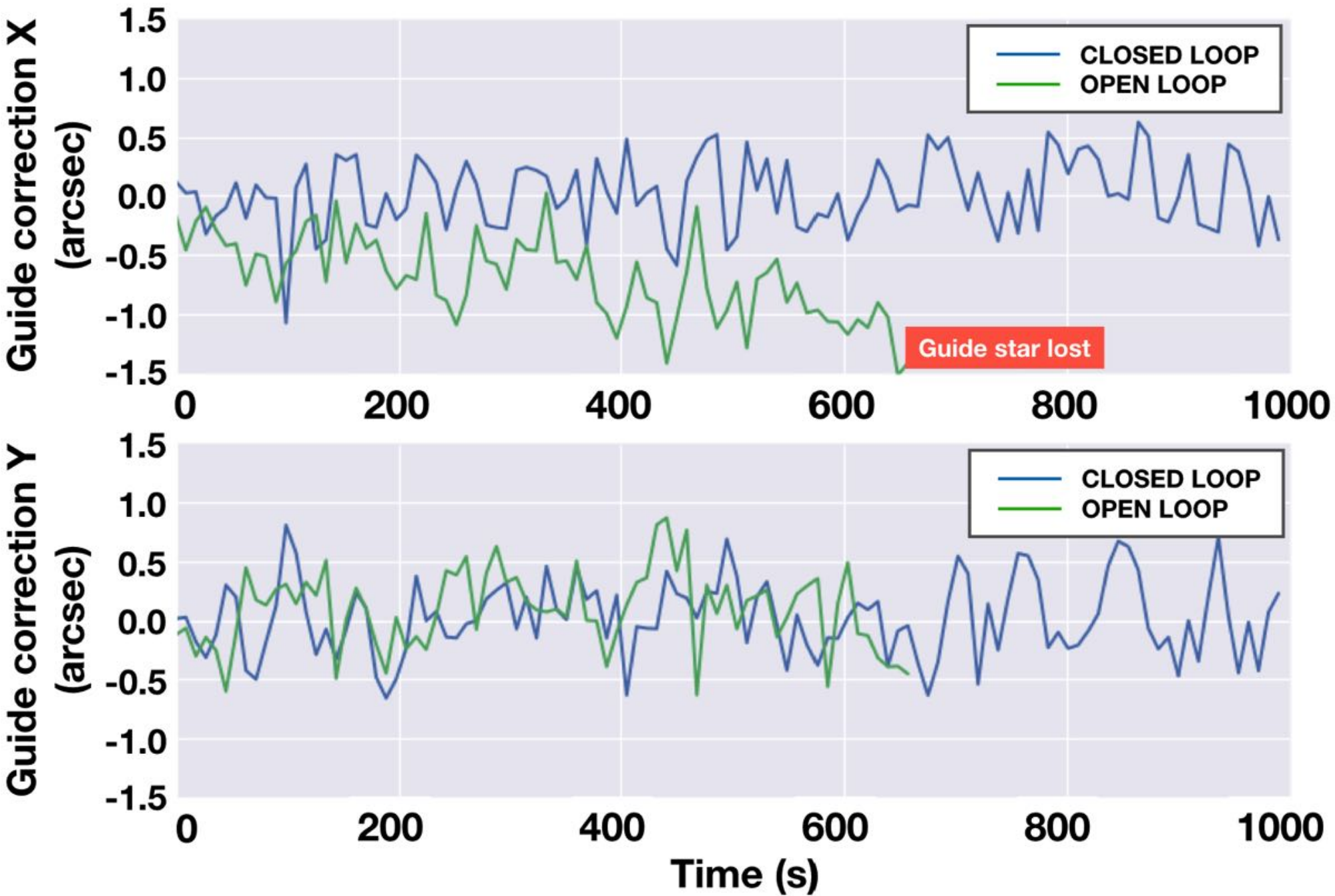}
\caption{ \label{fig:GuidingCorrection} 
Guiding Correction on X and Y as a function of time for open and closed loop configuration.}
\end{figure}

Figure~\ref{fig:GuidingCorrection} demonstrates how the guider corrects the telescope drift when guiding in closed loop. The figure shows the guiding correction computed in X and Y for the case where the signal is sent to the TCSi (closed loop) or when the correction is not applied. One can see that the tracking of the telescope can track the stars even without auto-guiding up to around 4 minutes but the stars are completely lost when longer exposures are made.

\subsection{Monitoring and Data Quality system}
\label{sec:Monitoring}

During observations, and before they are transferred to the PAU data center \citep{PAUOperations}, the FITS files produced by PAUCam have to pass a series of checks to monitor the quality of the images.
DRACO (see Section~\ref{sssec:draco}) is the satellite responsible of this task and consists of three layers:
a manager, that orchestrates the commands received by the OCS, defines which checks should be performed, and formats the response of the satellite; a series of algorithms to compute the analysis and the quick reduction, and a postgres relational database to organize the FITS files metadata and the quick reduction results.

The manager is receiving from the OCS the parameters of each FITS file that has been created: its location, file name and kind (calibration, scientific, test, focus,..). For each kind of file a specific series of checks to be performed is defined. The DRACO manager is responsible for sending the execution commands to check the files according to their kind.

The algorithms for the online analysis are written in python and organized in two groups: basic checks and quick reduction.

Basic tests are performed over all the raw FITS files and are aimed at checking the file integrity and the overscan values (electronic noise).  The file integrity test checks for the consistency of the information given in the header in terms of number of extensions and data type, and for the readability of the content.  The overscan test computes the mean, median absolute deviation and standard deviation values of the overscan region of each amplifier, and returns the values to the manager.

Quick reduction is performed only on scientific images and consists of more advanced tests which give the observer a better understanding of the quality of the data taken. The reduction procedure consists of running the Astromatic packages SExtractor \citep{1996A&AS..117..393B} and PSFEx  \citep{2011ASPC..442..435B} over each CCD, resulting from combining the data of the four amplifiers, after overscan and gain correction. The values returned are the mean background value, the number of detections, their mean ellipticity and the mean PSF for each CCD of a mosaic.

\begin{figure} [t]
\epsscale{1.17}
\plotone{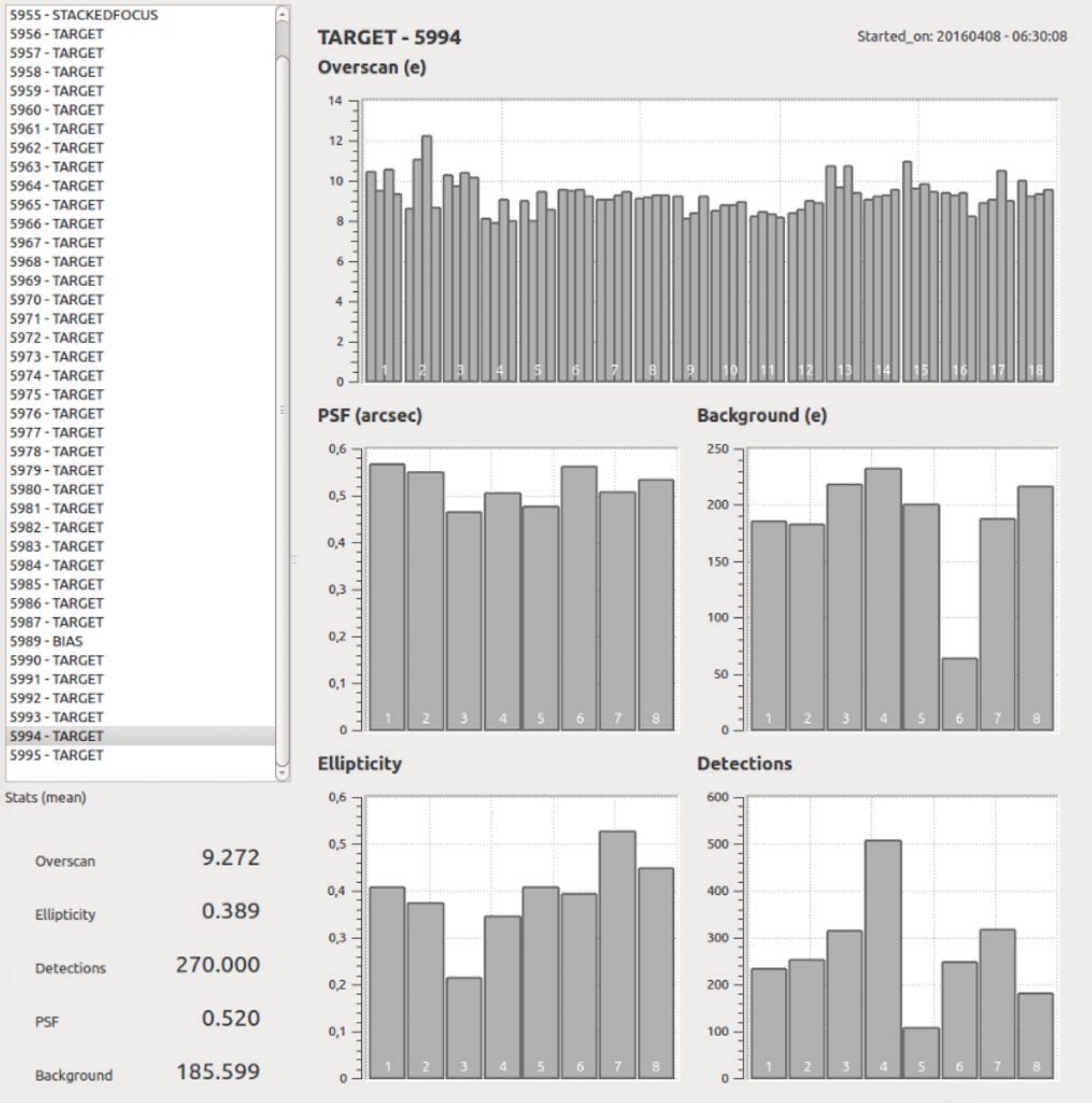}
\caption{ \label{fig:DRACODisplay} 
DRACO display for a TARGET exposure. The results of the overscan per amplifier, and the PSF, background, ellipticity and number of detections per CCD are shown, together with their mean values.
}
\end{figure}

The comparison to the reference values for every CCD (each of them with a different NB filter) in each single image is shown graphically. The manager is receiving the results from the overscan check and quick reduction, and formatting them to be displayed in the GUI monitor.  When a value is outside the valid range, the graph turns red and the operator is in charge to take the proper action. Figure~\ref{fig:DRACODisplay} shows the result of DRACO for a normal science exposure. It shows the results of the overscan per amplifier, PSF, background, ellipticity and number of detections per CCD. 

DRACO registers all the information related to the raw and reduced files as soon as the FITS files are created and the check is performed, organized in a relational database (postgres).
The DRACO database consists of 5 main tables: \emph{raw mosaic} and \emph{raw image} contain the metadata regarding the raw FITS files and the results of the basic checks; \emph{reduced mosaic} and \emph{reduced image} collect the results of the quick reduction; finally the \emph{observation set} table collects the information and eventually notes regarding group of mosaics taken the same night, by the same operator and under the same camera conditions.
The DRACO database is queried to monitor the evolution of the image parameters during the night (see Section~\ref{sec:logbook}). 

The quick reduction is the most time consuming process, but it has been designed to deliver results in real time. The hardware dedicated to the online analysis is a 12 core, $32 \unit{GB}$ computer, allowing DRACO quick reduction algorithms to work with multiple threads. The DRACO response is displayed to the operator just 1-2 minutes after the image has been taken.

DRACO has been designed such that it is easy to integrate new algorithms of analysis to run over PAUCam FITS files. As an example, the algorithm for the stacked focus analysis (Section~\ref{focusing}) has been integrated in the satellite, the results are formatted by the manager and displayed in the GUI in real time.

\section{DATA REDUCTION SYSTEM}
\label{sec:datareduction}

The data management of PAU is in charge of transferring, archiving, calibrating, measuring and distributing the observations required for the scientific analysis. All these tasks are supported by the Port d'Informaci\'o Cient\'ifica (PIC) data center, providing the infrastructure to manage the large volume of data generated by the camera \citep{PAUOperations}.

To allow the transfer of the $\sim 350 \unit{GB}$ of data produced every night, a dedicated machine with enough capacity to store up to 5 nights in case the connection is lost was installed at the WHT. The local archive at the WHT and the one at PIC are in sync and once the observer requests it, the transfer jobs to PIC begin. 

The DAQi system from PAUCam generates a YAML\footnote{\url{http://yaml.org/}} file with the metadata of each exposure set, such as the night or the name of the observer, along with the list of images and their adler32 checksums.

The actual data is transferred using the bbcp tool\footnote{\url{https://www.slac.stanford.edu/~abh/bbcp/}}, which also compares the checksums at the other end to ensure the integrity of the files.

The installed network bandwidth is $1\unit{Gb}$, although shared by all the telescopes in the Observatory. During regular operations we have observed a good transfer speed (around $20 \unit{MiB/s}$) and stability \citep{PAUOperations}. After around 3 hours, depending on the network traffic, all the data are transferred to PIC and the registering process begins, where all the exposures metadata is saved in the PAU data management (PAUdm) database. 

Once all data are safely stored in the main archive and in sync with the database, the nightly processing begins, performing a basic data reduction and astrometry, with a very specific photometric calibration. The reduced images with the calibrated zeropoints are also stored in the archive-database system. For community observers of PAUCam, this dataset is also made available via a webdav server, as the photometric calibration of narrow bands is a delicate non-standard procedure.

This processing is usually finished before the next observation night starts. Feedback is available for the observer that can select the images that do not comply with the needed quality requirements and re-schedule expositions for the next night.

Exclusively for the PAU survey, a multi-epoch and multi-band analysis pipeline (MEMBA) has been designed with the aim of providing accurate color photometry that delivers precise photometric redshifts of galaxies.

The details of this system and the processing of this unique dataset are extensively explained in \citet{DPPaper} and \citet{PAUCalibration}.

\section{COMMISSIONING AND FIRST RUN}
\label{sec:commissioning}

\begin{figure} [t]
\epsscale{1.16}
\plottwo{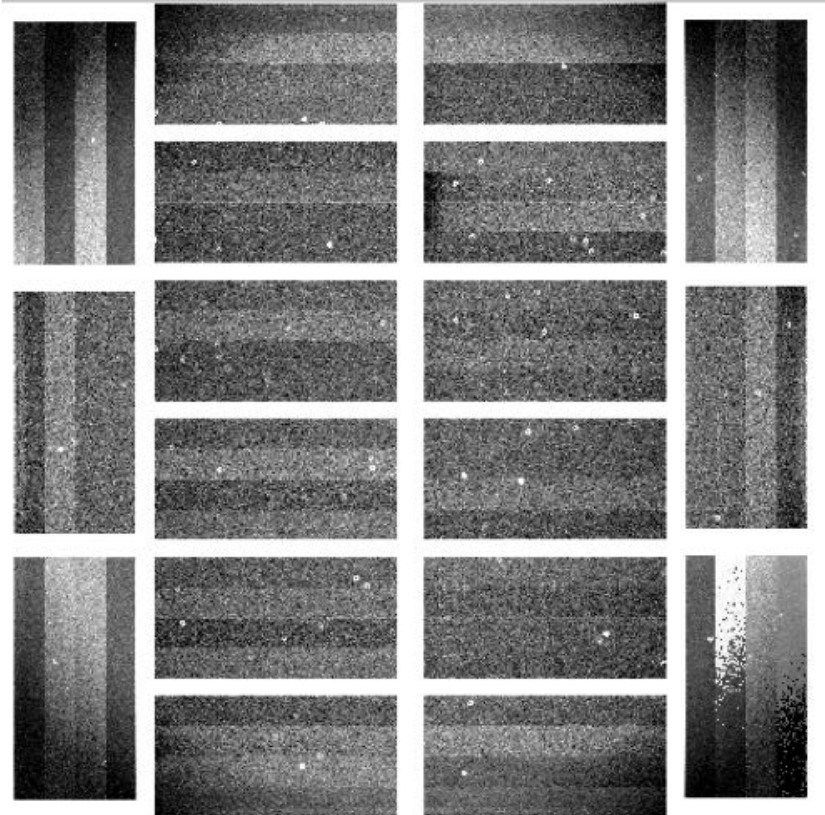}{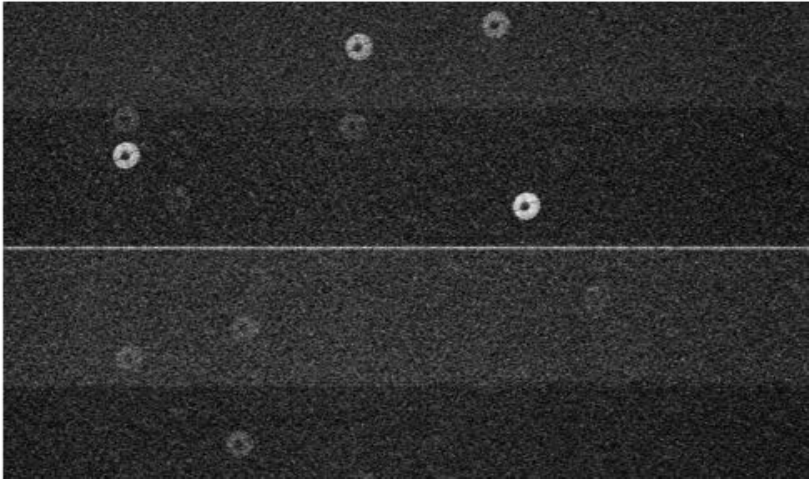}
\caption{ \label{fig:fitstimage} 
First Image taken with PAUCam on June 3rd, 2015. Left: full focal plane. Right: Detail with some unfocused stars. }
\end{figure} 

The PAU Camera saw first light on June 3rd, 2015. Figure~\ref{fig:fitstimage} shows the first image taken at the WHT. The image was out of focus and had a high read-out noise of around 70 electrons.
After some adjustment in the grounding scheme, the electronics noise decreased to 9-10 electrons. The camera has remained stable at this value, which is the limit we expect to have in the whole system. This first image also shows that the camera was not in focus.

\begin{figure} [t]
\epsscale{1.15}
\plotone{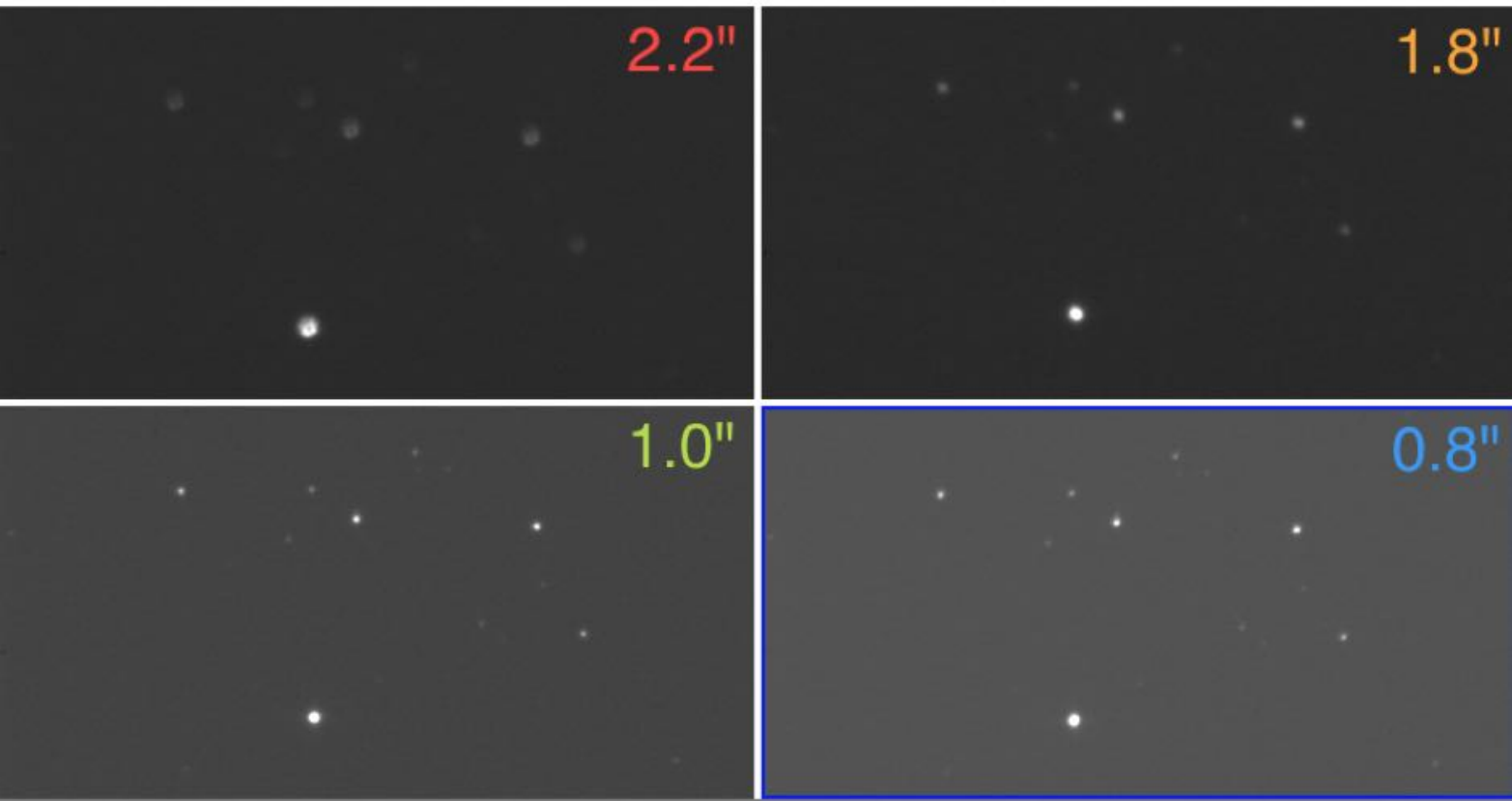}
\caption{ \label{fig:TipTiltimage} 
The same portion of the image taken in the tip-tilt tests at four different focus positions. The numbers show the measured FWHM of the detected stars.}
\end{figure}

To ensure that the focal plane was installed flat in the prime focus of the WHT, we performed a tip-tilt analysis during the commissioning, and in the later installations for the next operation runs. To carry out this analysis, a sequence of defocused images is taken, around the default focus position at $82.3 \unit{mm}$. Figure~\ref{fig:TipTiltimage} shows the same portion of the image taken in the tip-tilt tests at four different focus positions. The numbers show the measured FWHM of the detected stars.

\begin{figure} [t]
\epsscale{1.17}
\plotone{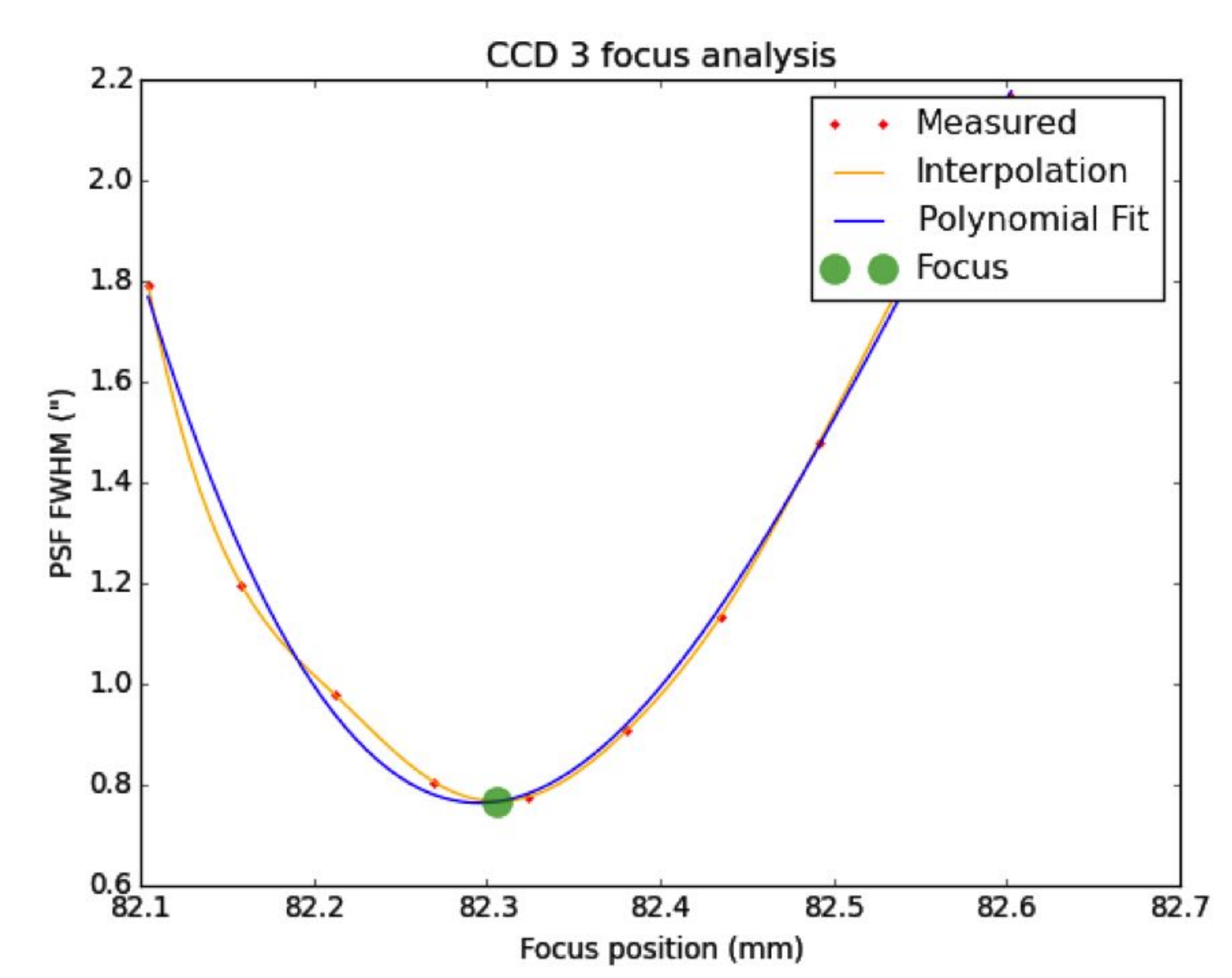}
\caption{ \label{fig:TipTiltPosition} 
Position of the best focus point for CCD3 during the tip-tilt analysis.}
\end{figure}

\begin{figure} [t]
\epsscale{1.17}
\plotone{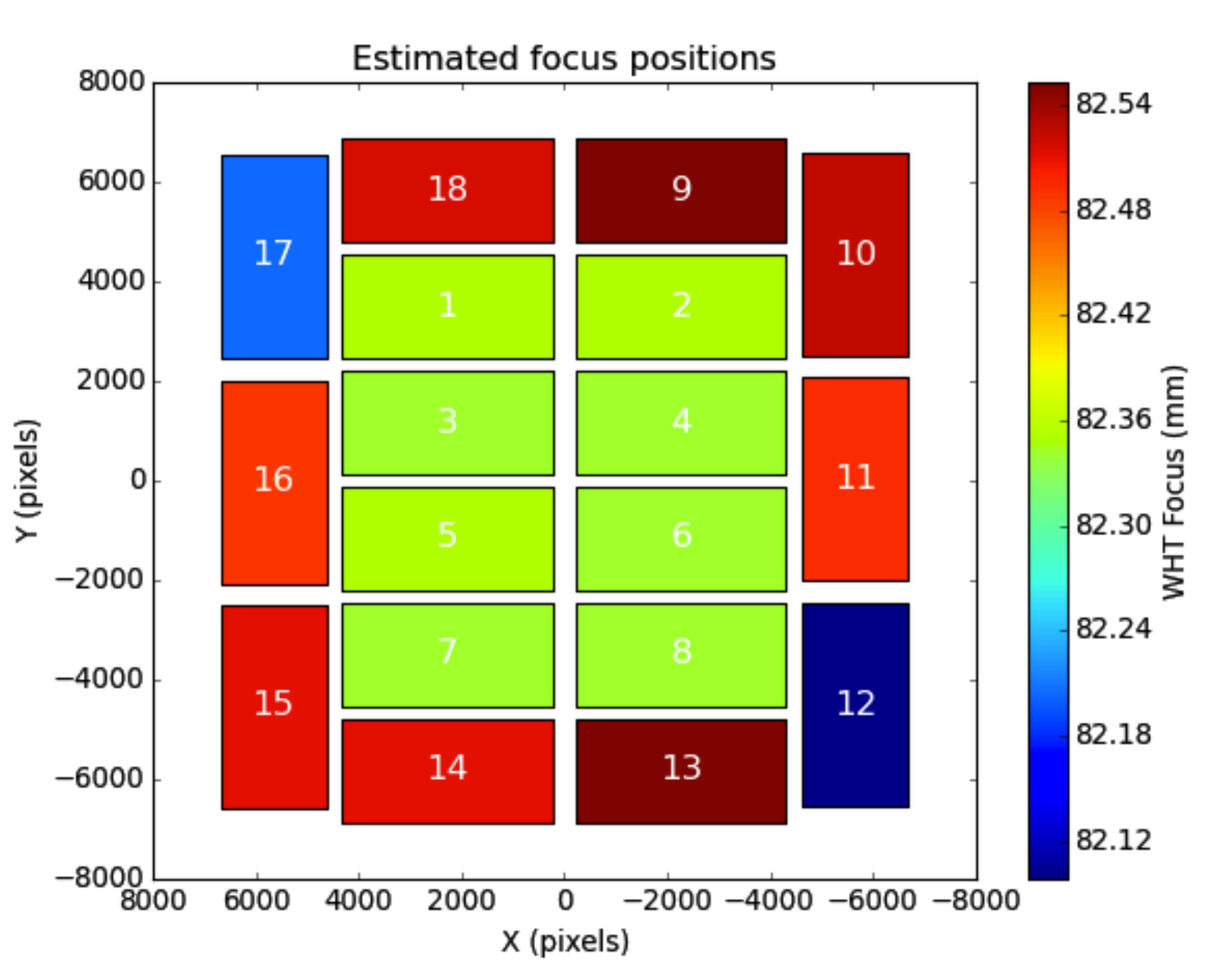}
\caption{ \label{fig:TipTiltFocalPlaneCorr} 
Average focus position for each CCD computed during the tip-tilt tests.}
\end{figure}

At each focus position the FWHM of the stars in each CCD is measured, providing an average PSF FWHM per CCD. Similarly to the focusing algorithm, the position of the focus (where the FWHM minimizes) is estimated in every CCD sequence (see an example for CCD3 in Figure~\ref{fig:TipTiltPosition}) providing 18 focus points in the focal plane. The results are presented in Figure~\ref{fig:TipTiltFocalPlaneCorr} showing the resulting focus position value for each CCD.

When the focal plane is completely flat to the optical path at the prime focus, the individual detector focus should lie at the same focus position. In case the camera does not sit completely flat in the telescope mount, 4 mounts (1 fix and 3 adjustable) are available to correct for the possible tip and tilt miss-positioning. 

\begin{figure}[t]
\epsscale{1.17}
\plotone{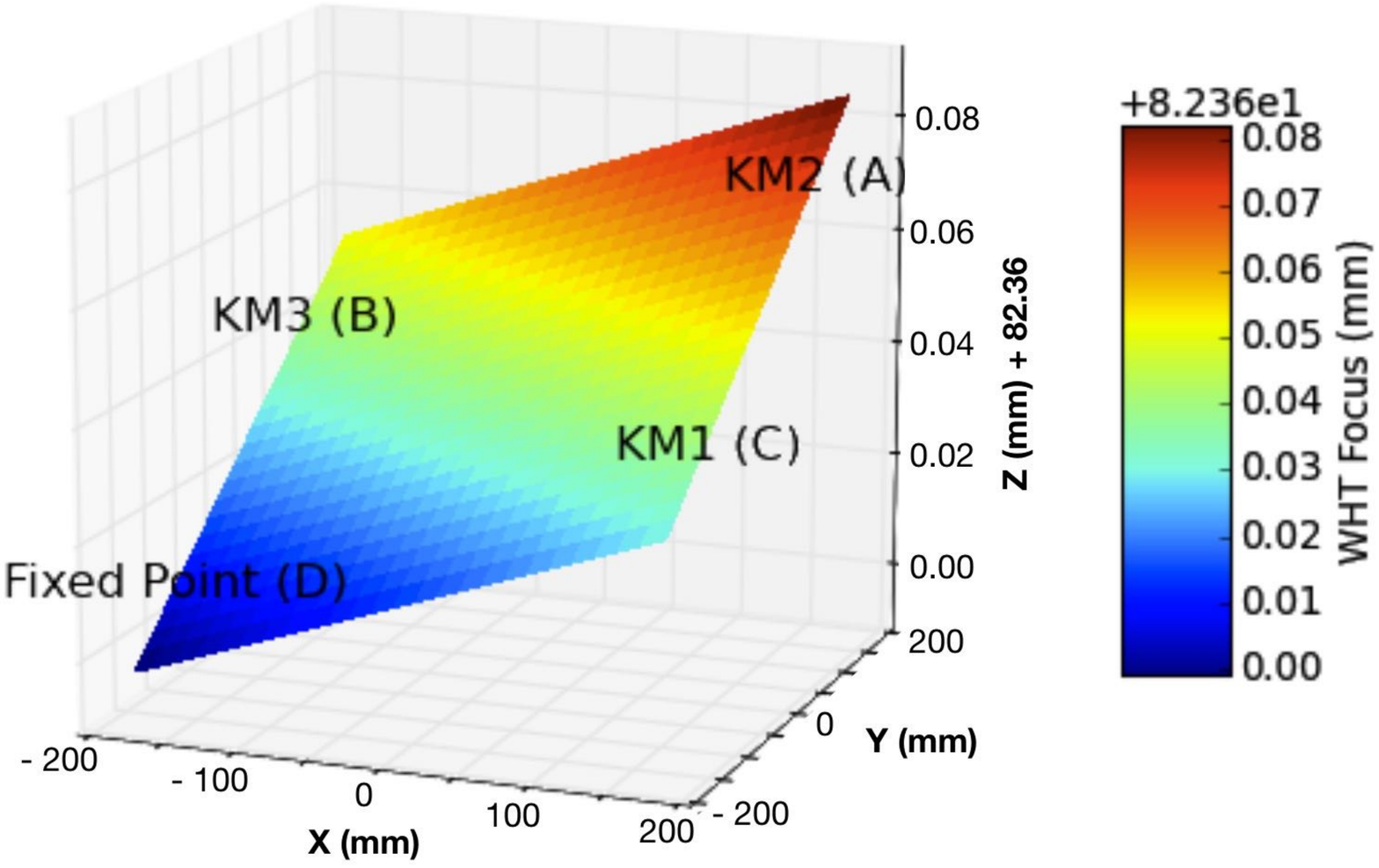}
\caption{ \label{fig:TipTiltPlane} 
Offset to be applied in every adjustable mount to have a completely flat focal plane.}
\end{figure}

To estimate the amount of tip and tilt, a 3D plane is fit to the focus positions (z) at the CCD positions (x,y). Following the mechanical drawings the position of the Kinematic Mounts are identified and the tip-tilt code returns the offset that needs to be applied for a flat focal plane (see Figure~\ref{fig:TipTiltPlane}). Due to the fact that there is a significant amount of distortion in the 10 outer detectors that may alter the result, only the 8 central detectors were used in the 3D fit. 

\begin{figure*}[t]
\epsscale{1.15}
\plotone{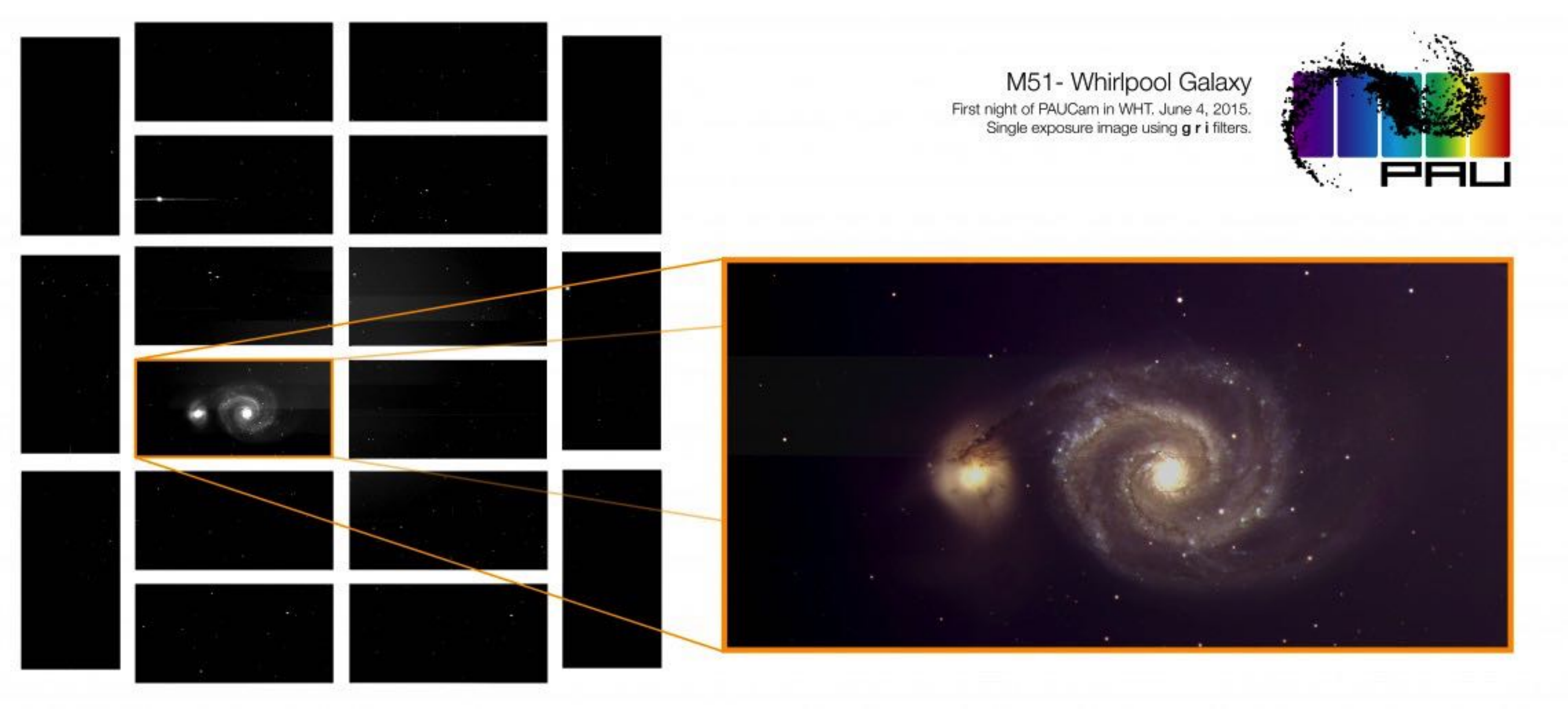}
\caption{ \label{fig:whirpool} 
Galaxy M51, known as the Whirlpool galaxy, situated at about 23 million light years from Earth. (Image acquired June 3rd, 2015).}
\end{figure*} 

\begin{figure} [t]
\epsscale{1.17}
\plotone{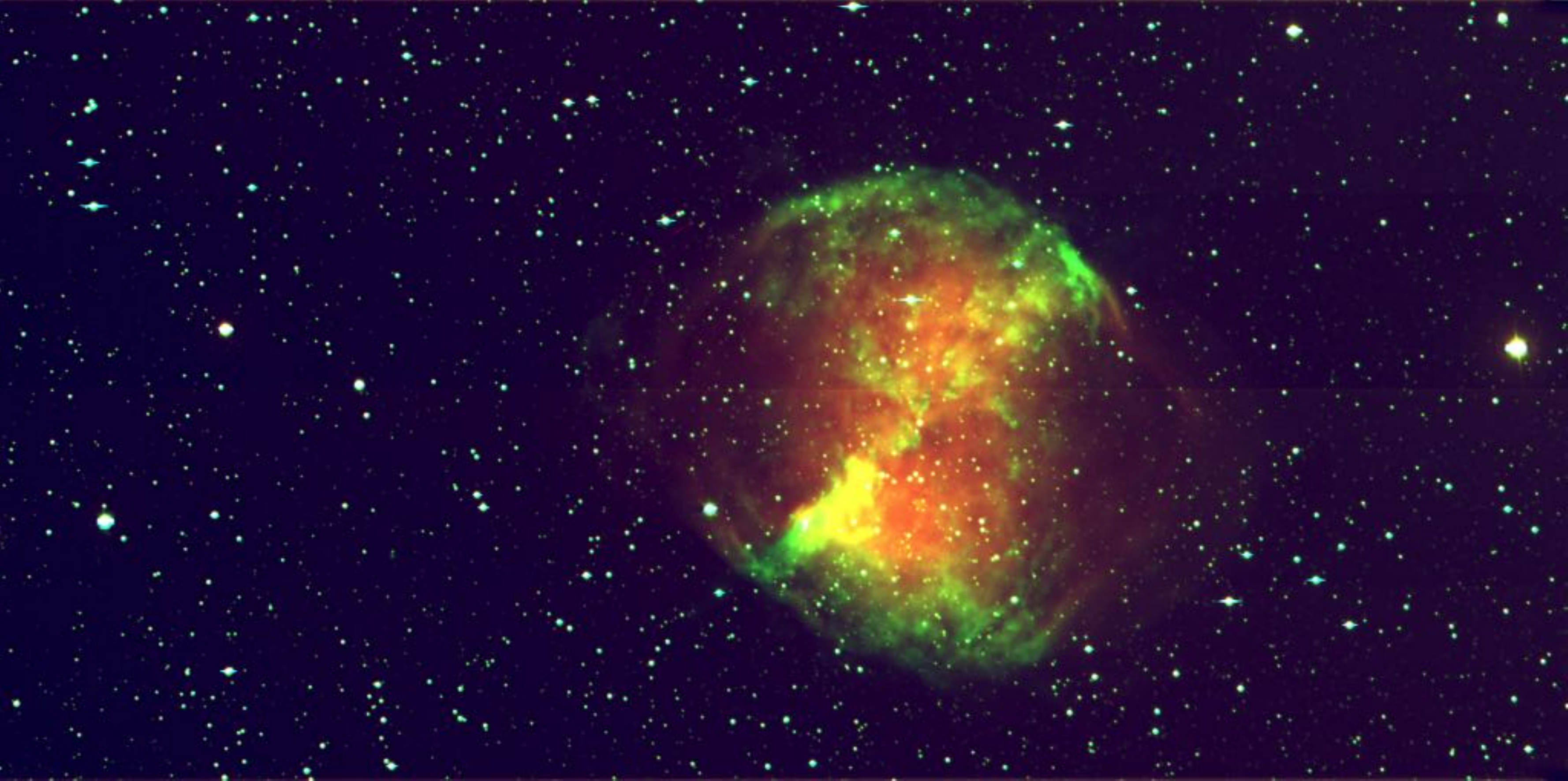}
\caption{ \label{fig:nebbula} 
Planetary nebula M27, known as the Dumbbell Nebula, situated in our own galaxy, the Milky Way, at a distance of about 1000 light years from Earth. (Image acquired June 3rd, 2015).}
\end{figure} 

Thanks to an accurate design, verification (Section~\ref{sec:mechanicalver}) and installation of the camera, all the analysis resulted into a very flat focal plane from the very first installation during commissioning, to the later installations of PAUCam in the next observing runs. The corrections suggested by this analysis in the 3 adjustable mounts where smaller than the tolerance that can be applied ($<10 \unit{\micron}$).

After a few more tests, during the first commissioning night we could take several images of known objects. Figure~\ref{fig:whirpool} shows the image taken of the Whirpool Galaxy, which occupies a complete CCD, therefore giving the scale of the PAUCam focal plane. Figure~\ref{fig:nebbula} shows a picture of the Dumbbell Nebula also taken during the first commissioning night.

Since its first installation, PAUCam has been dismounted and mounted several times per scheduled semester at the WHT. As of December 1st 2018, a total of 123.5 nights have been allocated for the PAU Survey Collaboration.  53\% of the allocated time could not be used either because of bad weather or poor atmospheric conditions. The data taking efficiency of PAUCam remained above 96\% during the whole period. Additionally, the PAUCam team has supported 21 nights for external observers not belonging to the PAU Survey Collaboration.

\section{SURVEY STRATEGY OPERATION}
\label{sec:operation}

\begin{figure*}[t]
\epsscale{1.17}
\plotone{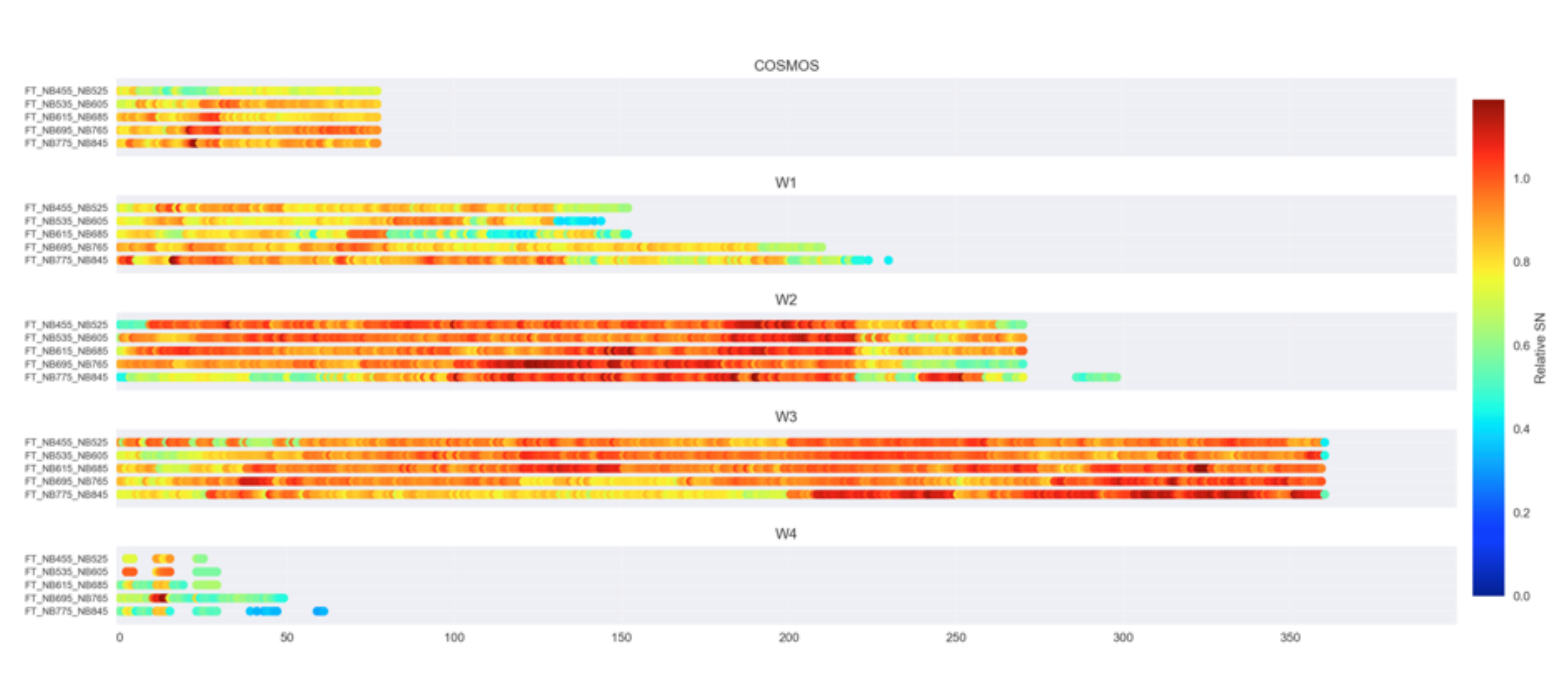}
\caption{ \label{fig:FieldsCoverage} 
Relative signal to noise of combined dithered exposures. For every field in the PAU Survey, there are five lines, corresponding to the five NB filter trays. The x axis shows the pointing number as defined in the observing database. A relative value of 1 corresponds to approximately a S/N $\sim$ 3 for extended sources at $i_{AB} \sim 22.5$ with all available dithers. This tool allows to monitor and homogenize the signal to noise to the desired goal across the survey area by adapting the exposure times as a function of sky conditions taking into account the S/N in the previous acquired dithers.}
\end{figure*} 

Cosmology surveys require observing large areas of the sky. This implies very large number of exposures, and in the case of the PAU Survey, with 40 narrow bands, this is even more extreme. Therefore, the operation and monitoring of the survey needs to allow for automated tools. To accomplish this task effectively, we have set up a database in the camera online system where we track the status of the survey and the pointings in each of the large fields PAU aims to observe. A single table called \emph{target} was added to the survey strategy database containing information such as sky coordinates, position angle, filter tray, default exposure time and current status. When a new field is created in the database for the first time, all targets are set to status “SCHEDULED”. A custom software in python allows the observer to request for pending targets in a specific field, for a specific filter tray and the system delivers a list of exposures that can be ingested into the PAU CCS via a NOD into the OCS (see Section~\ref{sec:OCS}). This allows the camera to process and observe sequences of multiple targets and observe automatically for hours. Once the targets are observed, the status is set to OBSERVED. Thanks to the quick feedback of the data management (see Section~\ref{sec:datareduction}), bad quality exposures (due to bad weather or any other reason) can be identified and re-scheduled to be observed during the next observing night.

Figure~\ref{fig:FieldsCoverage} is used to schedule the observations during the night. It shows the relative signal to noise for every pointing and tray for the 5 fields that the PAU Survey is planning to observe. The figure is updated automatically when a target is set to OBSERVED during the night. After the nightly processing, some targets can be removed and the observer can decide whether they need to be observed again to obtain the desired signal to noise ratio for those targets.


\begin{figure} [t]
\epsscale{1.2}
\plotone{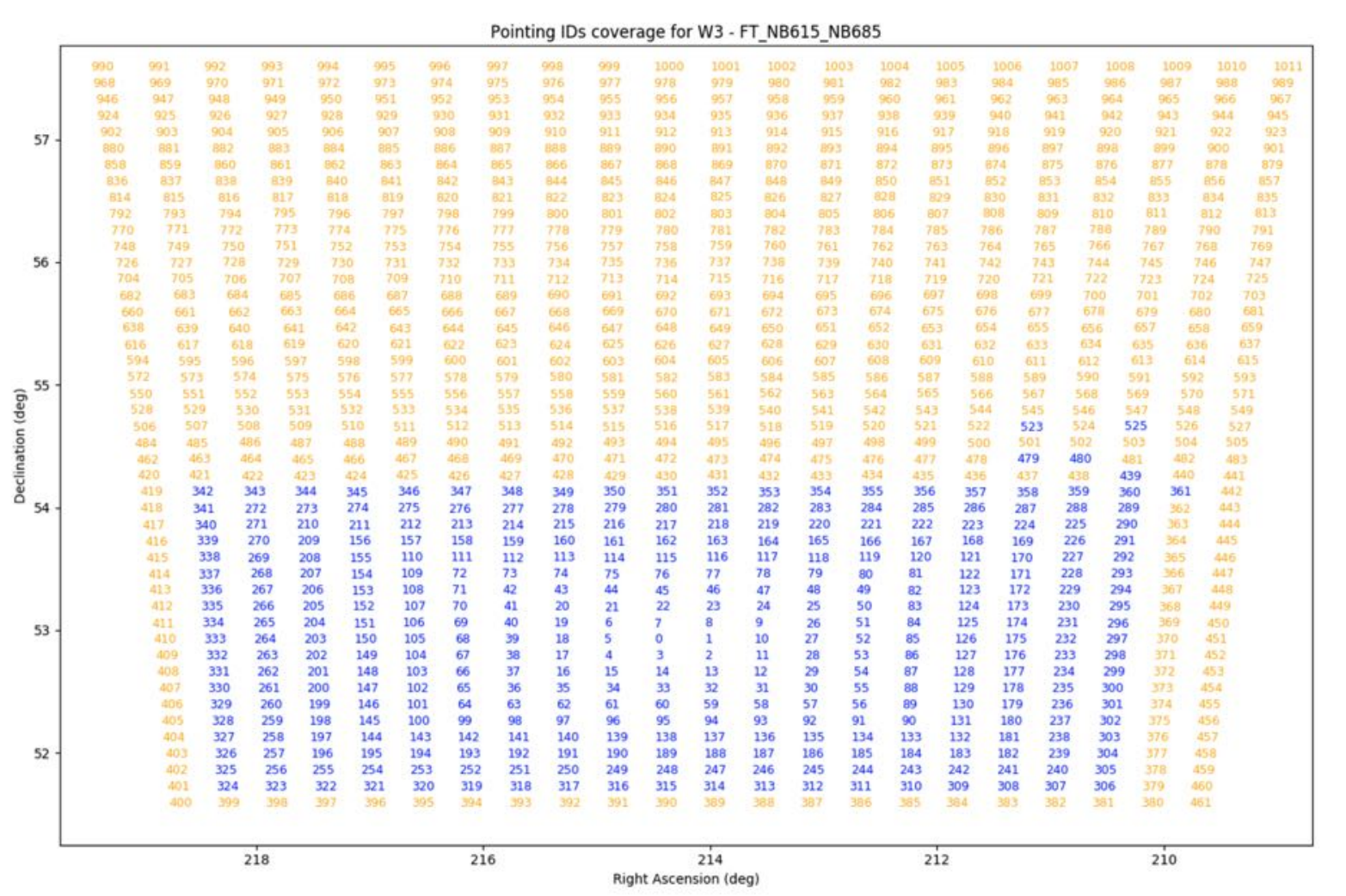}
\caption{ \label{fig:W3Coverage} 
Pointing number as function of the declination and right ascension coordinates for the W3 field. Blue pointings are the ones observed with PAUCam.}
\end{figure} 
   
\begin{figure} [t]
\epsscale{1.15}
\plotone{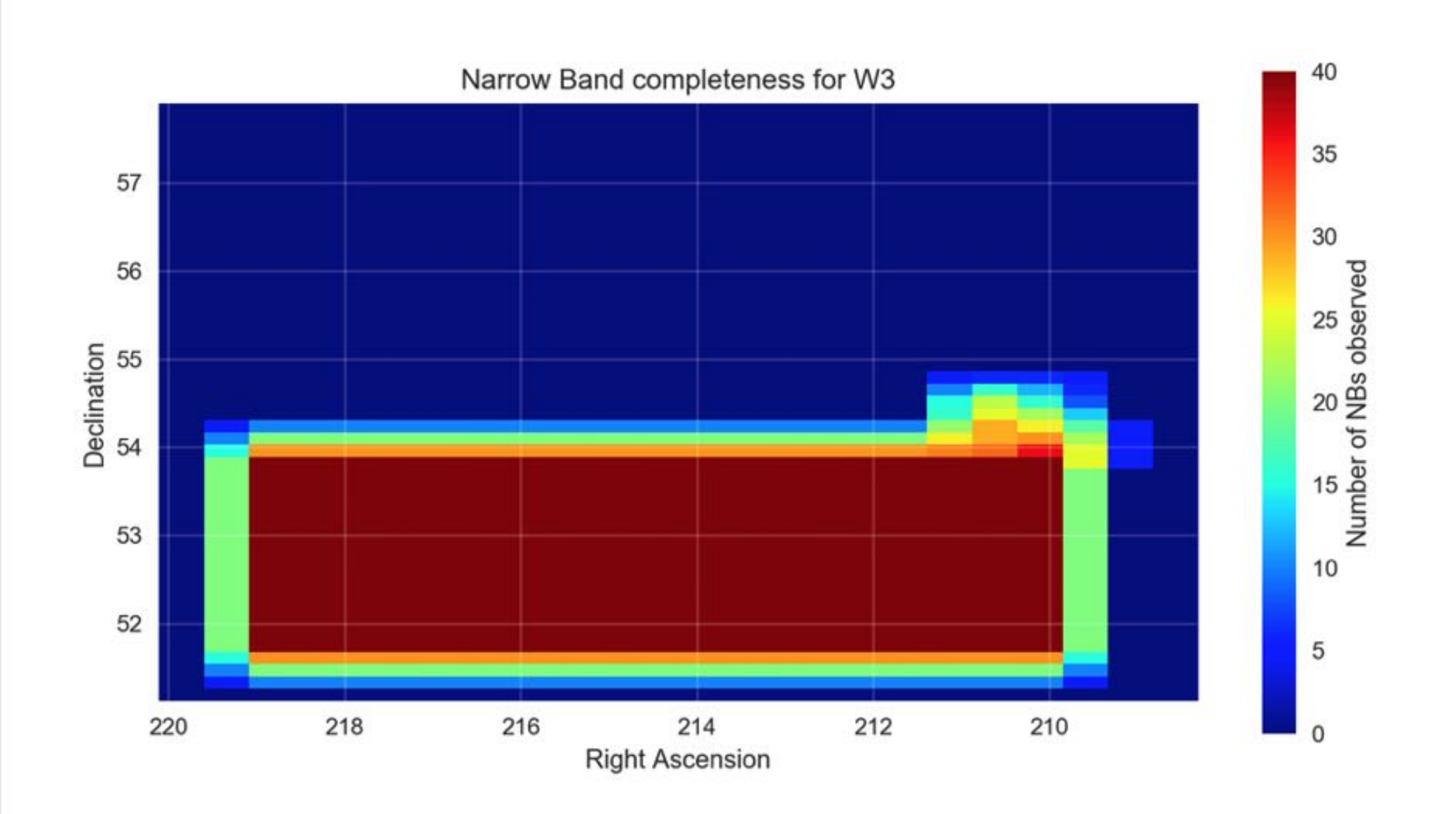}
\caption{ \label{fig:W3NBCoverage} 
Number of observed filters as function of the declination and right ascension coordinates for the W3 field.}
\end{figure} 
   

With the information in the survey strategy database, we can track the progress of the survey and obtain relevant information to schedule the observation during the night. For example, Figure~\ref{fig:W3Coverage} shows the declination and right ascension coordinates of the W3 field. We have ordered the different pointings starting from around the center of the field and then proceeding spirally outwards. Every pointing has the approximate size of a CCD. The blue pointings are those that we have observed up to now. Figure~\ref{fig:W3NBCoverage} shows another view of the coverage specifying the number of filters that have been observed for every position in the sky. More than $10\unit{deg^2}$ have been observed with the 40 Narrow Band filters in W3. In total, with all the fields, we have observed up to $15.5\unit{deg^2}$ in good conditions. An additional  $9.1\unit{deg^2}$ had to be observed again, mostly because of bad weather conditions.


\section{CAMERA PERFORMANCE}
\label{sec:performance}

\subsection{Detectors characterization}

Since its installation at the Prime Focus of the WHT in June 2015, we have been taking a Photon Transfer Curve (PTC) in each semester to check the behavior of the camera determining the current gain, the charge transfer efficiency (CTE), the linearity and other parameters of the camera. Monitoring the camera regularly we can determine its stability and update the map of the camera defects due to change or problems with the electronics and detectors.

\begin{figure}[t]
\epsscale{1.17}
\plotone{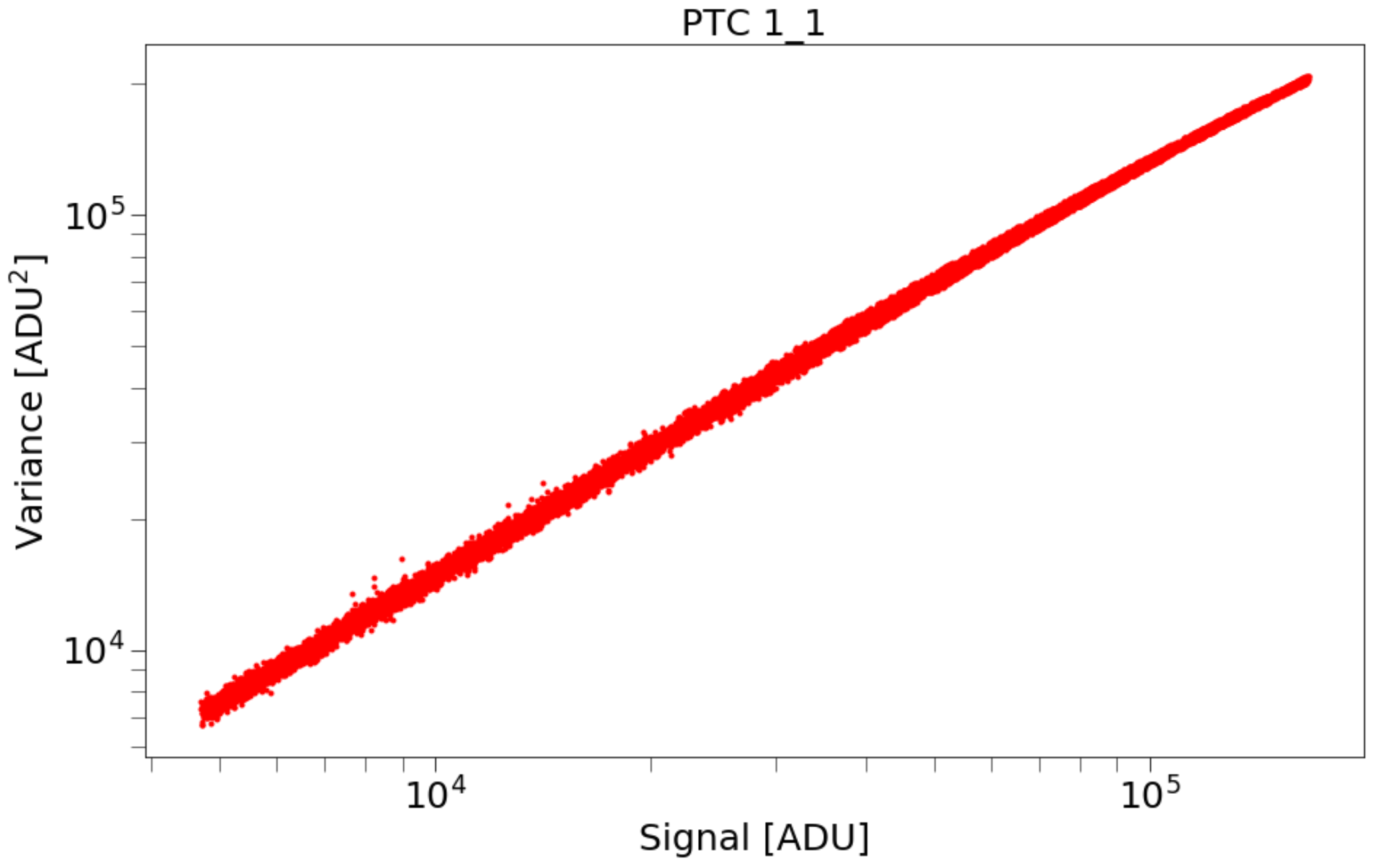}
\caption{ \label{fig:PTC_1_1} 
An example of a PTC obtained for the first amplifier of CCD 1. In this case we have removed from the figure the images that are saturated. For each pair of images, the mean and variance are calculated in square boxes of $50 \times 50$ pixels. Given the camera vignetting not all square boxes have the same counts for the same exposure time. This effect produces dispersion in the points presented.}
\end{figure}

Our PTC standard calibration set is composed of 32 dome flat images. Images are grouped in pairs of the same exposure time which varies logarithmically between 1 and 40 seconds. Each amplifier of each CCD is analyzed independently, obtaining 72 PTCs. Figure~\ref{fig:PTC_1_1} shows the PTC for the first amplifier of CCD number 1.

\begin{figure}[t]
\epsscale{1.17}
\plotone{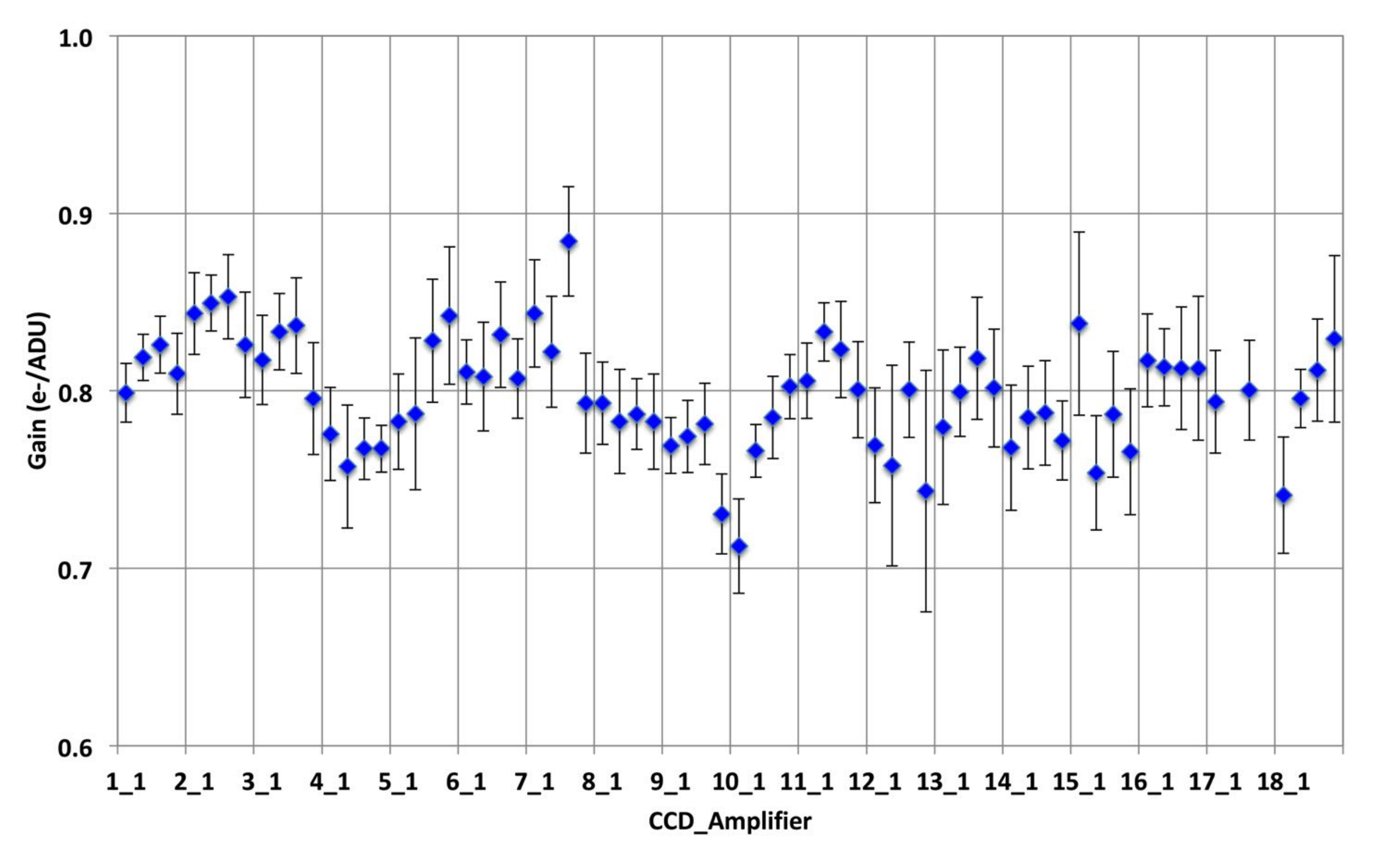}
\caption{ \label{fig:gain} 
Gain of each amplifier of PAUCam determined using all PTC measurements carried out from June 2015 until March 2018. The x-axis shows the CCD number and amplifier. There are a few gain values missing from the plot as these amplifiers are in a region of the focal plane highly vignetted and their determination has large errors.}
\end{figure}

With each PTC we can determine the gain for each amplifier of the camera. Figure \ref{fig:gain} presents the mean gain obtained from two PTC tests carried out in January 2018. Differences between amplifiers in the same CCD are expected because the electronics is different for each one.

\begin{figure}[t]
\epsscale{1.17}
\plotone{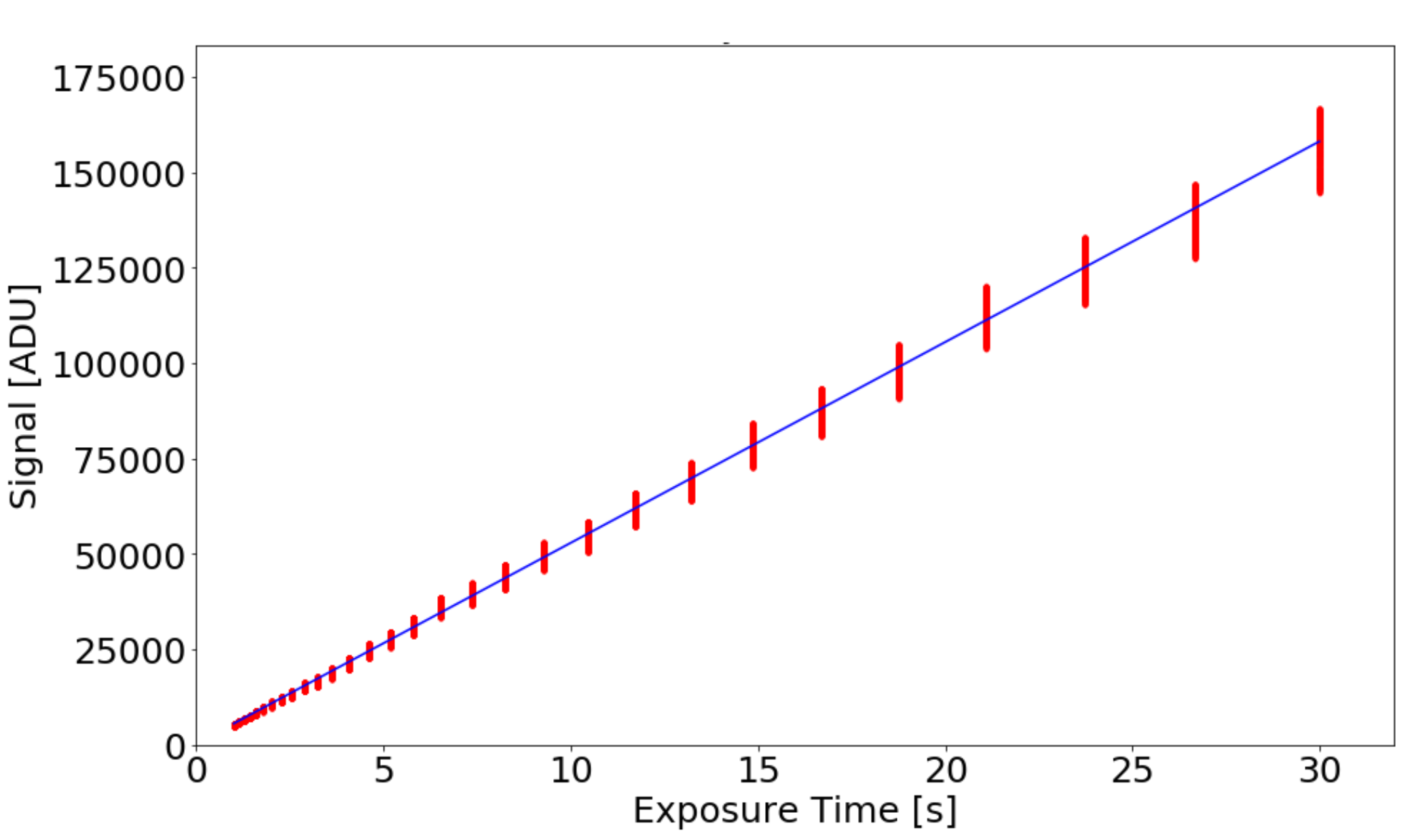}
\caption{\label{fig:lin} 
Response of the first amplifier of CCD 1 as a function of the exposure time. The dispersion in the counts at a given exposure time is due to the vignetting. Around an exposure time of 30 seconds the ADC is saturated, arriving to a constant signal.}
\end{figure}

The linearity curve for the first amplifier of CCD 1 is shown in Figure \ref{fig:lin},  presenting the signal versus the exposure time. After about 30 seconds the ADC is saturated. The  dispersion in each point in the graph is due to vignetting. The Photo-Response-Non-Uniformity (PRNU) of all CCDs is below 1\%.

\subsection{Image quality}

\begin{figure}[t]
\epsscale{1.2}
\plotone{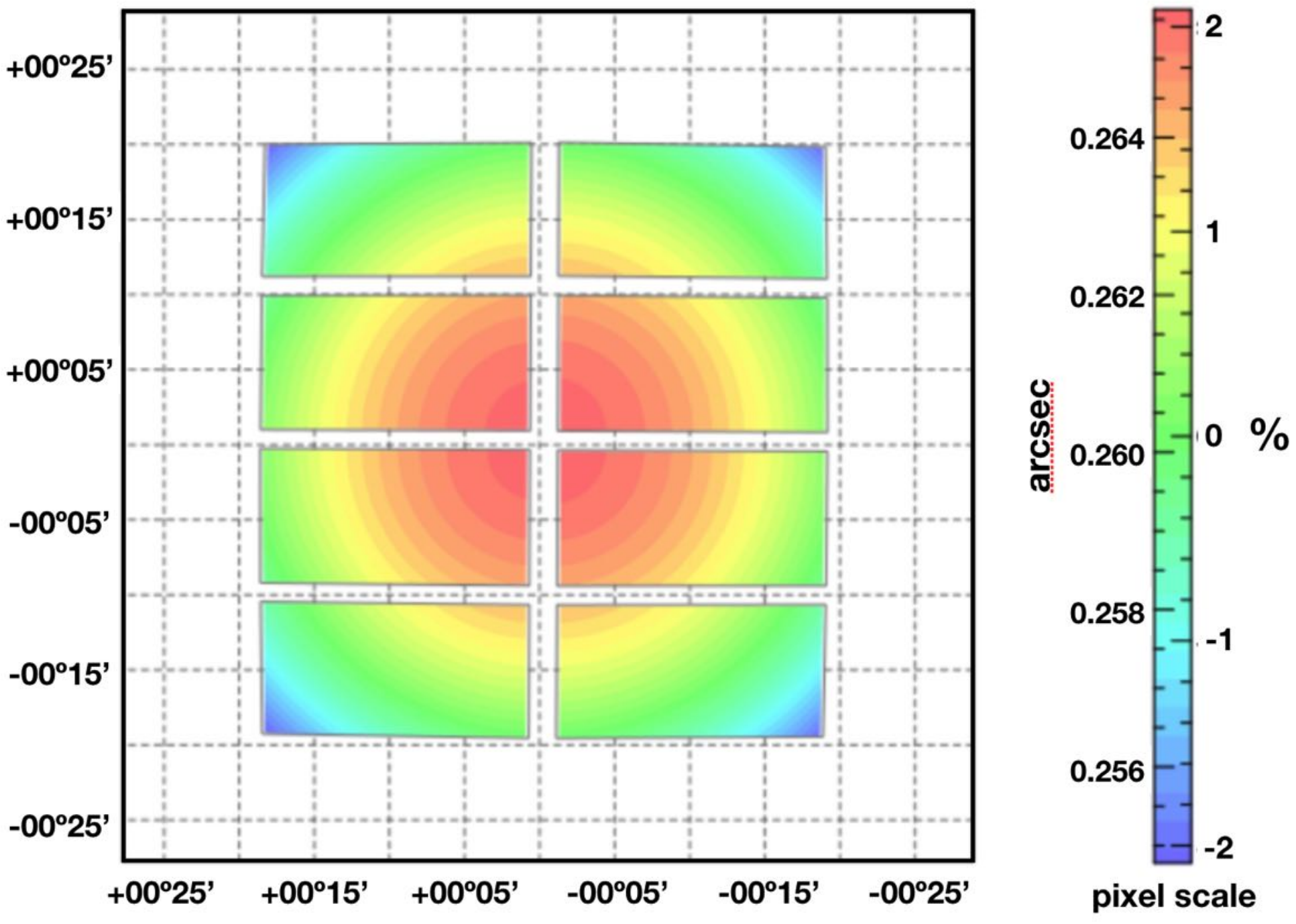}
\caption{ \label{fig:DistorsionMap} 
PAUCam focal plane distortion map in the eight central CCDs. 
}
\end{figure} 

The image quality delivered by PAUCam has the contributions of the optical system and that of the camera itself. The WHT prime focus corrector was not optimized to cover the wide field of view of the PAUCam focal plane. It delivers a non-flat focal plane introducing distortion in the PAUCam images. By now, we have collected enough images to be able to map this distortion accurately. 
Figure~\ref{fig:DistorsionMap} shows the average pixel scale across the eight central CCDs. This map has been calculated using reference stars during the astrometric solutions calculated in the main data reduction pipeline and provides an accurate representation of the distortion in the focal plane. The optics provide this smooth pixel scale pattern which is dominated by the prime focus corrector contribution. 

The best images we have measured with PAUCam have PSFs of around 0.55"-0.60" FWHM. The external seeing monitors measure FWHMs of around 0.30" when this happens. This would imply that the system introduces a contribution of approximately 0.50" to the observed PSF. 

\subsection{Astrometry}

\begin{figure}[t]
\epsscale{1.12}
\plotone{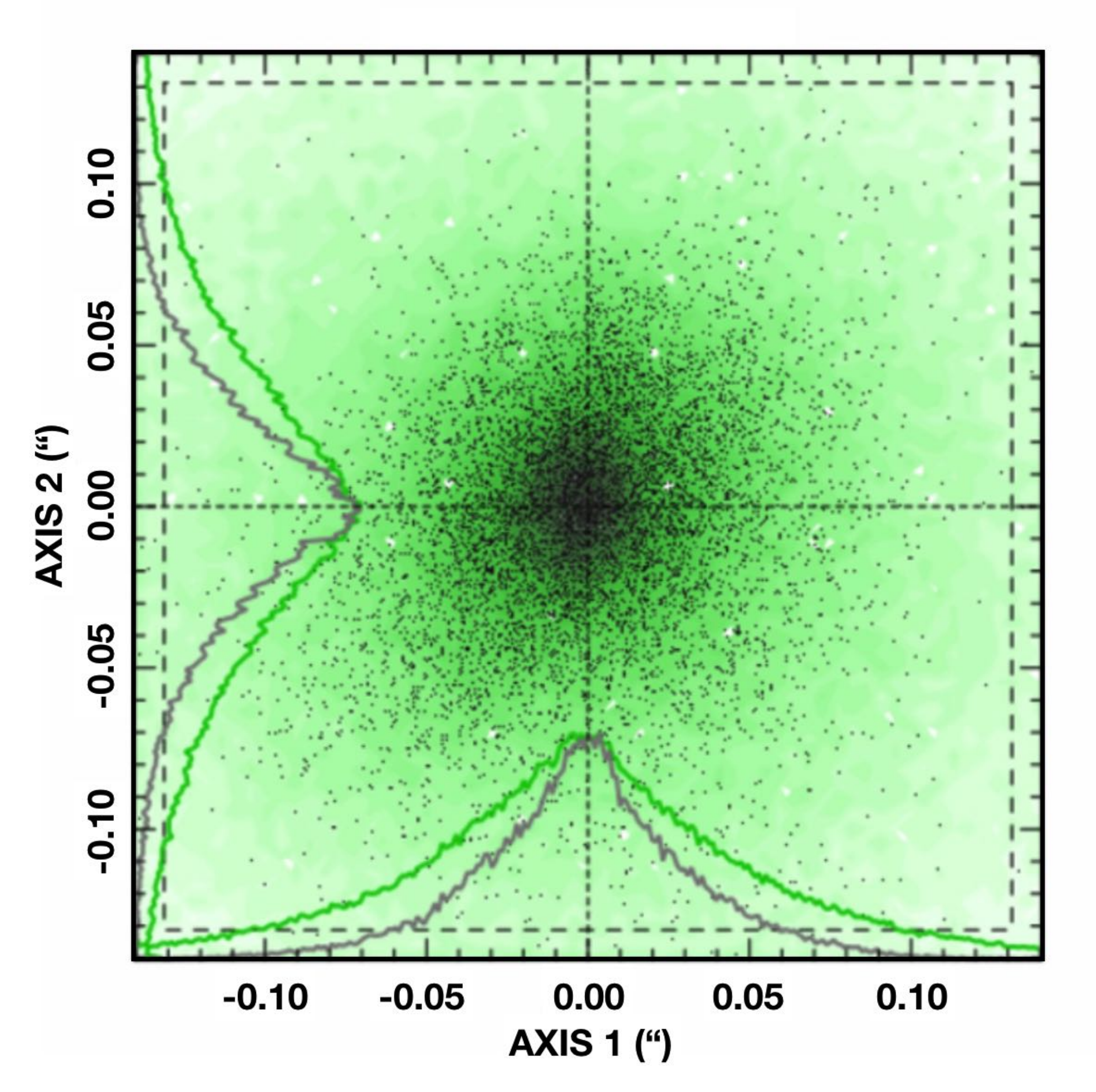}
\caption{ \label{fig:AstrometryResiduals}
Astrometry residuals bewteen repeated observations.
}
\end{figure} 

We calibrate our astrometry against external stellar catalogues. Currently, we use the Gaia DR2~ \citep{Gaia, Gaiadr2}. We also check our internal consistency with repeated observations of the same fields. Figure~\ref{fig:AstrometryResiduals} shows the internal astrometric residuals between overlapping exposures. We find an average RMS value below 50 mas, significantly smaller than the average pixel scale (shown with a dashed line). The accuracy achieved allows precise photometry even in the more distorted regions of the focal plane.

\subsection{Photometry}

\begin{figure}[t]
\epsscale{1.2}
\plotone{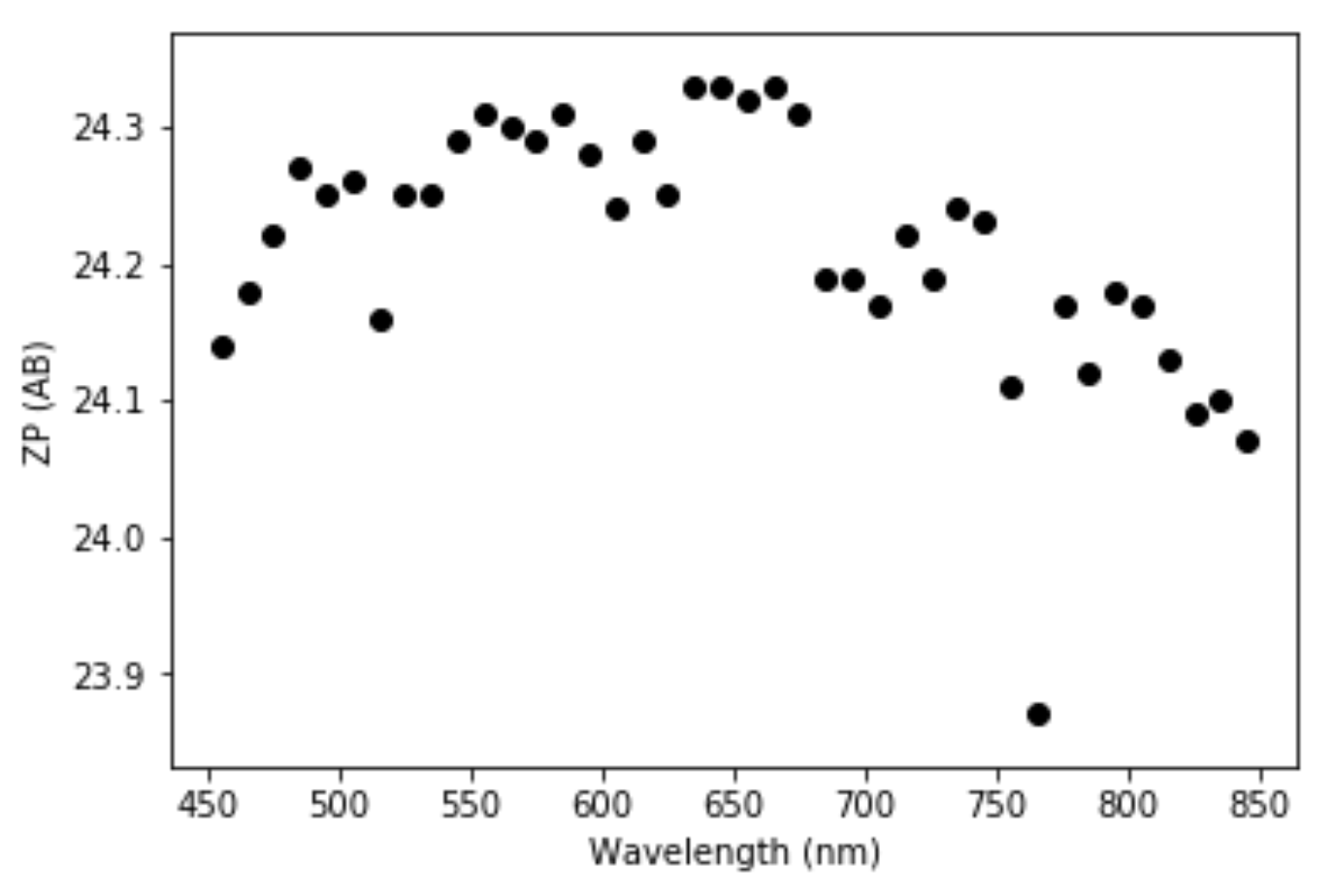}
\caption{ \label{fig:zeropoints} 
PAU Cam Zero points as a function of the middle wavelength of the narrow band filters.}
\end{figure}

PAUCam was designed to obtain accurate photometric redshifts for galaxies using photometry in narrow band filters. The relative photometric calibration is therefore very important to obtain robust spectral energy distributions of the galaxies. The photometric calibration is extensively discussed in \citet{PAUCalibration}. Here we outline the most salient points. The most contributing factors to the error on the photometry are the derivation of the zero points and the way the photometry is performed. The zero points (the conversion factor from electrons detected in the detector to fluxes or magnitudes) reflect the sensitivity of the whole system. We compute the zero point for each image from stellar photometry comparison to SDSS. We fit stellar templates to the SDSS photometry to compute stellar SEDs. We then synthesize the PAUCam narrow band photometry from the SED fit and compare the fluxes to the ones measured. We obtain the image zero point as the factor that we need to apply to convert the observed fluxes to the synthetic ones for each single image.  We also compute global zero points from spectrophotometric standards photometry. These need to include the observing conditions to be applied to each image. These zero points are the basis for our Exposure Time Calculator. Figure~\ref{fig:zeropoints} shows the computed zero points for the narrow band filters. The zero points for the broad band filters are similar to those of the Dark Energy Camera (DECam) at the CTIO $4 \unit{m}$  Blanco Telescope. This is expected as both camera have similar detectors and are mounted on similar size telescopes.

\begin{figure*} [t]
\epsscale{1.18}
\plotone{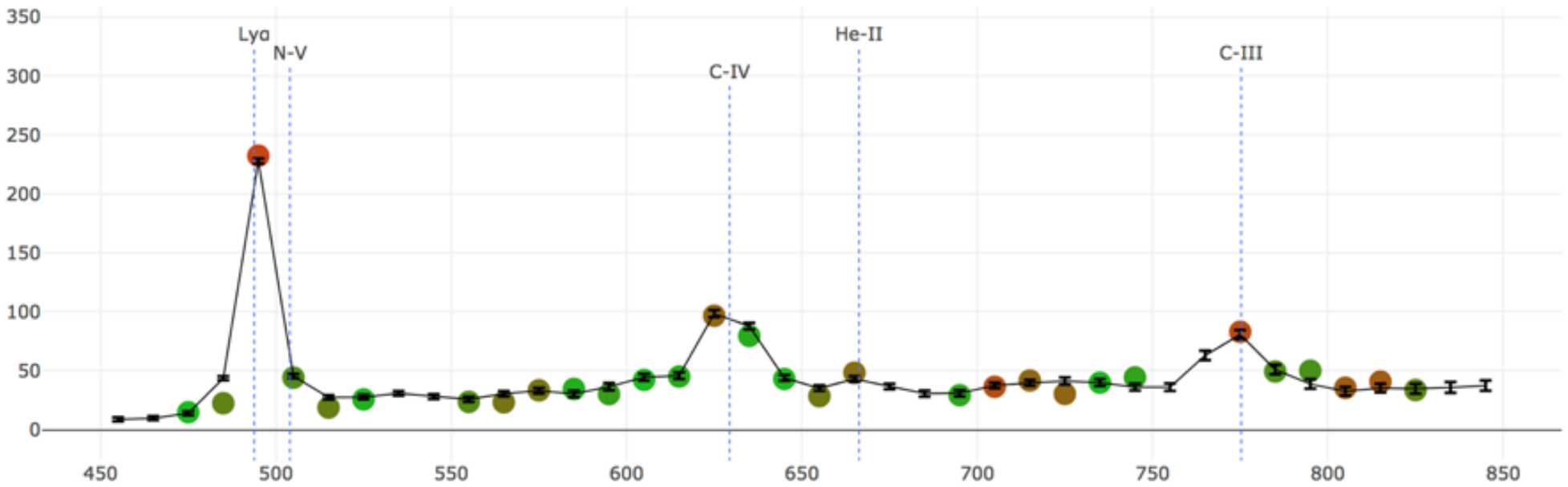}
\plotone{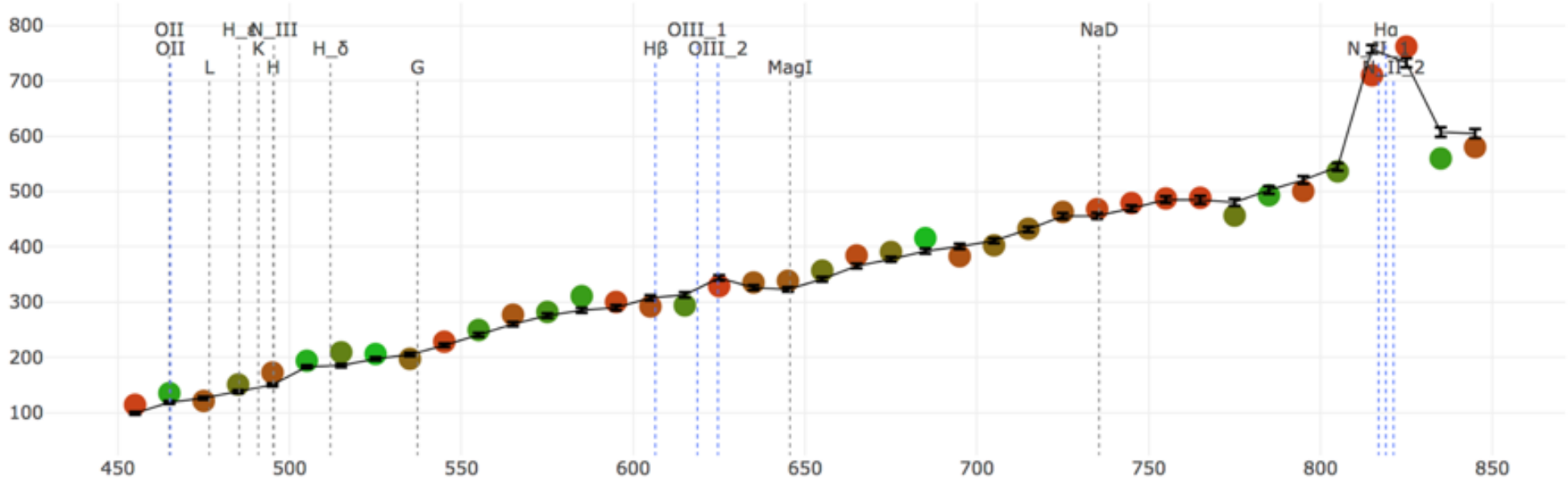}
\caption{ \label{fig:Galaxy} 
Spectral energy distributions of a QSO (top panel) and a emission line galaxy (bottom panel). Emission lines are clearly detected in both objects. The PAUCam measured fluxes are marked with circles while the SDSS spectra, binned at the narrow band resolution, are shown with a solid lines. 
}
\end{figure*}

\begin{figure*} [t]
\epsscale{1.18}
\plotone{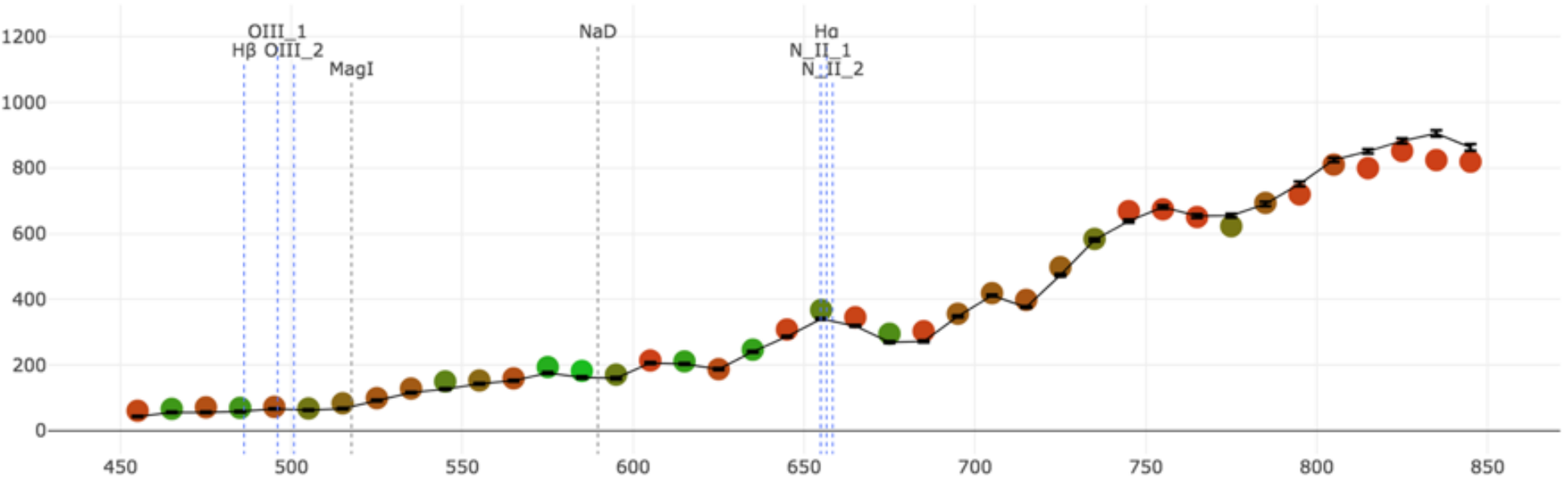}
\plotone{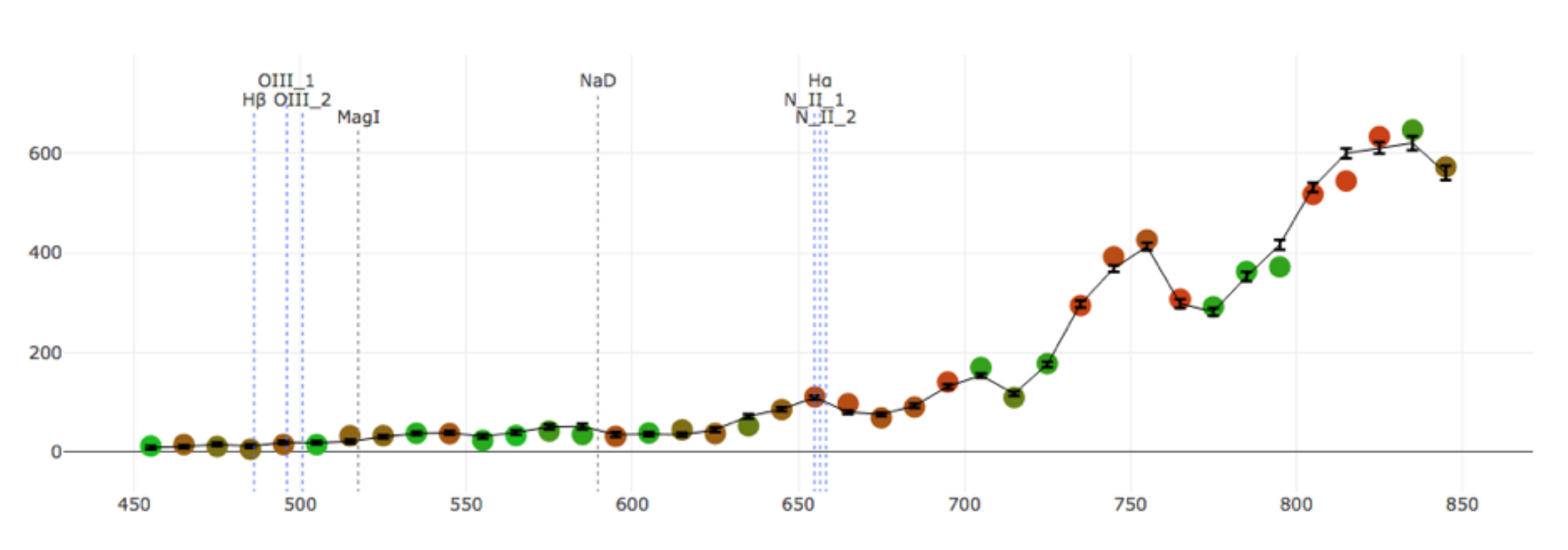}
\plotone{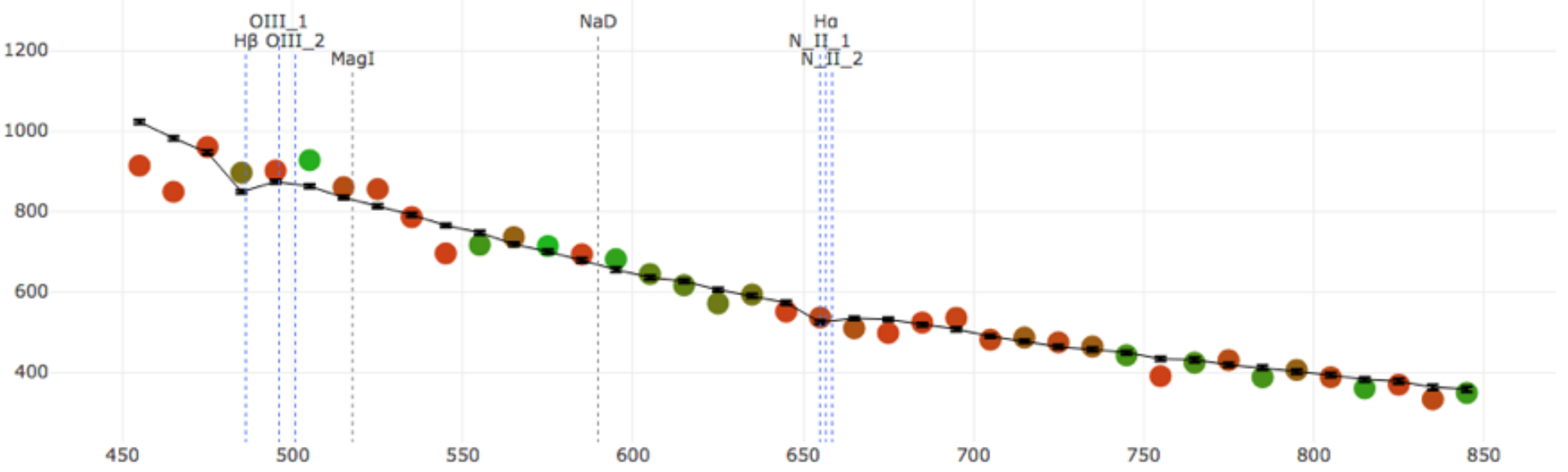}
\caption{ \label{fig:RedstarMag20} 
Spectral energy distributions of two late type stars (top and middle panels) and a blue star (bottom panel). The PAUCam measured fluxes are marked with circles while the SDSS spectra, binned at the narrow band resolution, are shown with a solid lines. 
}
\end{figure*}

The relative calibration of the image zero points of all filters is necessary to build reliable spectral energy distributions. Figures~\ref{fig:Galaxy} and~\ref{fig:RedstarMag20} show examples of quasar, galaxy and stellar spectral energy distributions. The PAUCam measurements (circles) are compared to the corresponding SDSS spectra  binned at the same resolution (solid line). The agreement between both measurement demonstrates the capabilities of narrow band imaging to obtain accurate spectral energy distributions.

The photometric calibration has allowed us to measure the photo-z accuracy in the COSMOS field \citep{Photoz} whose value is 
 $\sigma_{68} / (1+z) \simeq
0.0035$ for $i_{\mathrm{AB}} < 22.5$, which agrees with the design specifications of the PAU Camera.

\section{SUMMARY}
\label{sec:summary}

We have built a new instrument (PAUCam) covering the $\sim$1 degree FoV of the William Hershel Telescope prime focus with the aim to carry out a Narrow Band photometric survey (PAUS) and perform  galaxy intrinsic alignments, galaxy clustering and galaxy evolution studies. The PAU Camera saw first light in June 2015 and has been routinely observing at the WHT covering a total of $\sim 15.5 \unit{deg^2} $ in the COSMOS, W1, W2, W3 and W4 fields. Additionally the camera has also served other astronomical projects outside of the PAU Survey Collaboration.

To fulfill the scientific and installation requirements we have opted for several innovative technologies not commonly used in astronomical instrumentation. The camera is equipped with 40 Narrow Band filters ranging from $\sim\! 4500$ to $\sim\! 8500$ Angstroms installed in a movable filter exchange system inside the camera cryostat (in vacuum and cryogenic conditions). To minimize weight, the cryostat and most of the mechanical parts are made of carbon fiber. The camera is equipped with a dual cooling system to ensure it is operational the same day it is installed. All the front end electronics and camera services have been thoroughly designed to minimize the electronics noise and accommodate to the WHT infrastructure. The control and operation software has been designed to minimize the observer intervention, ensure the survey is carried out automatically and fast monitoring is available online. The entire design of PAUCam, the complete fabrication, except for very specific parts, and all the software were done in house, at the labs of the groups named in the authors list.

We have also built a system that allows the data to be transferred daily to PIC, where it is analyzed on a daily basis to ensure fast feedback during the observation periods.
After thorough calibration, the SEDs obtained with the camera reproduce SDSS spectra of the same objects, therefore demonstrating its capabilities to adequately sample SEDs. The photo-z precision has been measured to be  $\sigma_{68} / (1+z) \simeq
0.0035$ for $i_{\mathrm{AB}} < 22.5$, which is the design goal for the PAU Camera.


\acknowledgments

We thank the Management of the ING and the staff of the WHT for their invaluable help during the design, commissioning and operations of PAUCam. We also thank our colleagues in Bonn University, Durham University, Leiden Observatory, University College of London and ETH-Zurich, for their contributions to the ongoing PAU Survey (PAUS) project. The PAUCam Team was supported in part by the Mineco Consolider Ingenio 2010 project CSD2007-00060, and by the Mineco Grants CSD2007-00060, AYA2015-71825, ESP2015-66861, FPA2015-68048, SEV-2016-0588, SEV-2016-0597, and MDM-2015-0509, some of which include ERDF funds from the European Union. IEEC and IFAE are partially funded by the CERCA program of the Generalitat de Catalunya. The PAU data center is hosted by the Port d'Informació Científica (PIC), maintained through a collaboration of CIEMAT and IFAE, with additional support from Universitat Autònoma de Barcelona and ERDF funds.

%

\vspace{5mm}
\facility{ORM(WHT)}


\software{
		  PanView \citep{2010SPIE.7740E..1KH}
          SExtractor \citep{1996A&AS..117..393B}
          PSFEx \citep{2011ASPC..442..435B}
          RTree \citep{Rtree}
          }

\end{document}